\documentclass[11pt]{article}
\usepackage{amsfonts}
\usepackage{amsmath}
\usepackage{amssymb}
\usepackage{mathtools}
\usepackage{graphicx}
\usepackage{geometry}
\usepackage{listings}
\usepackage{mathrsfs}
\usepackage{natbib}
\usepackage{bm}
\usepackage{subcaption}
\usepackage{multirow}
\usepackage{hyperref}
\usepackage{booktabs, caption, makecell}

\usepackage{threeparttable}
\usepackage{enumitem}
\usepackage{color, colortbl}
\usepackage{float}

\usepackage[title]{appendix}
\definecolor{Gray}{gray}{0.5}

\usepackage{algorithm,algcompatible}
\algnewcommand\INPUT{\item[\textbf{Input:}]}%
\algnewcommand\OUTPUT{\item[\textbf{Output:}]}%
\newcommand{\email}[1]{\href{mailto:#1}{\normalfont\texttt{#1}}}

\usepackage{xr}
\makeatletter
\newcommand*{\addFileDependency}[1]{
  \typeout{(#1)}
  \@addtofilelist{#1}
  \IfFileExists{#1}{}{\typeout{No file #1.}}
}
\makeatother

\newcommand*{\myexternaldocument}[1]{%
    \externaldocument{#1}%
    \addFileDependency{#1.tex}%
    \addFileDependency{#1.aux}%
}
\myexternaldocument{supplementary}

\providecommand{\U}[1]{\protect\rule{.1in}{.1in}}

\newtheorem{theorem}{Theorem}
\newtheorem{corollary}{Corollary}
\newtheorem{lemma}{Lemma}

\newtheorem{remark}{Remark}
\newenvironment{proof}[1][Proof]{\noindent\textbf{#1.} }{\ \rule{0.5em}{0.5em}}
\numberwithin{equation}{section}
\begin{document}

\title{Mixture composite regression models with multi-type feature selection}

\author{Tsz Chai Fung\thanks{Address: RiskLab, ETH Zurich, 8092 Zurich, Switzerland. Email: \email{tszchai.fung@math.ethz.ch} (T.C. Fung); \email{mario.wuethrich@math.ethz.ch} (M.V. W\"{u}thrich).} \thanks{Address: Department of Risk Management and Insurance, Georgia State University, Atlanta, GA 30303. Email: \email{tfung@gsu.edu}.}
\and George Tzougas\thanks{Address: Department of Statistics, London School of Economics and Political Science, London WC2A 2AE, UK. Email: \email{g.tzougas@lse.ac.uk}.}\\
\and Mario V. W\"{u}thrich\footnotemark[1]
}

\maketitle

\abstract{The aim of this paper is to present a mixture composite regression model for claim severity modeling.
Claim severity modeling poses several challenges such as multimodality, tail-heaviness and systematic
effects in data.
We tackle this modeling problem by studying a mixture composite regression model for simultaneous modeling of
attritional and large claims, and for considering
systematic effects in both the mixture components as well as the mixing probabilities. 
For model fitting, we present a group-fused regularization approach that allows us for selecting the explanatory variables 
which significantly impact the mixing probabilities and the different mixture components, respectively. 
We develop an asymptotic theory for this regularized estimation approach, and fitting is performed
using a novel Generalized Expectation-Maximization algorithm. We exemplify our approach on a real motor insurance data set.}

\medskip


\section{Introduction}
Insurance claim severity modeling is a very challenging problem in actuarial science. Motivated by a Greek Motor Third Party Liability (MTPL) insurance data set which is further described in Section \ref{sec:data_descr:data}, we observe that insurance claim severity data sets often exhibit several peculiar characteristics:
Firstly, claim severity distributions are often multimodal, coming from the fact that there
are systematic effects in the data due to unobserved heterogeneity and latent factors such as different
claim types. 
Secondly, a claim severity distribution ranges over several magnitudes, from small attritional claims to large claim events, which often exhibits a heavy-tailed nature and a mismatch between body and tail behavior. 
Thirdly, insurance data are often accompanied by multiple types of policyholder attributes, including several continuous variables (e.g. driver's age), ordered categorical variables (e.g. sum insured categories) and nominal categorical variables (e.g. car brand). These variables may have different explanatory powers to different parts of the severity distribution.

For insurance pricing, reserving and risk management, it is crucial to have accurate descriptions of claim severity distributions, and to understand clearly the influence of policy attributes to the claim distribution. Therefore, it is essential to devise a claim severity modeling framework which possesses all of the following features to address the aforementioned modeling challenges:
\begin{enumerate}
\item \textbf{Distributional multimodality}: The model must enable sufficient flexibility to capture distributional multimodality.
\item \textbf{Tail-heaviness}: The severity distribution has to be heavy-tail in nature and it should allow for robust estimation of tail-heaviness.
\item \textbf{Covariates influence}: The model needs to capture the covariates influence on various parts of the severity distribution, including: (i) the probabilities which assign observations into various clusters or nodes (clustering probabilities), (ii) systematic effects in claim severity distributions conditioned on each cluster (body part) and (iii) the tail-heaviness of the distribution (tail part).
\item \textbf{Variable selection}: Realizing that not all variables are important, a variable selection strategy must also be employed to determine which variables are influential to which of the aforementioned three parts of the severity distributions. Moreover, the strategy has to be adapted to a multi-type variable setting.
\end{enumerate}

There are several actuarial research works in addressing each of the aforementioned modeling requirements. To capture distributional multimodality and tail-heaviness (points 1 and 2), there are three main streams of claim severity modeling approaches described as follows.

\begin{itemize}
\item \textbf{Finite mixture models} constitute an easily extendable model class for approximating general distribution functions in a semi-parametric way and accounting for unobserved heterogeneity. Notable actuarial contributions include e.g. \cite{LEE2010modeling}, \cite{tzougas2014optimal}, \cite{MILJKOVIC2016387} and \cite{FUNG2020MoESeverity}. Recent works such as \cite{tzougas2018bonus} and \cite{BLOSTEIN201935} combine both light-tailed and heavy-tailed distributions to cater for the mismatch between body and tail behavior. In this case, however, we will show empirically in Section \ref{sec:data_descr:fit} that the tail estimation is unrobust due to the overlapping density region between the body and the tail where the small claims can severely impact the estimated tail index.
\item \textbf{Composite or splicing models} comprise of light-tailed distributions up to a threshold for modeling moderate losses, and heavy-tailed distributions beyond the threshold for large losses, to potentially capture the mismatch between the body and the tail behavior. Composite models also address the tail unrobustness problem inherited from finite mixture models because there is no density overlap between body and tail parts. The relevant actuarial literature includes, for instance, \cite{cooray2005modeling}, \cite{scollnik2007composite}, \cite{pigeon2011composite}, \cite{scollnik2012modeling}, \cite{nadarajah2014new}, \cite{bakar2015modeling}, \cite{calderin2016modeling}, \cite{grun2019extending} and \cite{Parodi2020}.
\item \textbf{Combinations of finite mixtures and composite models} are proposed by \cite{reynkens2017modelling} who developed a global fitting strategy for a splicing model with an Erlang mixture distribution for the body and  a Pareto distribution for the tail.
\end{itemize}

To understand how the claim severity distribution is influenced by certain risk factors, covariates influence (point 3) has been extensively explored in actuarial literature using various types of severity regression models (\cite{nelder1972generalized}). We refer readers to \cite{frees2009regression} for an extensive summary.

Considering variable selection techniques (point 4), a popular approach is the use of penalty functions, such as LASSO, see \cite{tibshirani1996regression}, or SCAD, see \cite{fan2001variable}, to shrink unimportant regression coefficients to zero. In actuarial literature, \cite{jeong2020non} used non-convex regularization methods in order to obtain stable estimation of loss development factors in insurance claims reserving. In multi-type variable setting, \cite{devriendt2020sparse} is currently the only paper which considers multi-type feature selection under a Poisson regression framework for claim frequency modeling.

While the existing literature address some of the aforementioned claim severity modeling needs, we are still lacking a universal modeling framework, which not only provides versatility to fit a multimodal heavy-tailed severity distribution, but also explains the covariates' influence on multiple distributional parts with variable selections. As a result, the goal of this paper is to integrate, adapt and extend the existing modeling techniques, and devise a universal insurance claim severity modeling framework which simultaneously address all of the four modeling needs mentioned above. To this end, we make the following contributions:

Firstly, we introduce a mixture composite regression model for approximating claim severities based on the use of 
available covariate information. This extends the setup of \cite{reynkens2017modelling}, who used a finite mixture distribution for the body and a Pareto-type distribution for the tail without using covariates, by incorporating covariates impacts on all three parts of the severity distribution: clustering probabilities, body part and tail part. 

Secondly, we propose a group-fused regularization approach for variable selection. This approach allows us to select three different sets of variables which significantly impact the previously mentioned three parts of the claim severity distribution respectively. The set of variables chosen is homogeneous across all mixture components to preserve model interpretability. Furthermore, this approach enables regularization under multi-type variable settings.

Thirdly, we develop an asymptotic theory for the regularization approach. The following two main results theoretically justify the appropriateness of the proposed method: (i) The proposed method is consistent in terms of feature selection, 
in particular, as sample size goes to infinity, the proposed method will correctly merge and shrink regression coefficients across the various modeling parts. (ii) The parameters of the reduced model, after merging and shrinking the regression coefficients, are asymptotically normal with zero mean, and their variances are the same as the parameter uncertainties
obtained by fitting the same mixture composite regression model to the reduced model (e.g.~mean claim severity). The implication of the above two main results is that we can construct Wald-type confidence intervals and Efron percentile bootstrap confidence intervals of model parameters and other quantities of interest.

Finally, we present a novel Generalized Expectation-Maximization (GEM) algorithm for estimating the parameters of the proposed model with parameter regularization.
The GEM algorithm is demonstrated to perform satisfactorily when the mixture-Gamma Lomax composite regression model is fitted to a Greek MTPL dataset which inherits all the previously described features.

The remainder of this article proceeds as follows. In Section 2, we introduce the framework of the mixture composite regression model. Section 3 presents the feature selection approach which can be used for selecting important variables for explaining the claim severity distribution in the presence multi-type covariates.
In Section 4 we provide the theoretical foundations, such as consistency and asymptotic normality, upon which the proposed feature selection approach is based for merging and shrinking parameters correctly with high probability when the sample size is large. Furthermore, we develop Wald type and bootstrap two-sided confidence intervals for the parameters. The maximum likelihood estimation (MLE) procedure for our proposed model via the GEM algorithm is presented in Section 5. In Section 6, we describe the MTPL dataset that we use for our empirical analysis, and provide estimation and model comparison for various benchmark distributions. In Section 7, we fit the proposed mixture composite distribution and subsequently the mixture composite regression model with feature regularizations. Concluding remarks are given in Section 8, and other miscellaneous details are included in the Supplementary materials.

\section{Modeling framework} \label{sec:model}
This section summarizes the features that are incorporated in a regression modeling framework for addressing the challenges encountered in  claim severity datasets in general insurance.
In particular, motivated by the characteristics of the multimodal and heavy-tailed Greek MTPL insurance dataset
studied below, we propose the following mixture composite regression model.

Let $Y\in\mathbb{R}^{+}$ be the claim severity random variable, and let  $\bm{x}\in\mathbb{R}^{D}$ be the vector of covariate information\footnote{ Note also that all vectors are assumed to be column vectors.}. The density of the mixture composite regression model is given by

\begin{align} \label{eq:density}
h_{Y}(y;\bm{\alpha},\bm{\beta},\bm{\phi},\theta,\bm{\nu},\bm{x})
&=\sum_{j=1}^g\pi_j(\bm{x};\bm{\alpha})\frac{f(y;\exp\{\bm{\beta}_j^T\bm{x}\},\phi_j)}{F(\tau;\exp\{\bm{\beta}_j^T\bm{x}\},\phi_j)}1\{y\leq \tau\}\\\nonumber
&\qquad \qquad + ~\pi_{g+1}(\bm{x};\bm{\alpha})\frac{h(y;\theta,\exp\{\bm{\nu}^T\bm{x}\})}{1-H(\tau;\theta,\exp\{\bm{\nu}^T\bm{x}\})}1\{y>\tau\},
\end{align}
where $\pi_j(\bm{x};\bm{\alpha})$, $1\le j \le g+1$, are covariate-dependent component weights given by
\begin{equation}
\pi_j(\bm{x};\bm{\alpha})=\frac{\exp\{\bm{\alpha}_j^T\bm{x}\}}{\sum_{j'=1}^{g+1}\exp\{\bm{\alpha}_{j'}^T\bm{x}\}}, 
\end{equation}
with $\bm{\alpha}_{g+1}=\bm{0}$ for model identifiability, and $\bm{\alpha}=(\bm{\alpha}_1,\ldots,\bm{\alpha}_{g+1})\in\mathbb{R}^{D\times (g+1)}$. $f$ and $h$ are the body and tail density functions respectively, such that the first $g$ mixture components are specialized in capturing small to moderate claim amounts while the last component focuses on extreme claims. $F$ and $H$ are the corresponding cdfs.

In this paper, we specify $f$ and $h$ as Gamma (body) and Lomax (tail, also called Pareto type II) density functions given by, respectively,
\begin{equation} \label{eq:gamma}
f(y;\mu,\phi)=\frac{(\phi\mu)^{-1/\phi}}{\Gamma(1/\phi)}y^{1/\phi-1}e^{-y/(\phi\mu)},
\end{equation}
and 
\begin{equation} \label{eq:lomax}
h(y;\theta,\eta)=\frac{\eta\theta^{\eta}}{\left(y+\theta\right)^{\eta+1}}.
\end{equation}

The choice of Gamma density is motivated by its light-tailed and uni-modal characteristics to capture small to moderate claims. Also, mixture of Gammas provides sufficient flexibility to capture complex distributional structures like multimodality, thanks to the deneness property of Gamma mixture. The choice of the Lomax density for the tail is motivated by its polynomial tail characteristics with the tail index $\eta$ describing the tail-heaviness of the distribution. The analytical form of the truncated Lomax distribution also makes the model estimation procedures computationally desirable.
Note however that one may choose other plausible model specifications as long as $f$ is a unimodal light-tailed distribution while $h$ is a heavy-tailed distribution. To avoid distorting the focus of this paper and given that the fitting results of the real dataset (Section \ref{sec:application}) are satisfactory, we leave the comparisons among various model specifications as a future research direction.

Furthermore, $\bm{\beta}=(\bm{\beta}_1,\ldots,\bm{\beta}_g)\in\mathbb{R}^{D\times g}$ and $\bm{\nu}\in\mathbb{R}^{D}$ are the regression coefficients for the body and tail distributions, respectively. The proposed distribution is characterized by a splicing threshold $\tau>0$, which is predetermined using expert opinion via performing e.g.~extreme value analysis instead of treated it  as a parameter estimated by a likelihood approach; this is mainly motivated by estimation stability and is adopted by e.g. \cite{reynkens2017modelling}. 

The mean of $Y|\bm{x}$ is given by
\begin{equation}
E[Y|\bm{x}]=\sum_{j=1}^g\pi_j(\bm{x};\bm{\alpha})\frac{F(\tau;\exp\{\bm{\beta}_j^T\bm{x}\},\phi_j/(1+\phi_j))}{F(\tau;\exp\{\bm{\beta}_j^T\bm{x}\},\phi_j)}\exp\{\bm{\beta}_j^T\bm{x}\}+\pi_{g+1}(\bm{x};\bm{\alpha})\left(\frac{\theta+\tau}{\exp\{\bm{\nu}^T\bm{x}\}-1}+\tau\right).
\end{equation}
The composite model in Equation (\ref{eq:density}) can alternatively be regarded as a mixture of $g$ right truncated Gamma distributions for the body and a left truncated Lomax distribution for the tail. Each claim is classified to one of the $g+1$ subgroups ($g$ subgroups for body and one subgroup for tail) with probabilities $\{\pi_j(\bm{x};\bm{\alpha})\}_{j=1,\ldots,g+1}$, where each subgroup may correspond to a different claim sub-type. Regression coefficients $\bm{\alpha}$ explains the heterogeneity of the assignment probabilities across different claims, while $\bm{\beta}$ explain the systematic
effects of the claims within a given subgroup. The regression coefficients $\bm{\nu}$ for the tail distribution, on the other hand, capture the effect of covariates to the tail-heaviness of claims.

The motivation of introducing a composite model in Equation (\ref{eq:density}) instead of using a mixture-Gamma Lomax model is that there are no overlapping density regions between the body and tail distributions under the proposed framework. 
We will show in our motivating application in Section \ref{sec:data_descr} that this results in a more robust and stable estimation of the tail index, since it is not distorted by attritional claims from the body of the distribution. One should however note that the mixture probabilities connect tail and body regression parameter estimation, i.e., the proposed composite regression model does not decouple into independent estimation parts.

\section{Feature selection method} \label{sec:feature_sel}
In this section, we propose the group fused penalty approach to select the variables 
to describe the systematic effects in claim severities under a multi-type covariates setting. 
We will select three potentially different sets of variables that may influence, respectively, the subgroup probabilities, body and tail of the distribution. For the sake of model interpretability, we select the same set of variables for all
mixture components of the body of the data and the mixing probabilities.

Suppose there are $n$ independent claims $\bm{Y}=(Y_1,\ldots,Y_n)^T$, and denote their
realizations by $\bm{y}=(y_1,\ldots,y_n)^T$. For each claim $i=1,\ldots,n$, we have a covariates vector $\bm{x}_i=(x_{i1},\ldots,x_{iD})^T\in \mathbb{R}^D$ with $x_{i1}=1$ (for the intercept component). Define $\bm{X}=(\bm{x}_1,\ldots,\bm{x}_n)^T\in
\mathbb{R}^{n\times D}$ as the design matrix containing the covariate information of all $n$ observations. The observed data log-likelihood is given by
\begin{equation} \label{eq:lik}
\mathcal{L}_n(\bm{\Phi}):=\mathcal{L}_n(\bm{\Phi};\bm{y},\bm{X})=\sum_{i=1}^n\log h_{Y}(y_i;\bm{\alpha},\bm{\beta},\bm{\phi}, \theta,\bm{\nu},\bm{x}_i),
\end{equation}
where $\bm{\Phi}$ contains all model parameters. To incorporate variable selection, we propose a group fused regularization approach, where the penalty function for the regression parameters is as follows
\begin{equation} \label{eq:pen}
\mathcal{P}_n(\bm{\Phi})=
P_{\bm{\lambda}_1,n}(\bm{\alpha})
+P_{\bm{\lambda}_2,n}(\bm{\beta})
+P_{\bm{\lambda}_3,n}(\bm{\nu}),
\end{equation}
with $P_{\bm{\lambda}_1,n}(\bm{\alpha})$, $P_{\bm{\lambda}_2,n}(\bm{\beta})$ and $P_{\bm{\lambda}_3,n}(\bm{\nu})$ being the penalty functions on the regression parameters $\bm{\alpha}\in\mathbb{R}^{D\times (g+1)}$, $\bm{\beta}\in\mathbb{R}^{D\times g}$ and $\bm{\nu}\in\mathbb{R}^{D}$. These are given by
\begin{equation*}
P_{\bm{\lambda}_1,n}(\bm{\alpha})=\sum_{k=1}^{K_1}p_{1n}\left(\big\|\bm{c}_{1k}^T\bm{\alpha}\big\|_2;\lambda_{1kn}\right),\qquad
P_{\bm{\lambda}_2,n}(\bm{\beta})=\sum_{k=1}^{K_2}p_{2n}\left(\big\|\bm{c}_{2k}^T\bm{\beta}\big\|_2;\lambda_{2kn}\right),
\end{equation*}
\begin{equation} \label{eq:pen_func}
P_{\bm{\lambda}_3,n}(\bm{\nu})=\sum_{k=1}^{K_3}p_{3n}\left(\big|\bm{c}_{3k}^T\bm{\nu}\big|;\lambda_{3kn}\right),
\end{equation}
where $\|\cdot\|_2$ is the $L^2$-norm, $\lambda_{1kn}$, $\lambda_{2kn}$ and $\lambda_{3kn}$ are penalty tuning parameters, $p_{1n}$, $p_{2n}$ and $p_{3n}$ are concave non-decreasing penalty functions (which will be chosen proportional to the sample size $n$) and $K_1$, $K_2$ and $K_3$ correspond to the numbers of penalization terms. 
Finally, $\{\bm{c}_{lk}\}_{l=1,2,3}$ are predetermined vectors of penalty coefficients which allow for different types of penalties, including standard LASSO to shrink continuous variables, fused LASSO to merge regression coefficients of various ordinal categorical variables, and generalized fused LASSO to merge regression coefficients for nominal categorical variables. For a full description on constructing predetermined vectors for each type of variables (continuous, ordinal categorical and nominal categorical), we refer the reader to \cite{oelker2017uniform}, in the statistics literature, and \cite{devriendt2020sparse} in the actuarial literature.

The aim is to maximize the following objective function (penalized log-likelihood)
\begin{equation} \label{eq:lik_pen}
\mathcal{F}_n(\bm{\Phi})=\mathcal{L}_n(\bm{\Phi})-\mathcal{P}_n(\bm{\Phi}).
\end{equation}
We will use the following two commonly used penalty functions (for $l\in\{1,2,3\}$ and $\psi\geq 0$) to illustrate the usefulness of the proposed feature selection method:
\begin{itemize}
\item $L^1$-norm (Least Absolute Shrinkage and Selection Operator, LASSO) penalty: $p_{ln}(\psi;\eta)=n_l\eta\psi$;
\item \sloppy Smoothly Clipped Absolute Deviation (SCAD) penalty introduced by \cite{fan2001variable}: $p'_{ln}(\psi;\eta)=n_l\eta\left[1\{\psi\leq\eta\}+\frac{(a\eta-\psi)_{+}}{(a-1)\eta}1\{\psi >\eta\}\right]$, with $p_{ln}(0;\eta)=0$ and $a>2$ being a hyperparameter affecting the shape of the penalty function; we denote by $p'_{ln}(\psi;\eta)$ the first derivative of $p_{ln}(\psi;\eta)$ w.r.t.~$\psi$. Parameter $a$ is chosen as 3.7.
\end{itemize}
Here, we set $n_1=n$ for the total number of observations, $n_2:=n_b=\sum_{i=1}^n1\{y_i\leq\tau\}$ for the number of observations in the body, and $n_3:=n_t=\sum_{i=1}^n1\{y_i>\tau\}$ for  the number of observations in the tail.

Each of the parameters $\bm{\alpha}$ and $\bm{\beta}$ contains $g$ sets of regressors (one for each mixture component of the body, we initialize $\bm{\alpha}_{g+1}=\bm{0}$). For the sake of interpretability, the proposed group regularization method shrinks and merges regression coefficients of any variable uniformly across all mixture components, 
and the sets of variables does not vary across mixture components. Therefore, the proposed method allows 
us for choosing three sets of variables which significantly impact each of the three modeling parts -- 
the subgroup probabilities, the body part and the tail part of the severity distribution.

\begin{remark}
\normalfont Alternatively, one can adopt a fused penalization (ungrouped) instead of a grouped one in Equation (\ref{eq:pen_func}) for variable selection, resulting to different sets of variables being selected across mixture components. While this may provide more modeling flexibility, model interpretation will become more difficult when the number of mixture components $g$ becomes large, as we will expect slightly different shrinkage and mergence of variable levels across multiple mixture components.
\end{remark}

\section{Asymptotic properties} \label{sec:asym}
This section presents two asymptotic theorems regarding to the proposed mixture composite model with the feature selection method. Our motivation is two-fold: First, we want to theoretically justify the ability of the proposed feature selection approach in correctly merging and shrinking regression coefficients. Second, the theorems provide guidance to estimate model uncertainty. We will only present the key results and discuss their implications in this section. All construction details, including the assumptions and proof details, are postponed to Appendix \ref{apx:asym}. Suppose $Y_i$, given $\bm{x}_i$, is generated by the model of Equation (\ref{eq:density}) with true model parameter $\bm{\Phi}_0=(\bm{\alpha}_0,\bm{\beta}_0,\bm{\phi}_0,\theta_0,\bm{\nu}_0)$. We first have the following theorem:

\begin{theorem} \label{thm:consistency}
Assume \textbf{H1}-\textbf{H4} outlined in Appendix \ref{apx:asym_ass} hold for the penalty functions $p_{ln}(\psi;\lambda_{lkn})$. Let $\bm{V}_i=(Y_i,\bm{x}_i)$, $i=1,\ldots,n$, be a random sample from a density function $h(\bm{v};\bm{\Phi})$ that satisfies regularity conditions \textbf{R1}-\textbf{R5} outlined in Appendix \ref{apx:asym_reg}. Then, there exists a local maximizer $\hat{\bm{\Phi}}_n$ of the penalized log-likelihood function $\mathcal{F}_n(\bm{\Phi})$ for which $\|\hat{\bm{\Phi}}_n-\bm{\Phi}_0\|_2=O_P(n^{-1/2})$ as $n\rightarrow\infty$.
\end{theorem}

As discussed in Appendix \ref{apx:asym_ass}, both LASSO and SCAD penalty functions can be constructed to satisfy all assumptions \textbf{H1} to \textbf{H4}. Therefore, the above theorem says that the estimated parameters $\hat{\Phi}_n$ under the proposed model setup will converge to the true model parameters as $n\rightarrow\infty$. 

Apart from consistency, it is important to show sparsity of the proposed feature selection method, enabling consistent variable selection. To do so, we need to linearly transform the parameter space and formulate the asymptotic properties in the transformed space. We first define $\bm{C}_l=(\bm{c}_{l1},\ldots,\bm{c}_{lK_l})$ as a design matrix of penalty coefficients. Further, denote $\mathcal{Z}_1=\{k:\|\bm{c}_{1k}^T\bm{\alpha}_0\|_2=0\}$, $\mathcal{Z}_2=\{k:\|\bm{c}_{2k}^T\bm{\beta}_0\|_2= 0\}$ and $\mathcal{Z}_3=\{k:|\bm{c}_{3k}^T\bm{\nu}_0|= 0\}$, representing the regression coefficients to be merged or shrinked. W.l.o.g., we hereafter assume that under the true model, $\mathcal{Z}_l=\{1,2,\ldots,s_l\}$ for $l=1,2,3$, and we construct a reduced matrix $\bar{\bm{C}}_{\text{red},l}:=(\bm{c}_{l1},\ldots,\bm{c}_{lm_l})$ of the linearly independent vectors which span the space of the vectors $\{\bm{c}_{l1},\ldots,\bm{c}_{ls_l}\}$. Note that we always have $m_l\leq\min(s_l,P)$. Further, we construct linearly independent vectors $\bar{\bm{C}}_{\text{ind},l}:=(\bm{c}^*_{l,m_l+1},\ldots,\bm{c}^*_{l,D})$ which are also linearly independent of all vectors in $\bar{\bm{C}}_{\text{red},l}$. Then, define $\bar{\bm{C}}_l=(\bar{\bm{C}}_{\text{red},l},\bar{\bm{C}}_{\text{ind},l})$ which is a $D\times D$ full rank matrix, and define the transformed parameters $\bm{\alpha}^*=\bar{\bm{C}}_1^T\bm{\alpha}$, $\bm{\beta}^*=\bar{\bm{C}}_2^T\bm{\beta}$ and $\bm{\nu}^*=\bar{\bm{C}}_3^T\bm{\nu}$. Note that the transformed parameters can also be decomposed as $\bm{\alpha}^*=({\bm{\alpha}^{*}_{\text{red}}}^T,{\bm{\alpha}^{*}_{\text{ind}}}^T)^T$, $\bm{\beta}^*=({\bm{\beta}^{*}_{\text{red}}}^T,{\bm{\beta}^{*}_{\text{ind}}}^T)^T$ and $\bm{\nu}^*=({\bm{\nu}^{*}_{\text{red}}}^T,{\bm{\nu}^{*}_{\text{ind}}}^T)^T$, where $\bm{\alpha}^{*}_{\text{red}}=\bar{\bm{C}}_{\text{red},1}^T\bm{\alpha}$, $\bm{\alpha}^{*}_{\text{ind}}=\bar{\bm{C}}_{\text{ind},1}^T\bm{\alpha}$, $\bm{\beta}^{*}_{\text{red}}=\bar{\bm{C}}_{\text{red},2}^T\bm{\beta}$, $\bm{\beta}^{*}_{\text{ind}}=\bar{\bm{C}}_{\text{ind},2}^T\bm{\beta}$, $\bm{\nu}^{*}_{\text{red}}=\bar{\bm{C}}_{\text{red},3}^T\bm{\nu}$ and $\bm{\nu}^{*}_{\text{ind}}=\bar{\bm{C}}_{\text{ind},3}^T\bm{\nu}$. The above mathematical constructions allow us to re-write the penalized log-likelihood as a function of the transformed parameters $\bm{\Phi}^*:=(\bm{\alpha}^*,\bm{\beta}^*,\bm{\phi},\theta,\bm{\nu}^*)$ as follows:
\begin{equation}
\mathcal{F}_n^*(\bm{\Phi}^*)=\mathcal{F}_n(\bm{\Phi})=\mathcal{L}_n^*(\bm{\Phi}^*)-\mathcal{P}_n^*(\bm{\Phi}^*),
\end{equation}
with log-likelihood $\mathcal{L}_n^*(\bm{\Phi}^*)=\mathcal{L}_n(\bm{\Phi})$ and penalty term
\begin{equation}
\mathcal{P}_n(\bm{\Phi}^*)=
\sum_{k=1}^{K_1}p_{1n}\left(\big\|\tilde{\bm{c}}_{1k}^T\bm{\alpha}^*\big\|_2;\lambda_{1kn}\right)+
\sum_{k=1}^{K_2}p_{2n}\left(\big\|\tilde{\bm{c}}_{2k}^T\bm{\beta}^*\big\|_2;\lambda_{2kn}\right)+
P_{\bm{\lambda}_3,n}(\bm{\nu})=\sum_{k=1}^{K_3}p_{3n}\left(\big|\tilde{\bm{c}}_{3k}^T\bm{\nu}^*\big|;\lambda_{3kn}\right),
\end{equation}
where $\tilde{\bm{c}}_{lk}=\bar{\bm{C}}_l^{-1}\bm{c}_{lk}$, for $l=1,2,3$. Suppose that the true model parameters are given by $\bm{\Phi}^{*}_0:=(\bm{\alpha}^*_0,\bm{\beta}^*_0,\bm{\phi}_0,\theta_0,\bm{\nu}^*_0)$. This can be decomposed $\bm{\Phi}^{*}_0:=(\bm{\Phi}^{*}_{\text{red},0},\bm{\Phi}^{*}_{\text{ind},0})$, with $\bm{\Phi}^{*}_{\text{red},0}:=(\bm{\alpha}^*_{\text{red},0},\bm{\beta}^*_{\text{red},0},\bm{\nu}^*_{\text{red},0})$ and $\bm{\Phi}^*_{\text{ind},0}:=(\bm{\alpha}^*_{\text{ind},0},\bm{\beta}^*_{\text{ind},0},\bm{\phi}_0,\theta_0,\bm{\nu}^*_{\text{ind},0})$. Note that by construction $\bm{\Phi}^{*}_{\text{red},0}=\bm{0}$. Finally, denote $\hat{\bm{\Phi}}^*_n:=(\hat{\bm{\Phi}}^*_{\text{red},n},\hat{\bm{\Phi}}^*_{\text{ind},n})$ as the corresponding estimator of model parameters. We have the following theorem, which is an extension of the oracle property given by \cite{fan2001variable}:
\begin{theorem} \label{thm:oracle}
Assume that the conditions in Theorem \ref{thm:consistency} are fulfilled, and the conditions \textbf{H1}-\textbf{H4} hold for $p_{ln}(\psi;\lambda_{lkn})$. Then, for any $\sqrt{n}$-consistent local maximizer $\hat{\bm{\Phi}}^*_n$ of the regularized log-likelihood function $\mathcal{F}_n^{*}(\bm{\Phi}^*)$ as $n\rightarrow\infty$, we have:
\begin{enumerate}[font={\bfseries},label={(\alph*)}]
\item Consistency of feature selection: $P(\hat{\bm{\Phi}}^*_{\text{red},n}=\bm{0})\rightarrow 1$ as $n\rightarrow\infty$.
\item Asymptotic normality:
\begin{equation*}
\sqrt{n}\left\{\left[\mathcal{I}_{\text{ind}}^{*}(\bm{\Phi}^*_{\text{ind},0})-{\mathcal{P}_n^*}''(\bm{\Phi}^*_{\text{ind},0})/n)\right]\left(\hat{\bm{\Phi}}^*_{\text{ind},n}-\bm{\Phi}^*_{\text{ind},0}\right)+{\mathcal{P}^*}'(\bm{\Phi}^*_{\text{ind},0})/n\right\}\overset{d}{\rightarrow}N\left(\bm{0},\mathcal{I}_{\text{ind}}^{*}(\bm{\Phi}^*_{\text{ind},0})\right)
\end{equation*}
as $n\rightarrow\infty$, where $\mathcal{I}_{\text{ind}}^{*}(\bm{\Phi}^*_{\text{ind},0})$ is the Fisher information matrix and ${\mathcal{P}^*}'(\bm{\Phi}^*_{\text{ind},0})$ (respectively ${\mathcal{P}^*}''(\bm{\Phi}^*_{\text{ind},0})$) are the first (second) derivative of the penalty functions under the true (reduced) model after fixing $\bm{\Phi}^*_{\text{red},0}=\bm{0}$.
\end{enumerate}
\end{theorem}
The above theorem shows that when the sample size is large, the proposed feature selection method merges and shrinks parameters correctly with high probability. Moreover, the estimated parameters of the reduced model are asymptotically normal. While detailed discussions are leveraged to Remark \ref{apx:rmk:asym} of Appendix \ref{apx:asym:proof}, the adjustment and bias terms ${\mathcal{P}^*}''(\bm{\Phi}^*_{\text{ind},0})$ and ${\mathcal{P}^*}'(\bm{\Phi}^*_{\text{ind},0})$ are both asymptotically negligible when the penalty function is LASSO with an adaptive approach (which will be discussed in Section \ref{sec:est:tuning}) or SCAD. For large sample sizes, the estimated parameters are approximately unbiased and we may approximate the variance of the estimated transformed parameters as 
\begin{equation}
\widehat{\text{Var}}(\hat{\bm{\Phi}}_{\text{ind},n}^*)\approx \frac{1}{n}\left[\widehat{\mathcal{I}}_{\text{ind}}^{*}(\hat{\bm{\Phi}}^*_{\text{ind},n})\right]^{-1},
\end{equation}
where $\widehat{\mathcal{I}}_{\text{ind}}^{*}(\hat{\bm{\Phi}}^*_{\text{ind},n})$ is the sample Fisher information of the reduced model. In other words, parameter uncertainty under the proposed feature selection method is equivalent to that under the reduced model after selecting the variables. With this regards, the construction of confidence intervals (CI) of parameters is straightforward:
\begin{itemize}
    \item Wald-type CIs: Denote ${\psi}^{*}_{\text{ind},0,q}$ as the $q$-th element of $\bm{\Phi}^{*}_{\text{ind},0}$. 
    A two-sided Wald-type CI for ${\psi}^{*}_{\text{ind},0,q}$ is
    \begin{equation}
        \left[{\psi}^{*}_{\text{ind},0,q}-\frac{z_{1-\kappa/2}}{\sqrt{n}}\sqrt{\left[\widehat{\mathcal{I}}_{\text{ind}}^{*}(\hat{\bm{\Phi}}^*_{\text{ind},n})\right]^{-1}_{q,q}},{\psi}^{*}_{\text{ind},0,q}+\frac{z_{\kappa/2}}{\sqrt{n}}\sqrt{\left[\widehat{\mathcal{I}}_{\text{ind}}^{*}(\hat{\bm{\Phi}}^*_{\text{ind},n})\right]^{-1}_{q,q}}\right],
    \end{equation}
    where $z_{\kappa}$ is the $\kappa$-quantile of the standard normal distribution and $\left[\widehat{\mathcal{I}}_{\text{ind}}^{*}(\hat{\bm{\Phi}}^*_{\text{ind},n})\right]^{-1}_{q,q}$ is the $q$-th diagonal element of $\left[\widehat{\mathcal{I}}_{\text{ind}}^{*}(\hat{\bm{\Phi}}^*_{\text{ind},n})\right]^{-1}$. For other quantities of interest (e.g.~mean claim amounts), one may apply a delta method or simulate parameters from ${\cal N}(\hat{\bm{\Phi}}_{\text{ind},n}^*,\widehat{\text{Var}}(\hat{\bm{\Phi}}_{\text{ind},n}^*))$ to analytically or empirically approximate their CIs.
    \item Efron percentile bootstrap CIs: Consider a parametric bootstrap procedure which generates the bootstrap samples $\{(\bm{y}^{(b)},\bm{X})\}_{b=1,\ldots,B}$, where $\bm{y}^{(b)}$ is simulated from the reduced model with parameters $\bm{\Phi}^{*}_{\text{ind},0}$. For each $b=1,\ldots,B$, refit the bootstrap sample $(\bm{y}^{(b)},\bm{X})$ to the reduced model (the procedure which will be presented in Section \ref{sec:est}) to obtain bootstrap fitted parameters $\hat{\bm{\Phi}}^{*(b)}_{\text{ind},n}$. The Efron percentile bootstrap CI of a quantity of interest is then represented by its empirical quantiles based on the $B$ sets of bootstrap fitted parameters $\{\hat{\bm{\Phi}}^{*(b)}_{\text{ind},n}\}_{b=1,\ldots,B}$.
\end{itemize}

\section{Model estimation} \label{sec:est}
Direct optimization of the penalized log-likelihood in Equation (\ref{eq:lik_pen}) is difficult. Firstly, the log-likelihood $\log h_{Y}(y_i;\bm{\alpha},\bm{\beta},\bm{\phi}, \theta,\bm{\nu},\bm{x}_i)$ is the logarithm of a sum of $(g+1)$ mixture terms. Secondly, observe that $\log h_{Y}(y_i;\bm{\alpha},\bm{\beta},\bm{\phi}, \theta,\bm{\nu},\bm{x}_i)$ contains distribution
function $F(\tau;\exp\{\bm{\beta}_j^T\bm{x}_i\},\phi_j)$, which is not available in closed form; this is not the case for the Lomax distribution since $H(\tau;\theta,\exp\{\bm{\nu}^T\bm{x}_i\})$ has an analytical form. Thirdly, the penalty functions are not continuously differentiable. 

Model estimation of an ordinary splicing model is typically simple, thanks to the non-overlapping density parts between the body and tail, so that one may factor out the likelihood function and separately calibrate the three parts of distribution -- subgroup probability, body and tail. Nonetheless, under the regression framework outlined in Equation (\ref{eq:density}) with variable selection techniques embedded, the weight regression parameters $\bm{\alpha}$ share and interact across all $g$ body components and one tail component. Therefore, there is no straight-forward way to segregate the likelihood function and simplify the estimation procedure.

Motivated by the aforementioned computational challenges, this section presents the strategy to estimate the parameters and select important variables under the proposed modeling framework.

\subsection{Construction of complete data}
We first construct a hypothetical complete dataset. We present a modified 
version of the method introduced by \cite{FUNG2020MoECensTrun}. Define the complete data
\begin{equation}
\mathcal{D}^{\text{com}}=\{(y_i,\bm{z}_i,\bm{k}_i,\{\bm{y}_{ij}'\}_{j=1,\ldots,g})\}_{i=1,\ldots,n},
\end{equation}
with three extra elements defined as follows:
\begin{itemize}
\item $\bm{z}_i=(z_{i1},\ldots,z_{i(g+1)})$ is the realization of a categorical latent random vector $\bm{Z}_i=(Z_{i1},\ldots,Z_{i(g+1)})$ such that $Z_{ij}=1$ if the $i^{\text{th}}$ observation comes from the $j^{\text{th}}$ component of the proposed mixture distribution and $Z_{ij}=0$ otherwise; in fact, $\bm{Z}_i$ is one-hot encoding of the
selected mixture component that $Y_i$ is allocated to.
\item $\bm{k}_i=(k_{i1},\ldots,k_{ig})$ is the realization of $\bm{K}_i=(K_{i1},\ldots,K_{ig})$, where $K_{ij}$ is the number of missing sample points outside the truncation interval $(0,\tau)$ generated by the $j^{\text{th}}$ component of the $i^{\text{th}}$ observation for $j=1,\ldots, g$.
\item $\bm{y}_{ij}'=(y_{ij1}',\ldots,y_{ijk_{ij}}')$ is the realization of $\bm{Y}_{ij}'=(Y_{ij1}',\ldots,Y'_{ijk_{ij}})$, the missing sample points from the $j^{\text{th}}$ component of the $i^{\text{th}}$ observation for $j=1,\ldots, g$.
\end{itemize}
We assume that the cases $(Y_i,\bm{Z}_i,\bm{K}_i,\{\bm{Y}_{ij}'\}_{j=1,\ldots,g})$ are independent
in $1\le i \le n$. Moreover, we assume independence between $\bm{Z}_i$, $\bm{K}_i$, $Y_i$ and $\bm{Y}'_{ij}$, that $Y_{ij1}',\ldots,Y_{ijk_{ij}}'$ are i.i.d. given the covariates $\bm{x}_i$, and that $Y_i$ and $\bm{Y}'_{ij}$ are subgroup conditionally identically distributed. Furthermore, the components $K_{ij}$ of $\bm{K}_i$ are independent and follow the geometric distribution
\begin{equation} \label{eq:k_geom}
p(k_{ij};\bm{x}_i,\bm{\Phi})=[1-F(\tau;\exp\{\bm{\beta}_j^T\bm{x}_i\},\phi_j)]^{k_{ij}}F(\tau;\exp\{\bm{\beta}_j^T\bm{x}_i\},\phi_j), \qquad k_{ij}=0,1,\ldots.
\end{equation}
The complete data log-likelihood function is given by
\begin{align} \label{eq:com_ll}
\mathcal{L}_n^{\text{com}}(\bm{\Phi};\mathcal{D}^{\text{com}},\bm{X})
=&\sum_{i=1}^n\sum_{j=1}^{g+1}z_{ij}\Big(\log\pi_j(\bm{x}_i;\bm{\alpha})+\log f(y_i;\exp\{\bm{\beta}_j^T\bm{x}_i\},\phi_j)1\{y_i\leq\tau\}\nonumber\\
&\hskip5em+\log\frac{h(y_i;\theta,\exp\{\bm{\nu}^T\bm{x}_i\})}{1-H(\tau;\theta,\exp\{\bm{\nu}^T\bm{x}_i\})}1\{y_i>\tau\}\Big)\nonumber\\
&+\sum_{i=1}^n\sum_{j=1}^{g}\sum_{k=1}^{k_{ij}}z_{ij}\log f(y_{ijk}';\exp\{\bm{\beta}_j^T\bm{x}_i\},\phi_j).
\end{align}
This is easier to evaluate and optimize compared to Equation (\ref{eq:lik}) given that $H$ has an analytical form, which is the case for the Lomax distribution. The complete data penalized log-likelihood is given by
\begin{equation} \label{eq:lik_pen_com}
\mathcal{F}_n^{\text{com}}(\bm{\Phi})=\mathcal{L}^{\text{com}}_n(\bm{\Phi};\mathcal{D}^{\text{com}},\bm{X})-\mathcal{P}_n(\bm{\Phi}).
\end{equation}

\begin{remark}
\normalfont The choice of geometric distributions for $K_{ij}$ is motivated by the fact that it will allow for an efficient
fitting algorithm, as it will lead to nice cancellations. I.e.~this is an auxiliary tool that is computationally attractive;
for more information we refer to \cite{FUNG2020MoECensTrun}.
\end{remark}

\subsection{The GEM algorithm} \label{sec:est:GEM}
Parameter estimation is conducted using a Generalized Expectation-Maximazation (GEM) algorithm, where in the M-step is a modified version of the penalized iteratively re-weighted least squares (PIRLS) method proposed by \cite{oelker2017uniform}, where this method is also technically justified. Following \cite{oelker2017uniform}, due to the non-differentiability of the penalty function, we  perturb the penalty function in Equation (\ref{eq:pen}) as follows
\begin{equation} \label{eq:pen_eps}
\mathcal{P}_{\epsilon}(\bm{\Phi})=
P_{\bm{\lambda}_1,n,\epsilon}(\bm{\alpha})
+P_{\bm{\lambda}_2,n,\epsilon}(\bm{\beta})
+P_{\bm{\lambda}_3,n,\epsilon}(\bm{\nu}),
\end{equation}
where the $\epsilon$-perturbed penalty functions $P_{\bm{\lambda}_1,n,\epsilon}(\bm{\alpha})$, $P_{\bm{\lambda}_2,n,\epsilon}(\bm{\beta})$ and $P_{\bm{\lambda}_3,n,\epsilon}(\bm{\nu})$ are given by
\begin{equation*}
P_{\bm{\lambda}_1,n,\epsilon}(\bm{\alpha})=\sum_{k=1}^{K_1}p_{1k}\left(\big\|\bm{c}_{1k}^T\bm{\alpha}\big\|_{2,\epsilon};\lambda_{1kn}\right),\qquad
P_{\bm{\lambda}_2,n,\epsilon}(\bm{\beta})=\sum_{k=1}^{K_2}p_{2k}\left(\big\|\bm{c}_{2k}^T\bm{\beta}\big\|_{2,\epsilon};\lambda_{2kn}\right),
\end{equation*}
\begin{equation}
P_{\bm{\lambda}_3,n,\epsilon}(\bm{\nu})=\sum_{k=1}^{K_3}p_{3k}\left(\big|\bm{c}_{3k}^T\bm{\nu}\big|_{\epsilon};\lambda_{3kn}\right),
\end{equation}
with $\|\bm{w}\|_{2,\epsilon}=\left(\bm{w}\bm{w}^T+\epsilon\right)^{1/2}$ and $|w|_{\epsilon}=(w^2+\epsilon)^{1/2}$ for any vector $\bm{w}$ and scalar $w$. In the following, instead of maximizing Equation (\ref{eq:lik_pen_com}), we maximize the $\epsilon$-perturbed complete data penalized log-likelihood given by
\begin{equation}
\mathcal{F}_{n,\epsilon}^{\text{com}}(\bm{\Phi})=\mathcal{L}^{\text{com}}_n(\bm{\Phi};\mathcal{D}^{\text{com}},\bm{X})-\mathcal{P}_{n,\epsilon}(\bm{\Phi}).
\end{equation}
Now, $\mathcal{F}^{\text{com}}_{n,\epsilon}(\bm{\Phi})$ is continuously differentiable w.r.t.~any parameter and, hence, it is more computationally tractable. Furthermore, note that $\mathcal{P}_{n,\epsilon}(\bm{\Phi})\rightarrow\mathcal{P}_n(\bm{\Phi})$ and hence $\mathcal{F}_{n,\epsilon}^{\text{com}}(\bm{\Phi})\rightarrow\mathcal{F}^{\text{com}}_n(\bm{\Phi})$ as $\epsilon\rightarrow 0$, so choosing a very small $\epsilon>0$, the perturbation of the estimated parameters $\bm{\Phi}$ will be negligible. In simulation studies and real data analysis, we find that the choice of $\epsilon=10^{-10}$ works well.

\subsubsection{E-step}
In the $l^{\text{th}}$ iteration, the expectation of the complete data $\epsilon$-perturbed penalized log-likelihood is computed as follows:
\begin{align} \label{eq:Q_e}
Q_{\epsilon}(\bm{\Phi};\bm{y},\bm{X},\bm{\Phi}^{(l-1)})
&=E\left[\mathcal{F}^{\text{com}}_{n,\epsilon}(\bm{\Phi})\left|\bm{y},\bm{X},\bm{\Phi}^{(l-1)}\right]\right.\nonumber\\
&=\sum_{i=1}^n\sum_{j=1}^{g+1}z_{ij}^{(l)}\Big(\log\pi_j(\bm{x}_i;\bm{\alpha})+\log f(y_i;\exp\{\bm{\beta}_j^T\bm{x}_i\},\phi_j)1\{y_i\leq\tau\}\nonumber\\
&\hskip5em+\log\frac{h(y_i;\theta,\exp\{\bm{\nu}^T\bm{x}_i\})}{1-H(\tau;\theta,\exp\{\bm{\nu}^T\bm{x}_i\})}1\{y_i>\tau\}\Big)\nonumber\\
&\quad+\sum_{i=1}^n\sum_{j=1}^{g}k_{ij}^{(l)}z_{ij}^{(l)}\log \tilde{f}(\widehat{y_{ijk}'}^{(l)},\widehat{\log y_{ijk}'}^{(l)};\exp\{\bm{\beta}_j^T\bm{x}_i\},\phi_j)-\mathcal{P}_{\epsilon}(\bm{\Phi}),
\end{align}
where the updated quantities $z_{ij}^{(l)}$, $k_{ij}^{(l)}$, $\log \tilde{f}(\widehat{y_{ij}'}^{(l)},\widehat{\log y_{ij}'}^{(l)};\exp\{\bm{\beta}_j^T\bm{x}_i\},\phi_j)$, $\widehat{y_{ij}'}^{(l)}$ and $\widehat{\log y_{ij}'}^{(l)}$ are displayed in Equations (2.2) to (2.6) in the supplementary materials.

Note that $\widehat{\log y_{ij}'}^{(l)}$ is represented by a numerical integral, and hence, in general, it does not have an analytical solution. Here, we adopt a Stochastic EM approach, where for each $i=1,\ldots,n$ and $j=1,\ldots,g$ we simulate $\log Y_{ij}'^{(l)}$ from the conditional density of $Y_{ij}'^{(l)}$ given by
\begin{equation}
f_{Y_{ij}'^{(l)}}(y;\bm{\beta}_j^{(l-1)},\phi_j^{(l-1)},\bm{x}_i)
=\frac{f(y;\exp\{\bm{\beta}_j^{T(l-1)}\bm{x}_i\},\phi_j^{(l-1)})}{1-F(\tau;\exp\{\bm{\beta}_j^{T(l-1)}\bm{x}_i\},\phi_j^{(l-1)})}1\{y>\tau\}.
\end{equation}

\subsubsection{M-step}
In this step, we attempt to find a parameter update $\bm{\Phi}^{(l)}$ in such that 
we will receive a monotonicity $Q_{\epsilon}(\bm{\Phi}^{(l)};\bm{y},\bm{X},\bm{\Phi}^{(l-1)})\geq Q_{\epsilon}(\bm{\Phi}^{(l-1)};\bm{y},\bm{X},\bm{\Phi}^{(l-1)})$. While $Q_{\epsilon}(\bm{\Phi};\bm{y},\bm{X},\bm{\Phi}^{(l-1)})$ is now differentiable w.r.t.~any parameter, direct implementation of
an iteratively re-weighted least squares (IRLS) algorithm is challenging due to concavity of 
the penalty functions. In this section, we propose the use of convex quadratic approximation to the penalty functions, analogously to \cite{fan2001variable} and \cite{oelker2017uniform}, such that the implementation of 
an IRLS algorithm is feasible. We  approximate $\mathcal{P}_{n,\epsilon}(\bm{\Phi})$ by
\begin{equation}
\tilde{\mathcal{P}}_{n,\epsilon}(\bm{\Phi})
=\sum_{k=1}^{K_1}\tilde{p}_{1k}\left(\big\|\bm{c}_{1k}^T\bm{\alpha}\big\|_{2,\epsilon};\lambda_{1kn}\right)+\sum_{k=1}^{K_2}\tilde{p}_{2k}\left(\big\|\bm{c}_{2k}^T\bm{\beta}\big\|_{2,\epsilon};\lambda_{2kn}\right)
+\sum_{k=1}^{K_3}\tilde{p}_{3k}\left(\big|\bm{c}_{3k}^T\bm{\nu}\big|_{\epsilon};\lambda_{3kn}\right),
\end{equation}
where
\begin{align} \label{eq:a}
\tilde{p}_{1k}\left(\big\|\bm{c}_{1k}^T\bm{\alpha}\big\|_{2,\epsilon};\lambda_{1kn}\right)
&= p_{1k}\left(\big\|\bm{c}_{1k}^T\bm{\alpha}^{(l-1)}\big\|_{2,\epsilon};\lambda_{1kn}\right)+ \nonumber\\
&\qquad\frac{1}{2}\frac{\bm{c}_{1k}^T\bm{\alpha}\bm{\alpha}^T\bm{c}_{1k}-\bm{c}_{1k}^T\bm{\alpha}^{(l-1)}\bm{\alpha}^{T(l-1)}\bm{c}_{1k}}{\|\bm{c}_{1k}^T\bm{\alpha}^{(l-1)}\|_{2,\epsilon}}p_{1k}'\left(\big\|\bm{c}_{1k}^T\bm{\alpha}^{(l-1)}\big\|_{2,\epsilon};\lambda_{1kn}\right),
\end{align}
\begin{align} \label{eq:b}
\tilde{p}_{2k}\left(\big\|\bm{c}_{2k}^T\bm{\beta}\big\|_{2,\epsilon};\lambda_{2kn}\right)
&=p_{2k}\left(\big\|\bm{c}_{2k}^T\bm{\beta}^{(l-1)}\big\|_{2,\epsilon};\lambda_{2kn}\right)+ \nonumber\\
&\qquad\frac{1}{2}\frac{\bm{c}_{2k}^T\bm{\beta}\bm{\beta}^T\bm{c}_{2k}-\bm{c}_{2k}^T\bm{\beta}^{(l-1)}\bm{\beta}^{T(l-1)}\bm{c}_{2k}}{\|\bm{c}_{2k}^T\bm{\beta}^{(l-1)}\|_{2,\epsilon}}p_{2k}'\left(\big\|\bm{c}_{2k}^T\bm{\beta}^{(l-1)}\big\|_{2,\epsilon};\lambda_{2kn}\right),
\end{align}
\begin{equation} \label{eq:c}
\tilde{p}_{3k}\left(\big|\bm{c}_{3k}^T\bm{\nu}\big|_{\epsilon};\lambda_{3kn}\right)=p_{3k}\left(\big|\bm{c}_{3k}^T\bm{\nu}^{(l-1)}\big|_{\epsilon};\lambda_{3kn}\right)+\frac{1}{2}\frac{(\bm{c}_{3k}^T\bm{\nu})^2-(\bm{c}_{3k}^T\bm{\nu}^{(l-1)})^2}{\|\bm{c}_{3k}^T\bm{\nu}^{(l-1)}\|_{\epsilon}}p_{3k}'\left(\big|\bm{c}_{3k}^T\bm{\nu}^{(l-1)}\big|_{\epsilon};\lambda_{3kn}\right).
\end{equation}
The properties below justify the use of such approximations:
\begin{theorem} \label{thm:pen_approx}
As $\epsilon\rightarrow 0$, $\tilde{\mathcal{P}}_{n,\epsilon}(\bm{\Phi})$ majorizes $\mathcal{P}_{n,\epsilon}(\bm{\Phi})$, 
i.e.~$\tilde{\mathcal{P}}_{n,\epsilon}(\bm{\Phi}^{(l-1)})=\mathcal{P}_{n,\epsilon}(\bm{\Phi}^{(l-1)})$ and $\tilde{\mathcal{P}}_{n,\epsilon}(\bm{\Phi})\geq\mathcal{P}_{n,\epsilon}(\bm{\Phi})$ as $\epsilon\rightarrow 0$ for any $\bm{\Phi}\neq \bm{\Phi}^{(l-1)}$.
\end{theorem}
\begin{proof}
W.l.o.g., it suffices to prove that $\tilde{p}_{1k}\left(\big\|\bm{c}_{1k}^T\bm{\alpha}\big\|_{2,\epsilon};\lambda_{1kn}\right)$ majorizes $p_{1k}\left(\big\|\bm{c}_{1k}^T\bm{\alpha}\big\|_{2,\epsilon};\lambda_{1kn}\right)$. Firstly, it is trivial that $\tilde{p}_{1k}\left(\big\|\bm{c}_{1k}^T\bm{\alpha}^{(l-1)}\big\|_{2,\epsilon};\lambda_{1kn}\right)=p_{1k}\left(\big\|\bm{c}_{1k}^T\bm{\alpha}^{(l-1)}\big\|_{2,\epsilon};\lambda_{1kn}\right)$. Secondly, we notice that we have $\tilde{p}_{1k}\left(\big\|\bm{c}_{1k}^T\bm{\alpha}\big\|_{2,\epsilon};\lambda_{1kn}\right)\geq p_{1k}\left(\big\|\bm{c}_{1k}^T\bm{\alpha}\big\|_{2,\epsilon};\lambda_{1kn}\right)$ for any $\bm{\alpha}\neq \bm{\alpha}^{(l-1)}$ if for any $u^2\neq\bm{c}_{1k}^T\bm{\alpha}^{(l-1)}\bm{\alpha}^{T(l-1)}\bm{c}_{1k}=:u^{*2}$ we have
\begin{align} \label{eq: ineq}
\tilde{p}_{1k}\left((u^2+\epsilon)^{1/2};\lambda_{1kn}\right)
&:=p_{1k}\left((u^{*2}+\epsilon)^{1/2};\lambda_{1kn}\right)+ \nonumber\\ &\qquad\frac{1}{2}\frac{u^2-u^{*2}}{(u^{*2}+\epsilon)^{1/2}}p_{1k}'\left((u^{*2}+\epsilon)^{1/2};\lambda_{1kn}\right)
\geq p_{1k}\left((u^2+\epsilon)^{1/2};\lambda_{1kn}\right).
\end{align}
As $\epsilon\rightarrow0$, the above equation obviously holds for $u^{*}\neq0$ because $\tilde{p}_{1k}(u;\lambda_{1kn})$ is a convex function as opposed to that $p_{1k}(u;\lambda_{1kn})$ is a concave function with $\tilde{p}_{1k}(u^*;\lambda_{1kn})=p_{1k}(u^*;\lambda_{1kn})$. If $u^{*}=0$, then $\tilde{p}_{1k}\left((u^2+\epsilon)^{1/2};\lambda_{1kn}\right)\rightarrow\infty$ as $\epsilon\rightarrow\infty$ while $p_{1k}(u;\lambda_{1kn})<\infty$, so the result follows.
\end{proof}

\begin{corollary} \label{cor:pen_approx}
\sloppy As $\epsilon\rightarrow 0$, if $\tilde{Q}_{\epsilon}(\bm{\Phi}^{(l)};\bm{y},\bm{X},\bm{\Phi}^{(l-1)})\geq \tilde{Q}_{\epsilon}(\bm{\Phi}^{(l-1)};\bm{y},\bm{X},\bm{\Phi}^{(l-1)})$, then we have $Q_{\epsilon}(\bm{\Phi}^{(l)};\bm{y},\bm{X},\bm{\Phi}^{(l-1)})\geq Q_{\epsilon}(\bm{\Phi}^{(l-1)};\bm{y},\bm{X},\bm{\Phi}^{(l-1)})$, where $\tilde{Q}_{\epsilon}$ is the same as $Q_{\epsilon}$ except that the term $\mathcal{P}_{n,\epsilon}(\bm{\Phi})$ inside Equation (\ref{eq:Q_e}) is replaced by $\tilde{\mathcal{P}}_{n,\epsilon}(\bm{\Phi})$.
\end{corollary}
\begin{proof}
It follows by Theorem \ref{thm:pen_approx} that $\tilde{Q}_{\epsilon}$ is a minorizer of $Q_{\epsilon}$. Then the result follows by the ascending property of Minorization-Majorization (MM) algorithm.
\end{proof}

~

Next, we decompose $\tilde{Q}_{\epsilon}(\bm{\Phi};\bm{y},\bm{X},\bm{\Phi}^{(l-1)})$  into the following terms:
\begin{equation}
\tilde{Q}_{\epsilon}(\bm{\Phi};\bm{y},\bm{X},\bm{\Phi}^{(l-1)})
=S^{(l)}(\bm{\alpha})+T^{(l)}(\bm{\beta},\bm{\phi})+V^{(l)}(\theta,\bm{\nu}),
\end{equation}
where
\begin{align}
\label{eq:Q_e_S}
S^{(l)}(\bm{\alpha})&=\sum_{i=1}^n\sum_{j=1}^{g+1}z_{ij}^{(l)}\log\pi_j(\bm{x}_i;\bm{\alpha})-\sum_{k=1}^{K_1}\tilde{p}_{1k}\left(\big\|\bm{c}_{1k}^T\bm{\alpha}\big\|_{2,\epsilon};\lambda_{1kn}\right),
\\
\label{eq:Q_e_T}
T^{(l)}(\bm{\beta},\bm{\phi})
&=\sum_{i=1}^n\sum_{j=1}^g z_{ij}^{(l)}\left[\log f(y_i;\exp\{\bm{\beta}_j^T\bm{x}_i\},\phi_j)+k_{ij}^{(l)}\log \tilde{f}(\widehat{y_{ijk}'}^{(l)},\widehat{\log y_{ijk}'}^{(l)};\exp\{\bm{\beta}_j^T\bm{x}_i\},\phi_j)\right]\nonumber\\
&\qquad -\sum_{k=1}^{K_2}\tilde{p}_{2k}\left(\big\|\bm{c}_{2k}^T\bm{\beta}\big\|_{2,\epsilon};\lambda_{2kn}\right),
\\\label{eq:Q_e_V}
V^{(l)}(\theta,\bm{\nu})&=\sum_{i=1}^n \log\frac{h(y_i;\theta,\exp\{\bm{\nu}^T\bm{x}_i\})}{1-H(\tau;\theta,\exp\{\bm{\nu}^T\bm{x}_i\})}1\{y_i>\tau\}-\sum_{k=1}^{K_3}\tilde{p}_{3k}\left(\big|\bm{c}_{3k}^T\bm{\nu}\big|_{\epsilon};\lambda_{3kn}\right).
\end{align}
Update of $\bm{\alpha}^{(l-1)}$ to $\bm{\alpha}^{(l)}$ can be done by sequentially adopting the IRLS approach for $j=1,\ldots,g$:
\begin{equation} \label{eq:irls_alpha}
\bm{\alpha}_j \leftarrow \bm{\alpha}_j-\left(\frac{\partial^2S^{(l)}(\bm{\alpha})}{\partial\bm{\alpha}_j\partial\bm{\alpha}_j^T}\right)^{-1}\frac{\partial S^{(l)}(\bm{\alpha})}{\partial\bm{\alpha}_j},
\end{equation}
where the derivatives are presented in Equations  (2.7) and (2.8) of 
the supplementary material which are expressed in analytical forms.

Similarly, update of $\bm{\beta}^{(l-1)}$ to $\bm{\beta}^{(l)}$ using IRLS involves
\begin{equation} \label{eq:irls_beta}
\bm{\beta}_j \leftarrow \bm{\beta}_j-\left(\frac{\partial^2T^{(l)}(\bm{\beta},\bm{\phi}^{(l-1)})}{\partial\bm{\beta}_j\partial\bm{\beta}_j^T}\right)^{-1}\frac{\partial T^{(l)}(\bm{\beta},\bm{\phi}^{(l-1)})}{\partial\bm{\beta}_j},
\end{equation}
with the analytical forms of derivatives given by Equations (2.9) and (2.10) of the supplementary material.

After updating $\bm{\beta}$, we update $\phi_j^{(l-1)}$ to $\phi_j^{(l)}$ directly using
function \texttt{optimize} in \textbf{R}, which is found to involve little computational burden compared to the IRLS procedures above:
\begin{equation} \label{eq:irls_phi}
\phi_j^{(l)}=\underset{\phi_j>0}{\text{argmax}}~ T^{(l)}(\bm{\beta}_j^{(l)},\bm{\phi}).
\end{equation}
After that, the same IRLS procedure leads to an update of $\bm{\nu}^{(l-1)}$ to $\bm{\nu}^{(l)}$:
\begin{equation} \label{eq:irls_nu}
\bm{\nu} \leftarrow \bm{\nu}-\left(\frac{\partial^2V^{(l)}(\theta^{(l-1)},\bm{\nu})}{\partial\bm{\nu}\partial\bm{\nu}^T}\right)^{-1}\frac{\partial V^{(l)}(\theta^{(l-1)},\bm{\nu})}{\partial\bm{\nu}},
\end{equation}
with the analytical forms of derivatives given by Equations (2.11) and (2.12) of the supplementary material.

Finally, $\theta$ can be updated directly using the \texttt{optimize} function or the Newton-Raphson method, aiming to achieve
\begin{equation} \label{eq:irls_theta}
\theta^{(l)}=\underset{\theta>0}{\text{argmax}}~ V^{(l)}(\theta,\bm{\nu}^{(l)}).
\end{equation}
Because of Corollary \ref{cor:pen_approx}, the M-step ensures $Q_{\epsilon}(\bm{\Phi}^{(l)};\bm{y},\bm{X},\bm{\Phi}^{(l-1)})\geq Q_{\epsilon}(\bm{\Phi}^{(l-1)};\bm{y},\bm{X},\bm{\Phi}^{(l-1)})$ given a very small $\epsilon>0$. The GEM algorithm is iterated until the observed data $\epsilon$-perturbed penalized log-likelihood is improved by less than a threshold $10^{-2}$ or the maximum number of iterations of 200 is reached.

\subsection{Initialization procedures}
Initialization of parameters $\bm{\Phi}^{(0)}$ can be done using the clusterized method of moments (CMM) approach proposed by \cite{gui2018fitting}. It requires a $K$-means clustering method to assign observations $y_i$ with $y_i\leq\tau$ to one of the $g$ subgroups for the body, and observations $y_i$ with $y_i>\tau$ to the tail component. Then, we set initial parameters which match the first two moments for each mixture component, after initially fixing all the regression parameters be zero except for the intercepts. Refer to e.g.~Section 3.3.3 of \cite{FUNG2020MoECensTrun} for more details.

\subsection{Choice of the number of mixture components}
Usually, the choice of the number of mixture components $g$ of the body can be determined based on standard specification criteria, including Akaike's Information Criterion (AIC) and the Bayesian Information Criterion (BIC). However, for our motivating dataset, which will be described in Section 6, below, the AIC and BIC criteria would both lead to an excessively large number of components which would significantly impede the model interpretability. The main reason for obtaining a  large $g$ is that the claim severity distribution has many small nodes for smaller claim amounts (i.e.~less than $10,000$), as evidenced by Figure \ref{fig:density} in Section 6. Excessive fitting and modeling of such smaller claim amounts does not bring much insight from an insurance ratemaking perspective because such smaller claims could even be modelled by an empirical distribution. As a result, for this particular dataset, we adopt a qualitative method, which chooses $g$ as the minimum number of components required for the proposed model to capture all nodes above a claim severity threshold of $10,000$.

\subsection{Selection of variables} \label{sec:est:var_sel}
The proposed GEM algorithm with group fused penalty functions shrinks some regression coefficients to zero and merges some coefficients across different levels of a categorical variable. Being a variant of the PIRLS approach, the proposed algorithm also caters for a wide range of concave penalty functions. However, as pointed out by \cite{devriendt2020sparse}, the parameters obtained by the proposed algorithm are not exact. Therefore, in order to select the variables and reduce model complexity, after every model fit we need to perform an automatic adjustment algorithm to remove parameters very close to zero and merge the parameters when their values are very close to each other. 

Denote the fitted model parameter as $\hat{\bm{\Phi}}=(\hat{\bm{\alpha}},\hat{\bm{\beta}},\hat{\bm{\phi}},\hat{\theta},\hat{\bm{\nu}})$. Further, with slight abuse of notation, denote $\hat{\bm{\alpha}}_{p}$ as the $p^{\text{th}}$ row vector of $\hat{\bm{\alpha}}$, as opposed to $\hat{\bm{\alpha}}_{j}$ as the $j^{\text{th}}$ column vector of $\hat{\bm{\alpha}}$. Similarly, denote $\hat{\bm{\beta}}_{p}$ as the $p^{\text{th}}$ row vector of $\hat{\bm{\beta}}$. Also, let $\hat{z}_{ij}$, $\hat{k}_{ij}$, $\widehat{y_{ijk}'}$ and $\widehat{\log y_{ijk}'}$ be the $z_{ij}^{(l)}$, $k_{ij}^{(l)}$, $\widehat{y_{ijk}'}^{(l)}$ and $\widehat{\log y_{ijk}'}^{(l)}$ obtained by the E-step using the fitted parameters. Define the partial log-likelihood functions $S(\bm{\alpha})$, $T(\bm{\beta},\bm{\phi})$ and $V(\theta,\bm{\nu})$, respectively, for the mixing probabilities, body distributions and tail distribution analogously to Equations (\ref{eq:Q_e_S}) to (\ref{eq:Q_e_V}) as follows:
\begin{align}
\label{eq:plik_S}
S(\bm{\alpha})&=\sum_{i=1}^n\sum_{j=1}^{g+1}\hat{z}_{ij}\log\pi_j(\bm{x}_i;\bm{\alpha})-\sum_{k=1}^{K_1}p_{1k}\left(\big\|\bm{c}_{1k}^T\bm{\alpha}\big\|_{2,\epsilon};\lambda_{1kn}\right)=:S_0(\bm{\alpha})-P_{\bm{\lambda}_1}(\bm{\alpha}),
\\\label{eq:plik_T}
T(\bm{\beta},\bm{\phi})
&=\sum_{i=1}^n\sum_{j=1}^g \hat{z}_{ij}\left[\log f(y_i;\exp\{\bm{\beta}_j^T\bm{x}_i\},\phi_j)+\hat{k}_{ij}\log \tilde{f}(\widehat{y_{ijk}'},\widehat{\log y_{ijk}'};\exp\{\bm{\beta}_j^T\bm{x}_i\},\phi_j)\right]-\nonumber\\
&\qquad\sum_{k=1}^{K_2}p_{2k}\left(\big\|\bm{c}_{2k}^T\bm{\beta}\big\|_{2,\epsilon};\lambda_{2kn}\right)=:T_0(\bm{\beta},\bm{\phi})-P_{\bm{\lambda}_2}(\bm{\beta}),
\\\label{eq:plik_V}
V(\theta,\bm{\nu})&=\sum_{i=1}^n \log\frac{h(y_i;\theta,\exp\{\bm{\nu}^T\bm{x}_i\})}{1-H(\tau;\theta,\exp\{\bm{\nu}^T\bm{x}_i\})}1\{y_i>\tau\}-\sum_{k=1}^{K_3}p_{3k}\left(\big|\bm{c}_{3k}^T\bm{\nu}\big|_{\epsilon};\lambda_{3kn}\right)
=:V_0(\theta,\bm{\nu})-P_{\bm{\lambda}_3}(\bm{\nu}).
\end{align}

The general principle of the automatic adjustment algorithm is to fine tune the regression parameters, so the regression parameters (which are close to zero or very close to each other) are shrinked or merged in exact. Since fine tuning of parameters would lead to another source of error, the automatic adjustment algorithm needs to ensure that the likelihood-based quantities displayed above would not change significantly due to fine-tuning. We leverage the step-by-step algorithm to Section 2.2 of the supplementary materials.

\subsection{Tuning of hyperparameters} \label{sec:est:tuning}
The remaining problem is to select appropriate tuning parameters $\bm{\lambda}:=(\bm{\lambda}_1,\bm{\lambda}_2,\bm{\lambda}_3)$ which control the model complexity and hence select variables useful for explaining different parts of 
the claim severity distributions. The current theory in Section \ref{sec:asym} only provides guidance on the order of $\bm{\lambda}$, but in application it is obvious that grid search on $\bm{\lambda}$ is computationally prohibitive because of the curse of dimensionality. As a result, we adopt an adaptive-standardization approach similar to \cite{devriendt2020sparse}, where we restrict $\lambda_{1kn}=w_{1k}\lambda_1$, $\lambda_{2kn}=w_{2k}\lambda_2$ and $\lambda_{3kn}=w_{3k}\lambda_3$. Here, 
\begin{equation}
w_{1k}=w_{1k}^{(\text{ad})} w_k^{(\text{st})},\quad
w_{2k}=w_{2k}^{(\text{ad})} w_k^{(\text{st})},\quad
w_{3k}=w_{3k}^{(\text{ad})} w_k^{(\text{st})},\quad\text{with}~~
w_k^{(\text{st})}=\frac{p_{\bm{G}}-1}{r_{\bm{G}}}\sqrt{\frac{n_{p_1}+n_{p_2}}{n}},
\end{equation}
where $w_{1k}^{(\text{ad})}=\|\bm{c}_{1k}^T\hat{\bm{\alpha}}\|_2^{-1}$, $w_{2k}^{(\text{ad})}=\|\bm{c}_{2k}^T\hat{\bm{\beta}}\|_2^{-1}$ and $w_{3k}^{(\text{ad})}=|\bm{c}_{3k}^T\hat{\bm{\nu}}|^{-1}$ are the adaptive terms, and $w_k^{(\text{st})}$ is the standardization term. Note here that the estimated parameters $\hat{\bm{\alpha}}$, $\hat{\bm{\beta}}$ and $\hat{\bm{\nu}}$ are obtained on the fitting procedures obtained in Section \ref{sec:est:GEM} starting with very small tuning parameters $\bm{\lambda}$ (or even $\bm{\lambda}=0$). $p_1$ and $p_2$ are the two categories that the $k^{\text{th}}$ penalty term is attempting to merge for categorical variables, and $(n_{p_1},n_{p_2})$ are the number of observations being classified to those respective categories. $p_{\bm{G}}$ is the number of categories for the respective explanatory variable, while $r_{\bm{G}}$ is the number of penalty terms for the respective explanatory variable. Note that $r_{\bm{G}}=p_{\bm{G}}-1$ for ordinal variables and $r_{\bm{G}}=p_{\bm{G}}(p_{\bm{G}}-1)/2$ for nominal variables. For continuous variables, we set $w_k^{(\text{st})}=1$ instead. The adaptive weights facilitate more efficient shrinkage or merger of regression coefficients, achieving the oracle property presented by \cite{zou2006adaptive}. The standardization weights, on the other hand, address the issues of level imbalances and imbalances of numbers of terms on an explanatory variable involved in the penalty functions. 

After specifying the weights, we can perform a grid search on $(\lambda_1,\lambda_2,\lambda_3)$ to find optimal tuning parameters. Motivated by the likelihood-based deviance approach by \cite{khalili2010new}, we propose the following method, which allows us doing the separate grid search for $\lambda_1$, $\lambda_2$ and $\lambda_3$.

After obtaining the fitted model parameters $\bm{\Phi}$ starting with a small $\bm{\lambda}$, we compute the estimated latent variables $\hat{z}_{ij}$, $\hat{k}_{ij}$, $\widehat{y_{ijk}'}$ and $\widehat{\log y_{ijk}'}$ outlined in Section \ref{sec:est:var_sel} and assume that they are fixed during the whole process of grid searching. Then, for each $\lambda_1$, $\lambda_2$ and $\lambda_3$ within separate specified (one-dimensional) grids, we refit the models by maximizing the (unpenalized) partial log-likelihood functions $S_0(\bm{\alpha})$ in Equation (\ref{eq:plik_S}) (which only requires iterating Equations (\ref{eq:irls_alpha})), $T_0(\bm{\beta},\bm{\phi})$ in Equation (\ref{eq:plik_T}) (iterating Equations (\ref{eq:irls_beta}) and (\ref{eq:irls_phi})) and $V_0(\theta,\bm{\nu})$ in Equation (\ref{eq:plik_V}) (iterating Equations (\ref{eq:irls_nu}) and (\ref{eq:irls_theta})). We denote the resulting fitted parameters as $\hat{\bm{\alpha}}(\lambda_1)$, $(\hat{\bm{\beta}}(\lambda_2), \hat{\bm{\phi}}(\lambda_2))$ and $(\hat{\theta}(\lambda_3),\hat{\bm{\nu}}(\lambda_3))$. This avoids repeating the whole GEM procedure over a multidimensional grid of $(\lambda_1,\lambda_2,\lambda_3)$ which is computationally prohibitive. Optimal $(\lambda_1,\lambda_2,\lambda_3)$ can be determined by various choices of criteria, where in this paper we will present partial AIC (pAIC), partial BIC (pBIC) and $K$-fold cross-validation (CV). For pAIC or pBIC approach, we define
\begin{equation}
\text{pAIC}_1(\lambda_1)=-2 S_0(\hat{\bm{\alpha}}(\lambda_1))+2\mathcal{N}_1(\lambda_1);\quad
\text{pBIC}_1(\lambda_1)=-2 S_0(\hat{\bm{\alpha}}(\lambda_1))+\mathcal{N}_1(\lambda_1)\log n,
\end{equation}
\begin{equation}
\text{pAIC}_2(\lambda_2)=-2 T_0(\hat{\bm{\beta}}(\lambda_2),\hat{\bm{\phi}}(\lambda_2))+2\mathcal{N}_2(\lambda_2);\quad
\text{pBIC}_2(\lambda_2)=-2 T_0(\hat{\bm{\beta}}(\lambda_2),\hat{\bm{\phi}}(\lambda_2))+\mathcal{N}_2(\lambda_2)\log n_b,
\end{equation}
\begin{equation}
\text{pAIC}_3(\lambda_3)=-2 V_0(\hat{\theta}(\lambda_3),\hat{\bm{\nu}}(\lambda_3))+2\mathcal{N}_3(\lambda_3);\quad
\text{pBIC}_3(\lambda_3)=-2 V_0(\hat{\bm{\beta}}(\lambda_3),\hat{\bm{\phi}}(\lambda_3))+\mathcal{N}_3(\lambda_3)\log n_t,
\end{equation}
where $\mathcal{N}_1(\lambda_1)$, $\mathcal{N}_2(\lambda_2)$ and $\mathcal{N}_3(\lambda_3)$ are the effective number of parameters (i.e.~the number excluding zeroes and redundant parameter values) for $\hat{\bm{\alpha}}(\lambda_1)$, $(\hat{\bm{\beta}}(\lambda_2),\hat{\bm{\phi}}(\lambda_2))$ and $(\hat{\theta}(\lambda_3),\hat{\bm{\nu}}(\lambda_3))$, respectively. Recall that $n_b$ and $n_t$ are the number of observations allocated to body and tail components, respectively. Now, $\lambda_1$, $\lambda_2$ and $\lambda_3$ can be chosen once at a time through minimizing the pAICs or pBICs. 

For $K$-fold CV, we partition the data into $K$ disjoint folds and measure the performance on each fold after training the remaining $K-1$ folds. The performance metric we use in this paper is called ``partial deviance" given by
\begin{equation}
p\mathcal{D}_1(\lambda_1)=-2 S_0(\hat{\bm{\alpha}}(\lambda_1));\quad
p\mathcal{D}_2(\lambda_2)=-2 T_0(\hat{\bm{\beta}}(\lambda_2),\hat{\bm{\phi}}(\lambda_2));
\quad
p\mathcal{D}_3(\lambda_3)=-2 V_0(\hat{\theta}(\lambda_3),\hat{\bm{\nu}}(\lambda_3)).
\end{equation}
For $l=1,2,3$, the optimal $\hat{\lambda}_l$ is the largest one such that the corresponding partial deviance is within one standard deviation of its minimum.

As the regularization functions make the estimated parameters of the fitted model biased towards zero, it is important to collapse the regression parameters $(\hat{\bm{\alpha}}(\hat{\lambda}_1),\hat{\bm{\beta}}(\hat{\lambda}_2),\hat{\bm{\nu}}(\hat{\lambda}_3))$ and covariate matrix $\bm{X}$, and re-estimate the model where the penalties are excluded (i.e. $\bm{\lambda}=\bm{0}$), using the full GEM algorithm outlined by Section \ref{sec:est:GEM}. This will also update the estimated latent variables for better accuracy. A related approach can be found by \cite{devriendt2020sparse}.

\section{Motivating dataset: Greece MTPL claim amounts}  \label{sec:data_descr}

In this section, we present a dataset which motivates the proposed modeling and feature selection framework described in
above. The characteristics of the dataset are first described in Section \ref{sec:data_descr:data}. Then, we fit some state-of-the-art models to the dataset in Section \ref{sec:data_descr:fit} to show the necessity of adopting the proposed modeling framework.

\subsection{Data description} \label{sec:data_descr:data}
The dataset for our study was kindly provided by a major insurance company operating in Greece. It consists of 64,923 motor third-party liability (MTPL) insurance policies with non-zero property claims for underwriting years 2013 to 2017. The sample comprised of policyholders with complete records; i.e., with the availability of all explanatory variables under consideration, and with at least one reported accident over the five underwriting years. These explanatory variables are summarized in Table \ref{tab:variables}. 
\begin{table}[H]
\centering
\resizebox{\textwidth}{!}{%
\begin{threeparttable}
\begin{tabular}{lllll}
Name & Short Description & Categories & Type & Categories Description \\ \hline
DriverAge & Driver's age & 18-74 & Continuous\tnote{*} & From 18 to 74+ years old \\
\arrayrulecolor{Gray}\hline
VechicleBrand & Automobile brand & B1 - B31 & Unordered & 31 different brands \\
\arrayrulecolor{Gray}\hline
CC & Car cubism & 0 - 18 & Ordered & 19 different categories \\
\arrayrulecolor{Gray}\hline
\multirow{3}{*}{PolicyType} & \multirow{3}{*}{Policy Type} & C1 & \multirow{3}{*}{Ordered} & Economic type - only MTPL coverage \\
 &  & C2 &  & Middle type - includes other types \\
 &  & C3 &  & Expensive type - own coverage \\
\arrayrulecolor{Gray}\hline
FHP & Automobile horsepower & 1 - 13 & Ordered & 13 categories of horsepower \\
\arrayrulecolor{Gray}\hline
\multirow{3}{*}{InsuranceDuration} & \multirow{3}{*}{Insurance duration} & ID1 & \multirow{3}{*}{Ordered} & Up to 5 years \\
 &  & ID2 &  & From 6 to 10 years \\
 &  & ID3 &  & Greater than 10 years \\
\arrayrulecolor{Gray}\hline
\multirow{2}{*}{PaymentWay} & \multirow{2}{*}{Payment way} & C1 & \multirow{2}{*}{Unordered} & Cash \\
 &  & C2 &  & Credit card \\
\arrayrulecolor{Gray}\hline
Region & City population & 1-2; 4-14 & Unordered & 13 Administrative Regions of Greece \\
\arrayrulecolor{Gray}\hline
\multirow{3}{*}{VehicleAge} & \multirow{3}{*}{Vehicle age} & C1 & \multirow{3}{*}{Ordered} & New car (up to 7 years old) \\
 &  & C2 &  & Middle (from 8 to 15 years old) \\
 &  & C3 &  & Old (greater than 15 years old) \\
\arrayrulecolor{Gray}\hline
\multirow{3}{*}{SumInsured} & \multirow{3}{*}{Sum insured} & C1 & \multirow{3}{*}{Ordered} & Up to 5,000 Euros \\
 &  & C2 &  & Between 5,001 and 10,000 Euros \\
 &  & C3 &  & Greater than 10,000 Euros\\
\hline
\end{tabular}
\begin{tablenotes}\footnotesize
\item[*] While driver's age is by nature a continuous variable, in data analysis of this paper we will treat it as an ordered categorical variable with 57 levels instead.
\end{tablenotes}
\end{threeparttable}
}
\caption{Descriptions of the explanatory variables.}
\label{tab:variables}
\end{table}

\begin{figure}[H]
\begin{center}
\begin{subfigure}[h]{0.49\linewidth}
\includegraphics[width=\linewidth]{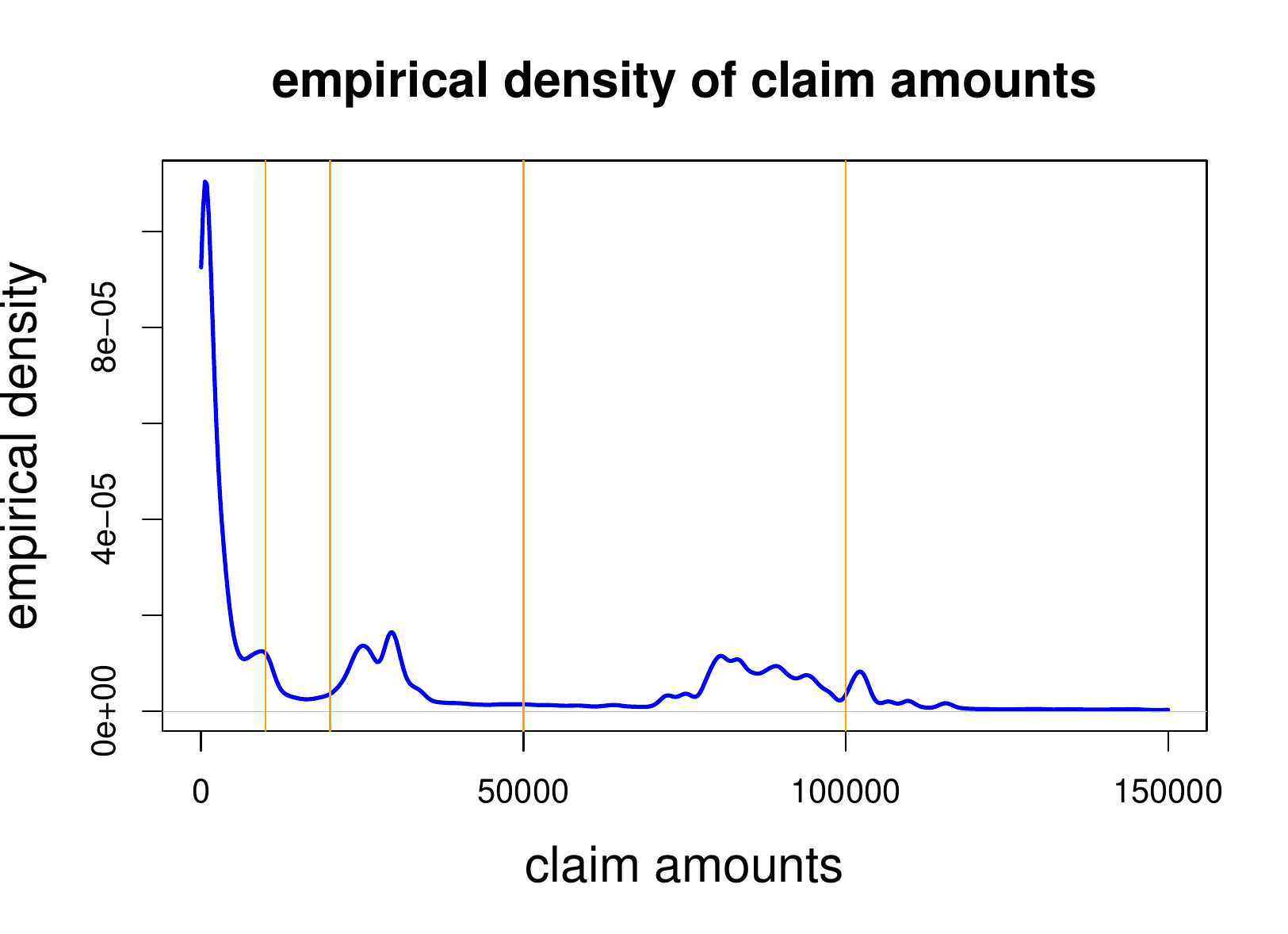}
\end{subfigure}
\hfill
\begin{subfigure}[h]{0.49\linewidth}
\includegraphics[width=\linewidth]{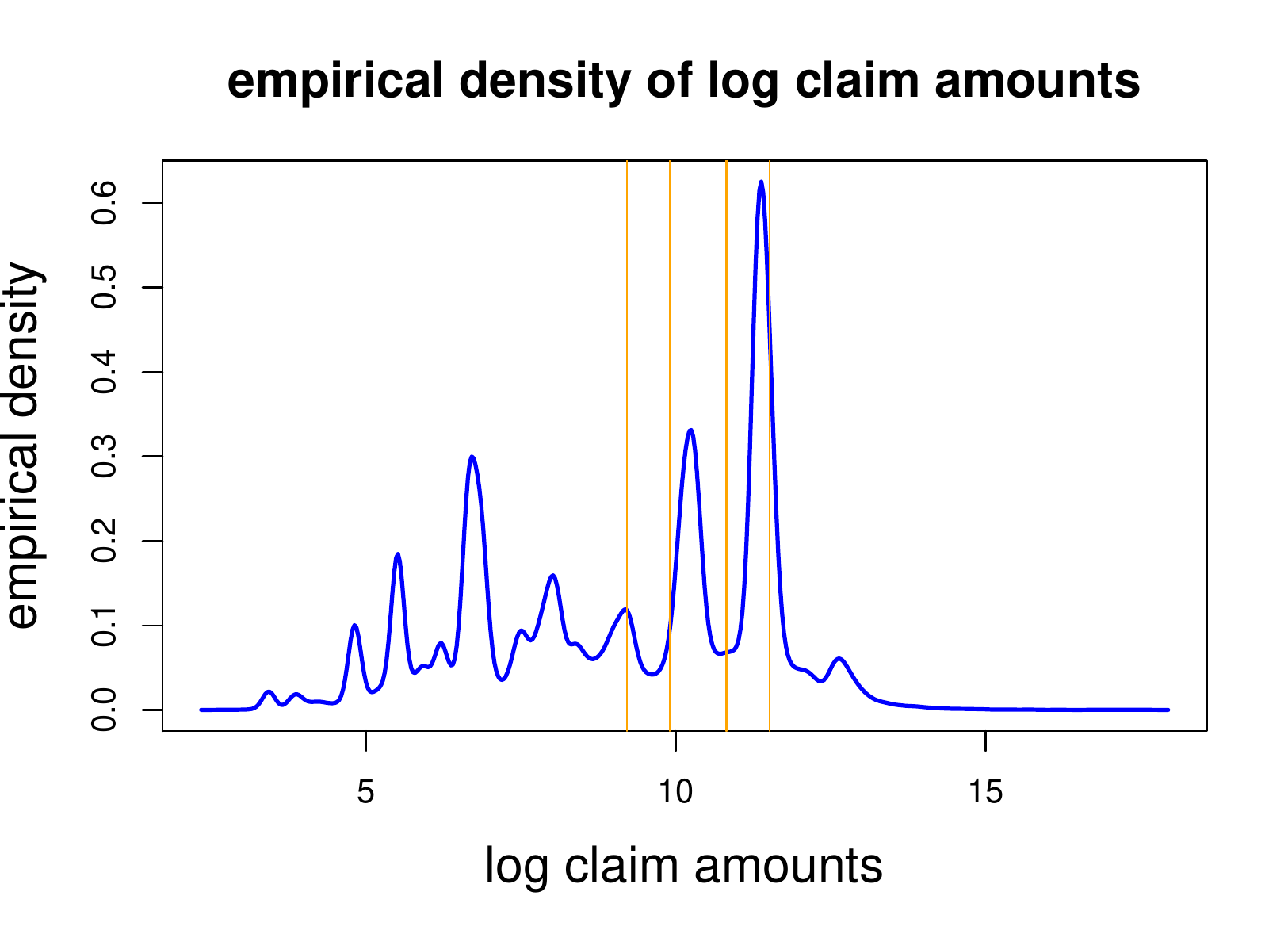}
\end{subfigure}
\end{center}
\caption{Empirical density of claim amounts (left panel) and log claim amounts (right panel); the orange vertical lines represent amounts of 10,000, 20,000, 50,000 and 100,000 respectively.}
\label{fig:density}
\end{figure}

\begin{figure}[!h]
\begin{center}
\begin{subfigure}[h]{0.49\linewidth}
\includegraphics[width=\linewidth]{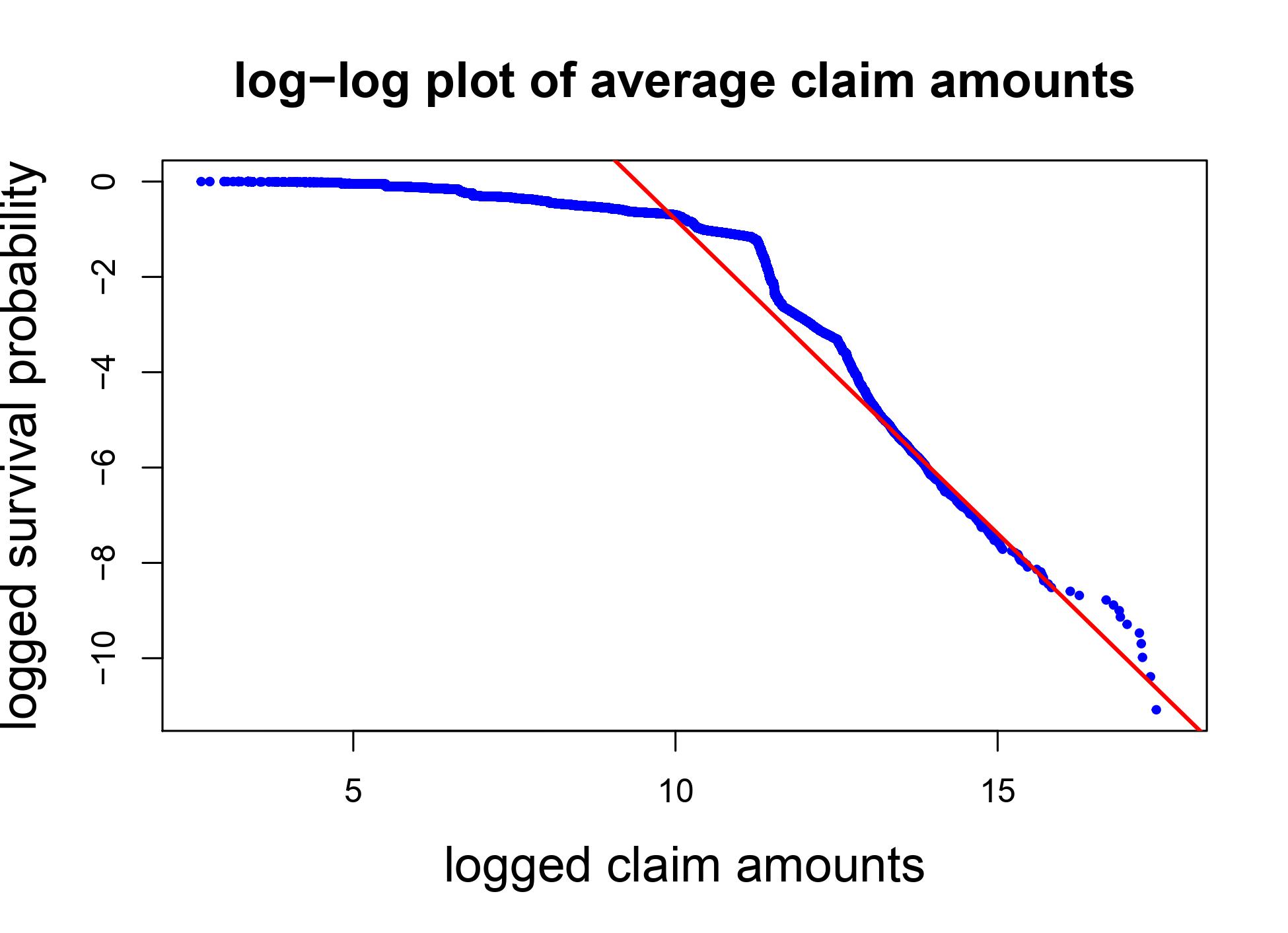}
\end{subfigure}
\begin{subfigure}[h]{0.49\linewidth}
\includegraphics[width=\linewidth]{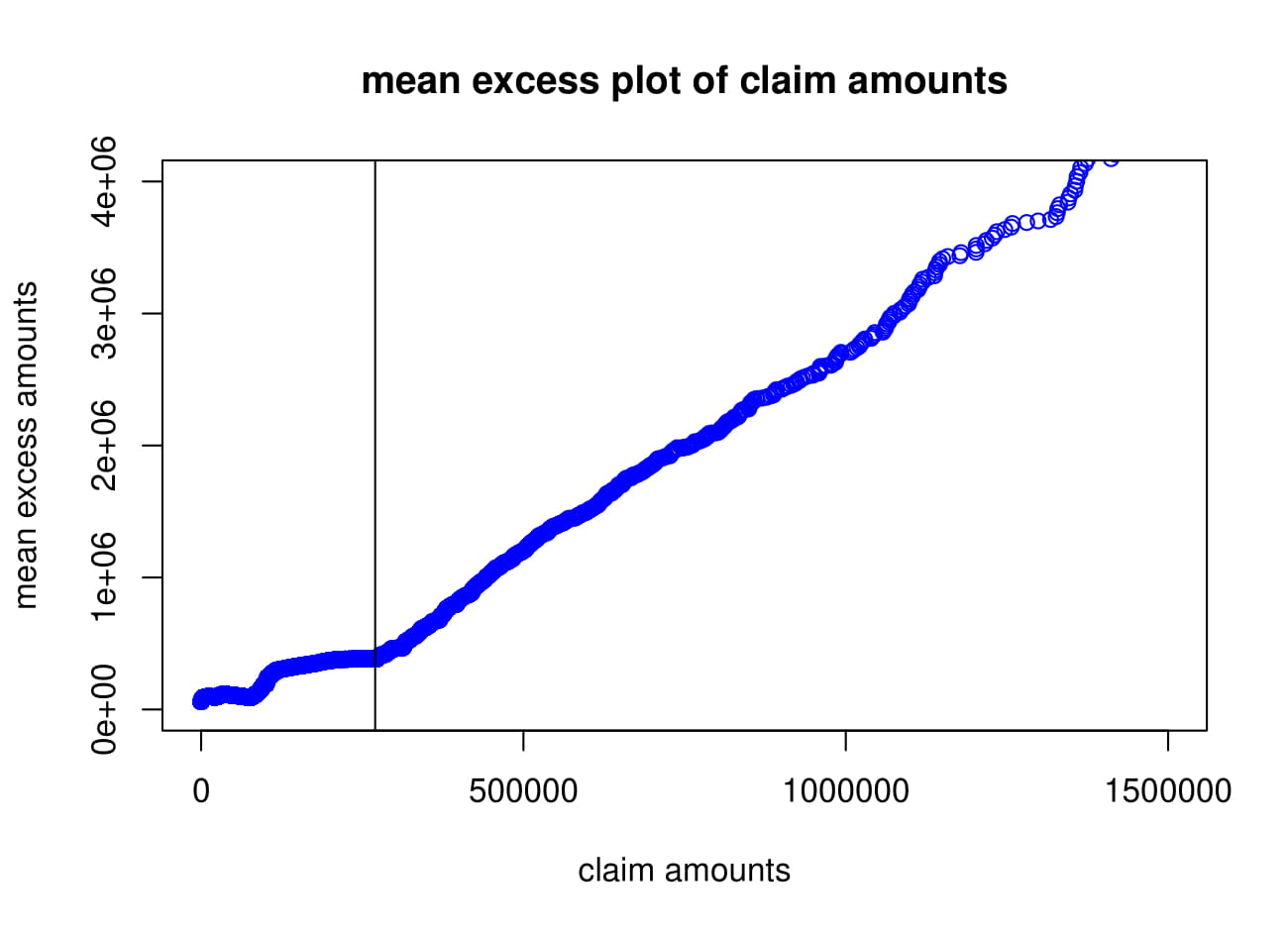}
\end{subfigure}
\end{center}
\caption{Left panel: log-log plot of the claim amounts; right panel: mean excess plot.}
\label{fig:loglogplot}
\end{figure}

\begin{figure}[!h]
\begin{center}
\begin{subfigure}[h]{0.49\linewidth}
\includegraphics[width=\linewidth]{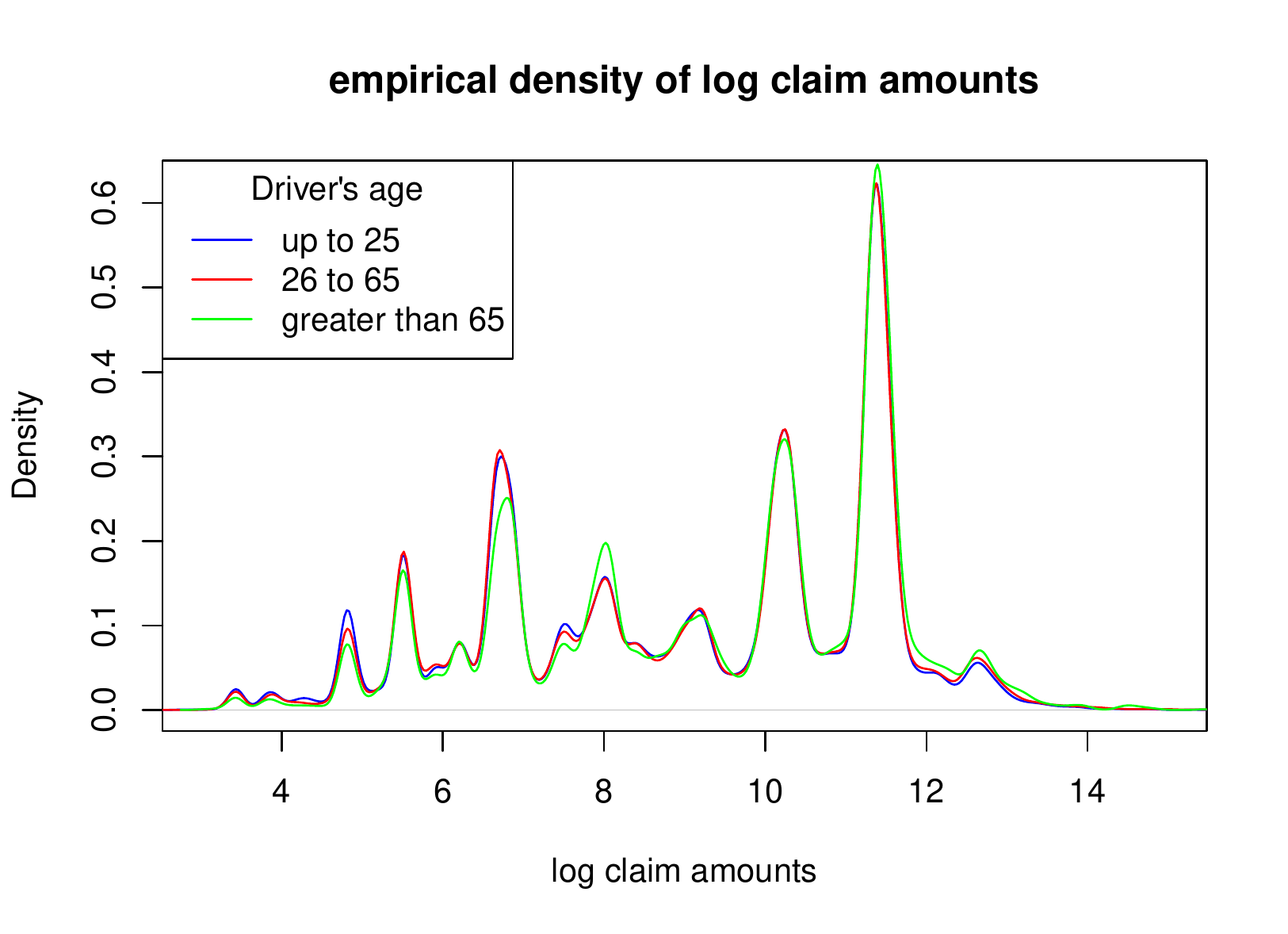}
\end{subfigure}
\hfill
\begin{subfigure}[h]{0.49\linewidth}
\includegraphics[width=\linewidth]{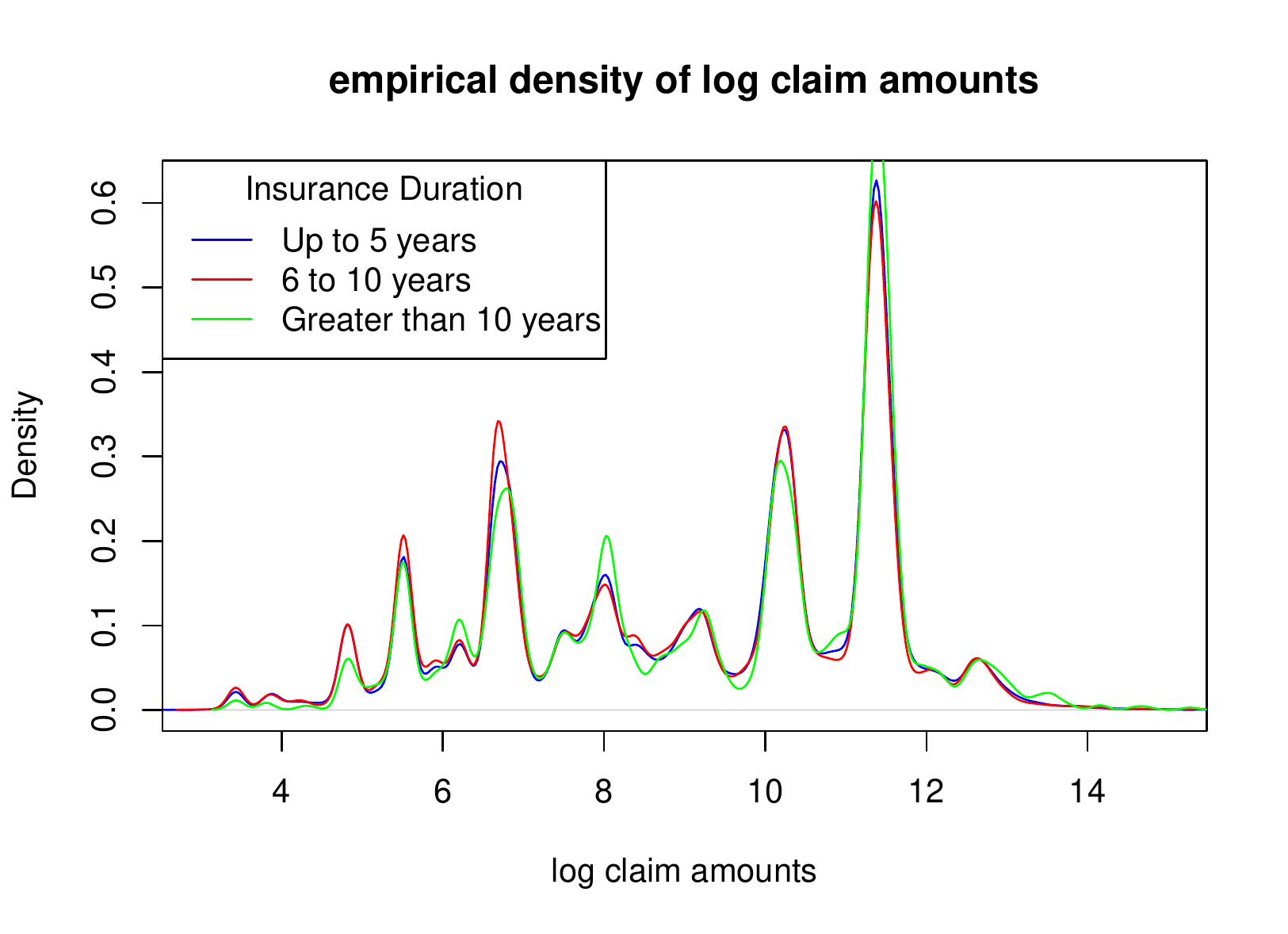}
\end{subfigure}
\hfill
\begin{subfigure}[h]{0.49\linewidth}
\includegraphics[width=\linewidth]{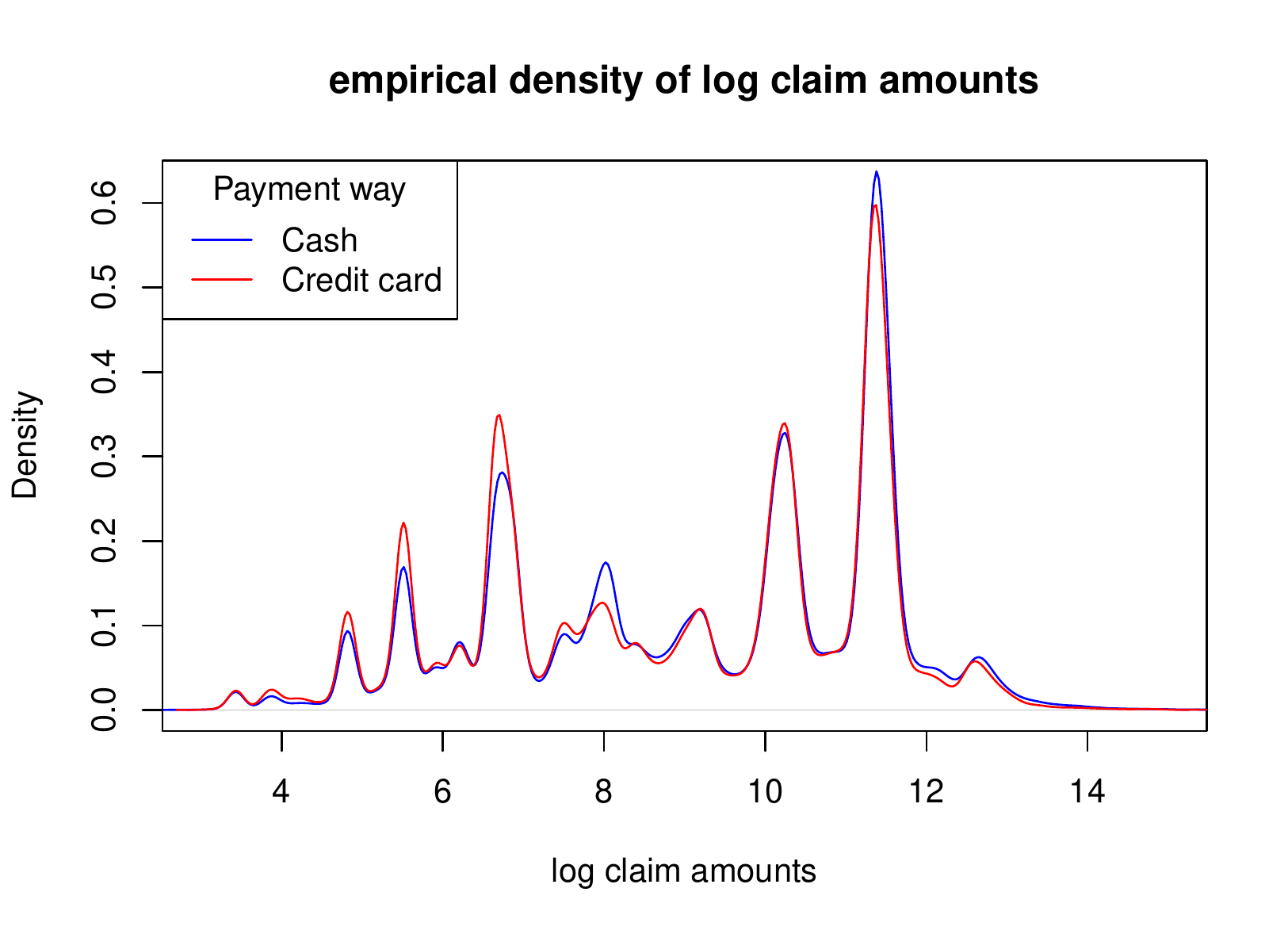}
\end{subfigure}
\hfill
\begin{subfigure}[h]{0.49\linewidth}
\includegraphics[width=\linewidth]{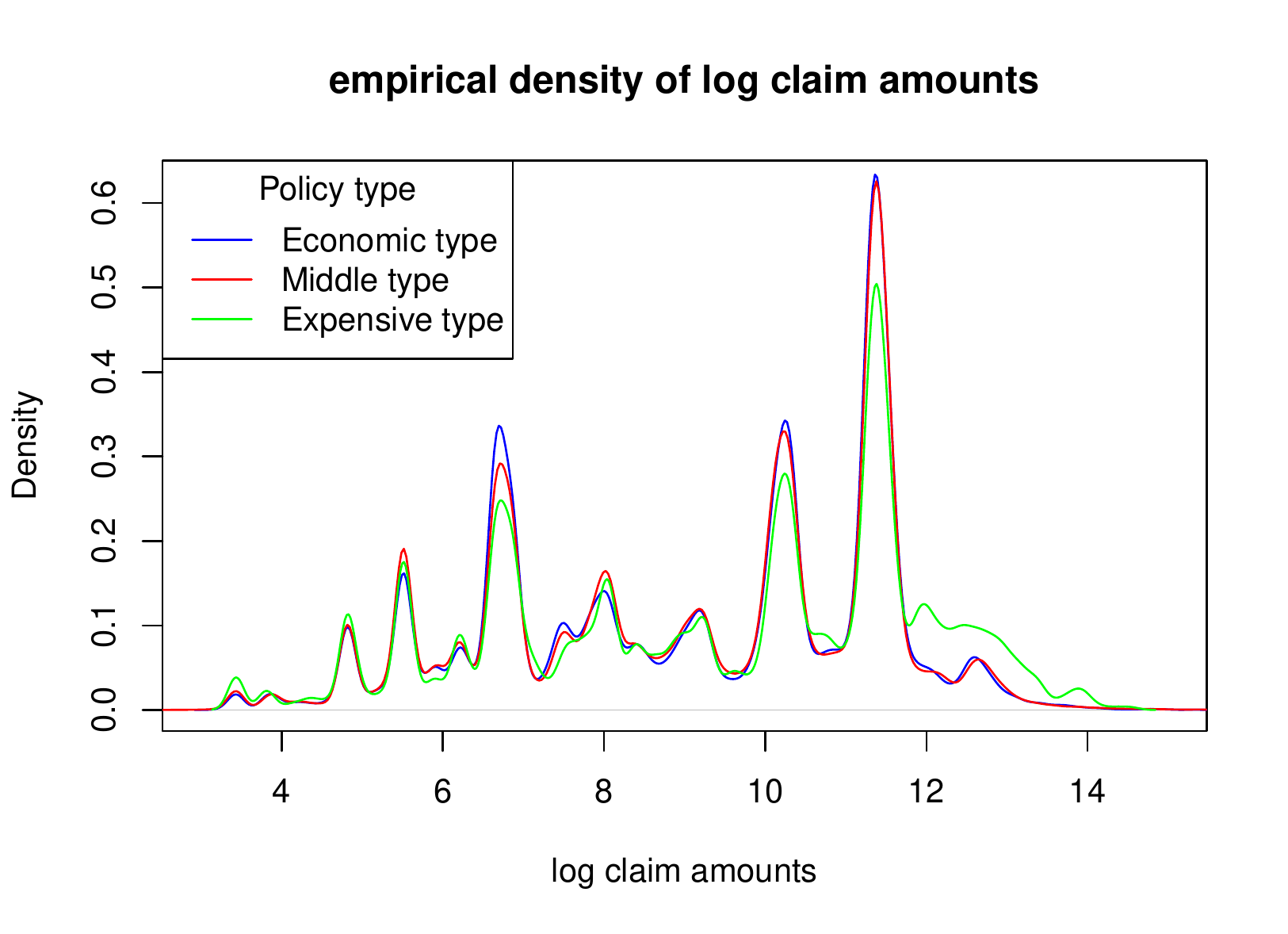}
\end{subfigure}
\end{center}
\caption{Marginal empirical density plots of log claim amounts for selected covariate components.}
\label{fig:density_x}
\end{figure}

An exploratory analysis was carried out in order to identify the challenges that need to be surmounted for efficiently modeling these property damage claim costs based on the subset of explanatory variables with the highest predictive power. Firstly, as we observe from Figures \ref{fig:density} and \ref{fig:loglogplot}, the empirical claim severity distribution is multimodal and heavy-tailed. In particular, the empirical density plot of the claim amounts in the left panel of Figure \ref{fig:density} shows that there are at least three major nodes or clusters in the empirical density function: one for small claim severities of $<$10,000, one for claim severities of about 30,000 and one for claim severities of about 80,000--100,000. Additionally, the density for the log claim amounts in the right panel of Figure \ref{fig:density} reveals even more complex distribution characteristics illustrated by many small peaks of the density function, especially for small claim sizes. Furthermore, regarding the heavy-tailed nature of the data, the log-log plot in the left panel of Figure \ref{fig:loglogplot} seems asymptotically linear (an asymptotic red straight line is fitted) with slope of roughly $-1.3$ (this represents the tail index $\alpha$ of the empirical distribution). Furthermore, the mean excess plot which is depicted in the right panel of Figure \ref{fig:loglogplot} appears linear when the claim size exceeds a threshold of around 270,000 (black vertical line), with asymptotic slope of 3.35 which also suggests $\alpha\approx 1.3$. Secondly, as far as the explanatory power of the variables is concerned, we studied the influence of the explanatory variables to the claim amounts through plotting the empirical density plots across each level of each variables. The results for the variables Driver's age, Insurance duration, Payment way and Policy type are displayed in Figure \ref{fig:density_x}. From Figure \ref{fig:density_x}, we see that some variables have some apparent effects on the peak (or probability) of each cluster instead of the position of each cluster. For example, from the bottom right panel, the policy with ``expensive type" has a higher probability assigned to the tail cluster and lower probabilities assigned to the remaining body clusters. Finally, it should be noted that considering all explanatory variables to be categorical, the 10 explanatory variables lead to 137 covariates in total. Therefore, since, as was previously mentioned, the impact of covariates on the claim severity distribution could be multi-fold (e.g.~covariates may affect cluster assignment probabilities, average claim severity given a particular cluster and/or tail-heaviness), an appropriate regression model for this Greek MTPL dataset should contain multiple regressors. Obviously, this will lead to a large number of parameters without parameter regularizations which can potentially result in an over-fitting problem and impede model interpretations. Therefore, these issues outline the importance of variable selection.

\subsection{Preliminary model fitting} \label{sec:data_descr:fit}
In this subsection we first explore probability distributions which may appropriately fit the distribution of claim amounts, ignoring the effects of covariates. As we observed multiple nodes and heavy-tailed characteristics of the claim amount distribution, it is natural to consider a finite mixture model with both light- and heavy-tailed mixture components to capture such characteristics. Motivated by \cite{BLOSTEIN201935} who propose a  finite mixture of various classes of distributions, a plausible benchmark is a mixture-Gamma Lomax model where the claim severity $Y$ is modelled by a density function

\begin{equation} \label{eq:density_mixture}
h_{Y}(y;\bm{\pi},\bm{\mu},\bm{\phi},\theta,\eta)
=\sum_{j=1}^g\pi_jf(y;\mu_j,\phi_j)+\pi_{g+1}h(y;\theta,\gamma),
\end{equation}
where $\bm{\pi}=(\pi_1,\ldots,\pi_{g+1})$ are the mixture probabilities, $\bm{\mu}=(\mu_1,\ldots,\mu_{g})$ and $\bm{\phi}=(\phi_1,\ldots,\phi_g)$ are the mean and dispersion parameters. $f$ is the Gamma density function for modeling the body, and $h$ is the Lomax density function for modeling the tail given by Equations (\ref{eq:gamma}) and (\ref{eq:lomax}) respectively.

Observing three major nodes in the density shown in the left panel of Figure \ref{fig:density}, we first start with the above model with $g=3$ components for the body. Summary statistics are shown in Table \ref{tab:fit_stat}, which compares the above model to several other classical unimodal models, including the Gamma (GA), Weibull (WEI), Weibull type three (WEI3), Generalized Gamma (GG) and Generalized Pareto distributions (GP) and a nonparametric maximum likelihood estimation (NPMLE) of a mixing distribution for mixtures of Exponential distributions. The probability
distribution functions of the WEI, GG, GP and the NPMLE for Exponential mixtures are given by Equations (1.1) to (1.5) of the supplementary material. 

The results show that the the NPMLE and in particular the three component mixture-Gamma Lomax model fit much better than all other preliminary models except for the mixture-Gamma Lomax case, revealing that a mixture-based model to capture distributional multimodality is necessary and important. However, one drawback of the latter mixture model is the instability of the estimation of the implied tail index $\alpha_0:=\exp\{\nu_0\}$. Model fitting has been tested across various numbers of Gamma components $g$ and we examine how robust the estimates of tail heaviness across different $g$'s is. The results are shown in Table \ref{tab:fit2}. We see that the implied tail index $\alpha_0$ fluctuates greatly from smaller than $1.5$ to greater than $1.8$ across $g$, which does not make sense in practice because $g$ should control the body part of severity distribution only and bring very little impact on the estimated tail index. The main reason of seeing such an undesirable phenomenon is that the Lomax distribution, which is designed to capture the tail distribution, also calibrates to the body of the distribution and the MLE approach is found not very stable in estimating the tail parameter. This motivates the use of the composite model proposed in Section \ref{sec:model}, where the tail component only interacts with the body via the mixture probability. Also, note that while both AIC and BIC suggest a bigger number of components for the body (the optimal $g$ goes way beyond 15), this mainly reflects improvements of fitting small claims below 3,000. This should not be over-weighted because exact prediction of these small claims is less relevant in pricing, while excessive model complexity may impede interpretability. As a result, AIC and BIC may be less appropriate in determining the number of mixture components under this dataset.

\begin{table}[!h]
\centering
\begin{tabular}{l|rrrrrr}
  \multicolumn{1}{l|}{} & \multicolumn{1}{l}{GA} & \multicolumn{1}{l}{WEI} & \multicolumn{1}{l}{GG} & \multicolumn{1}{l}{GP} & \multicolumn{1}{l}{NPMLE} & \multicolumn{1}{l}{3-Gamma Lomax} \\ \hline
DF & 2 & 2 & 3 & 2 & 9 & 11 \\
log-likelihood & -743,608 & -740,360 & -740,248 & -748,596 & -732,695  & \textbf{-723,447} \\
AIC & 1,487,221 & 1,480,725 & 1,480,503 & 1,497,197 & 1,465,409 & \textbf{1,446,917} \\
BIC & 1,487,239 & 1,480,743 & 1,480,530 & 1,497,215 & 1,465,491 & \textbf{1,447,017}
\end{tabular}
\caption{Distributional fitting results among different model candidates. GA: Gamma; WEI: Weibull; GG: Generalized Gamma; GP: Generalized Pareto; NPMLE: NPMLE for Exponential mixtures.}
\label{tab:fit_stat}
\end{table}

\begin{table}[!h]
\centering
\begin{tabular}{l|rrrrrrr}
\multicolumn{1}{c|}{} & \multicolumn{7}{c}{number of body components $g$ for $g$-Gamma Lomax distribution} \\
\multicolumn{1}{c|}{} & \multicolumn{1}{c}{2} & \multicolumn{1}{c}{3} & \multicolumn{1}{c}{5} & \multicolumn{1}{c}{6} & \multicolumn{1}{c}{8} & \multicolumn{1}{c}{12} & \multicolumn{1}{c}{15} \\ \hline
DF & 8 & 11 & 17 & 20 & 26 & 38 & 47 \\
tail index $\alpha_0$ & 1.8591 & 1.5839 & 1.7958 & 1.4969 & 1.7074 & 1.6250 & 1.4771 \\
log-likelihood & -740,000 & -723,447 & -719,375 & -719,352 & -718,799 & -717,294 & -713,341 \\
AIC & 1,480,016 & 1,446,917 & 1,438,784 & 1,438,744 & 1,437,650 & 1,434,663 & 1,426,776 \\
BIC & 1,480,089 & 1,447,017 & 1,438,939 & 1,438,925 & 1,437,886 & 1,435,008 & 1,427,203
\end{tabular}
\caption{Distributional fitting results of $g$-Gamma Lomax finite mixture models among different $g$.}
\label{tab:fit2}
\end{table}

\section{Fitting results} \label{sec:application}
In this section, we analyze the performance under the proposed mixture composite model with multi-type feature regularization
for the covariates.

\subsection{Distributional fitting}
\begin{figure}[!h]
\begin{center}
\begin{subfigure}[h]{0.49\linewidth}
\includegraphics[width=\linewidth]{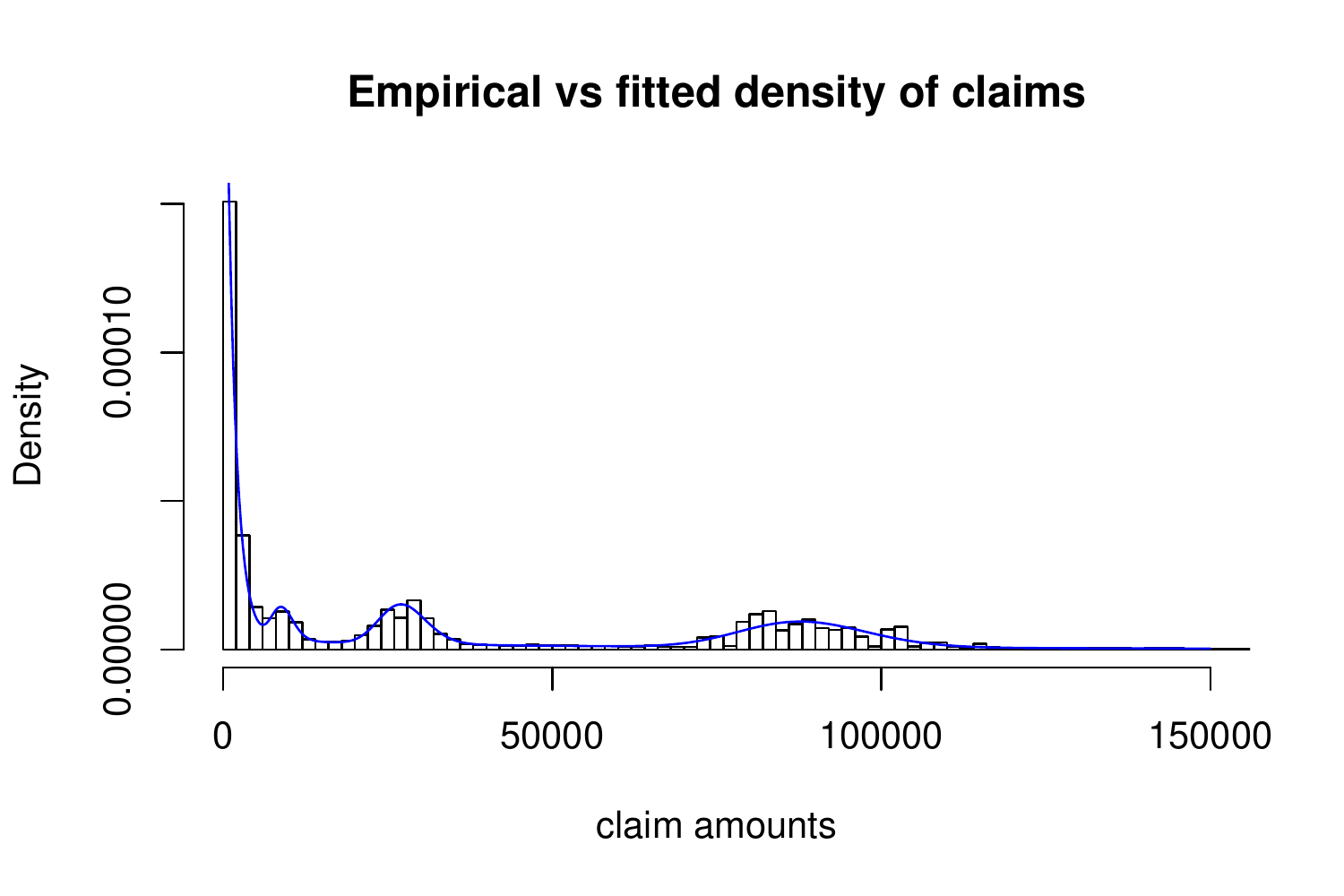}
\end{subfigure}
\hfill
\begin{subfigure}[h]{0.49\linewidth}
\includegraphics[width=\linewidth]{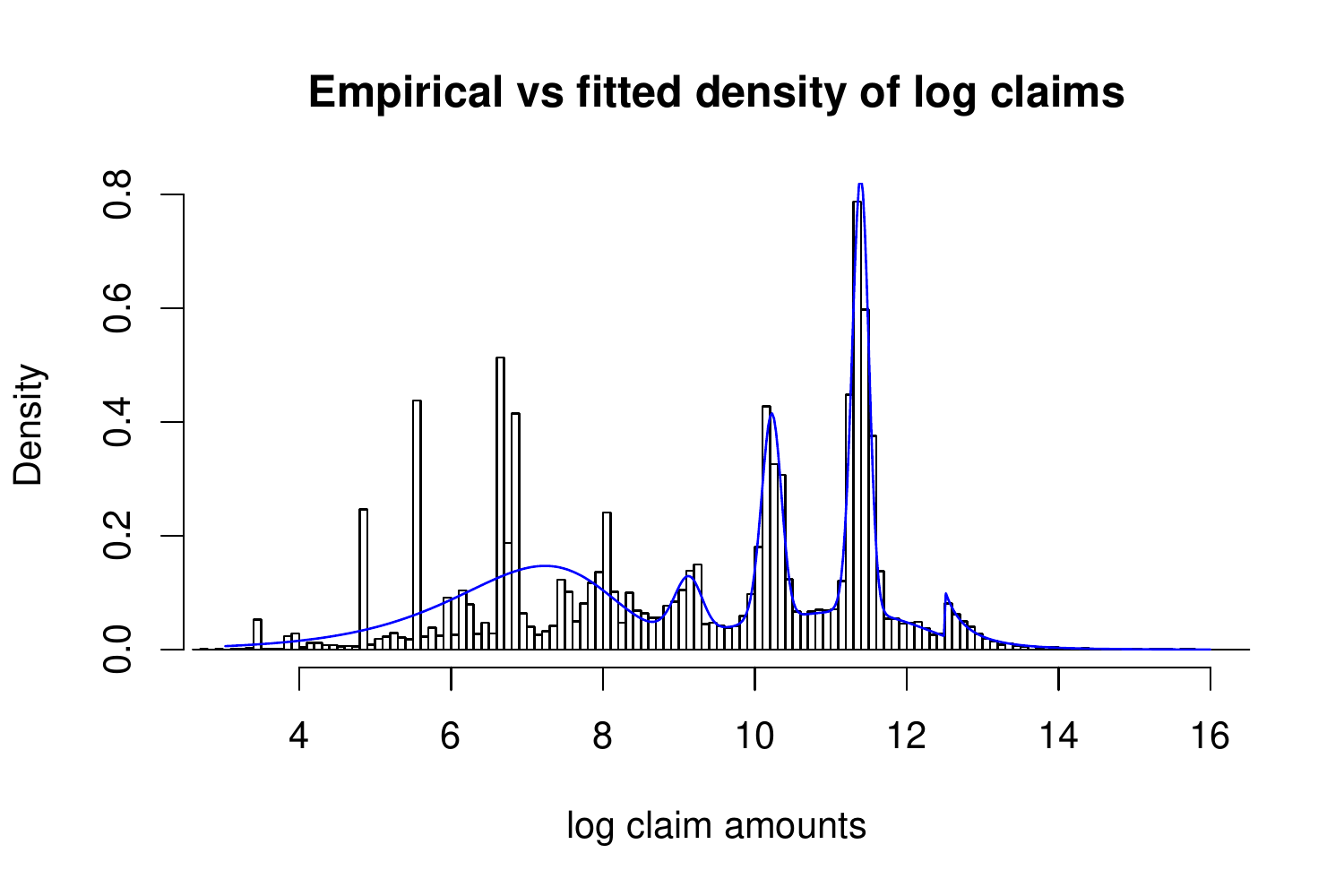}
\end{subfigure}
\end{center}
\caption{Empirical vs.~fitted density of claim amounts (left panel) and log claim amounts (right panel).}
\label{fig:disn:density}
\end{figure}

\begin{figure}[!h]
\begin{center}
\begin{subfigure}[h]{0.49\linewidth}
\includegraphics[width=\linewidth]{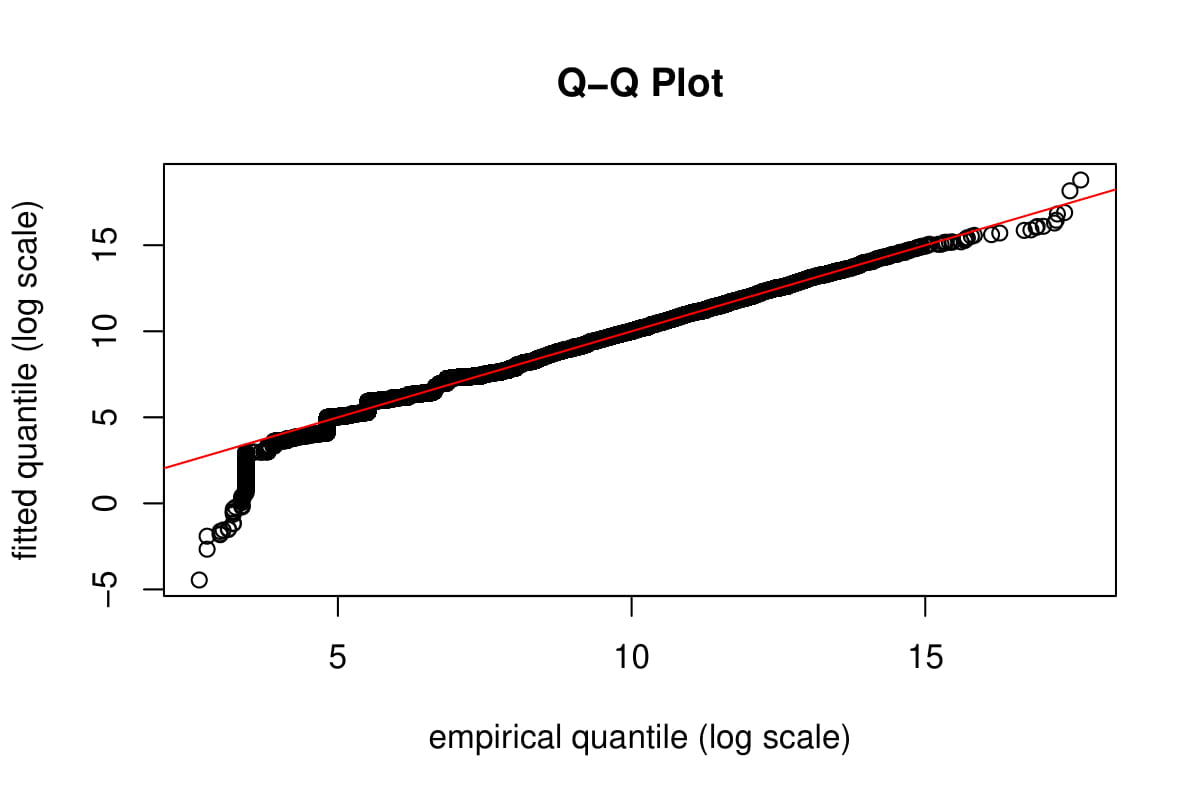}
\end{subfigure}
\hfill
\begin{subfigure}[h]{0.49\linewidth}
\includegraphics[width=\linewidth]{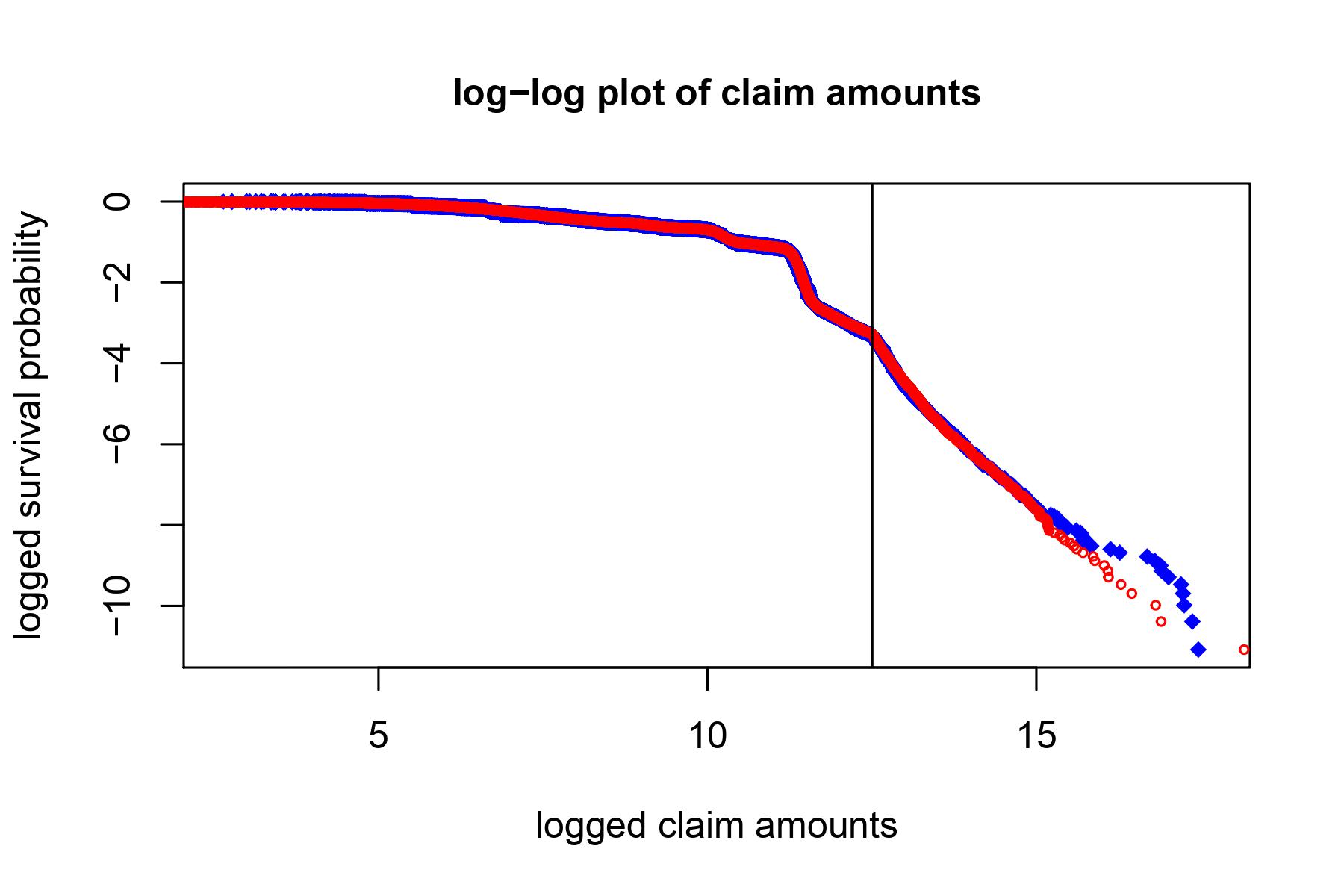}
\end{subfigure}
\end{center}
\caption{Left panel: Q-Q plot; right panel: empirical (blue dots) vs.~fitted (red dots) log-log plot of claim amounts.}
\label{fig:disn:qq}
\end{figure}

As in the preliminary analysis, we first fit the distribution of claim amounts under the proposed modeling framework, without considering covariates. Notice that in the mean excess plot of claim amounts (right panel of Figure \ref{fig:loglogplot}) under the preliminary analysis, the plot becomes linear beyond claim severity of $270,000$ indicated by the vertical line of the plot. As a result, a reasonable choice of the splicing threshold is $\tau=270,000$. After fitting the proposed model across various choices of the number of body components $g$, we find that $g=5$ is the minimum number of components required to capture all the density nodes above a claim severity of $10,000$. The summary statistics of the fitted model is presented in Table \ref{tab:disn:summary_fit}, the fitted versus empirical density plots are shown in Figure \ref{fig:disn:density}, and the Q-Q and log-log plots of claim sizes are illustrated in Figure \ref{fig:disn:qq}. The model estimated tail index of 1.3817 roughly resembles that estimated by the asymptotic slope of the log-log plot which is 1.3 (left panel of Figure \ref{fig:loglogplot}). Also, as expected we find that the model estimated tail index is robust across various choices of $g$. The density plots indicate that the fitted distribution captures all nodes representing a larger amount of claims, with multiple small nodes for smaller claims explained smoothly by one single component (to be precise, by the subgroup $j=1$ indicated by Figure \ref{tab:disn:summary_fit}). From the Q-Q plot, we see that the fitting performance is satisfactory except for very small claims ($y<100$) which are less relevant from an insurance pricing perspective. The fitted versus empirical log-log plot also indicates satisfactory fitting performance for the tail part. The fitted log-likelihood is $-719,309.1$, with $\text{AIC}=1,438,652$ and $\text{BIC}=1,438,640$, which is even slightly superior compared to the 5-Gamma Lomax distribution illustrated in Table \ref{tab:fit2}.

\begin{table}[!h]
\centering
\begin{tabular}{l|rrrrrr}
\hline
subgroup $j$ & \multicolumn{1}{c}{1} & \multicolumn{1}{c}{2} & \multicolumn{1}{c}{3} & \multicolumn{1}{c}{4} & \multicolumn{1}{c}{5} & \multicolumn{1}{c}{6} \\ \hline
classification probability & 0.3790 & 0.0450 & 0.1184 & 0.2099 & 0.2111 & 0.0365 \\
subgroup mean & 1,339 & 9,184 & 27,527 & 88,818 & 71,439 & 616,666 \\
subgroup dispersion $\phi_j$ & 0.0331 & 0.9985 & 0.0119 & 1.3734 & 0.0169 &  \\
tail index $\exp\{\hat{\nu}\}$ &  &  &  &  &  & 1.3817 \\ \hline
\end{tabular}
\caption{Summary of the fitted mixture-Gamma Lomax composite distribution.}
\label{tab:disn:summary_fit}
\end{table}

\subsection{Effects of the covariates}
We now include all variables described in Table \ref{tab:variables} and fit our proposed mixture composite regression model with LASSO and SCAD regularizations. Since all variables are included as categorical covariates, there is a total of $D=138$ parameters for each set of regressors. The grid searches are performed on $\lambda_1\in\{0.1n,0.2n,\ldots,6553.6n\}$, $\lambda_2\in\{0.1n_b,0.2n_b,\ldots,6553.6n_b\}$ and $\lambda_3\in\{0.1n_t,0.2n_t,\ldots,6553.6n_t\}$ to find optimal tuning parameters. The fitting performances for different model settings (without regression vs.~with regression), penalty settings (without regularization vs.~with regularization) and model selection criteria (pAIC, pBIC or CV with one standard deviation rule) are summarized in Table \ref{tab:reg:summary_performance}. With a large number of covariates, we first note from the table that regularization of regression coefficients is a must, or else some parameters would diverge to very large values (due to overfitting), causing the algorithm to collapse eventually because of numerical instability (spurious solutions). As a result, for a full model as a benchmark for comparison, we need to apply a weak LASSO penalty, which sets very small $\lambda_l>0$ ($l=1,2,3$) such that no covariates are removed or merged. We next investigate the effect of the model selection criteria to the resulting fitted model. For both LASSO and SCAD penalties chosen as regularization function, pAIC results to very large models with a total of $\mathcal{N}=809$ parameters for LASSO and $\mathcal{N}=613$ for SCAD, indicating that many variables have predictive power on explaining all parts (body, tail and subgroup probabilities) of the claim severity distribution. The large number of predictors, however, makes the fitted models very difficult to interpret. Also, the selected model severities can vary greatly across various choices of initializations or grids for tuning parameters, because we find that model sizes within a range of about 150 to 1,000 parameters all have very similar AICs. In contrast, pBIC heavily penalizes the regression parameters and leads to a very small fitted model which chooses very few or even no variables useful to describe any parts of the distribution.

\begin{table}[!h]
\centering
\begin{tabular}{lrrrr}
\hline
 \multicolumn{1}{c}{Model selection criteria} & \multicolumn{1}{c}{\# parameters} & \multicolumn{1}{c}{log-likelihood} & \multicolumn{1}{c}{AIC} & \multicolumn{1}{c}{BIC} \\ \cline{1-5} 
$\mathcal{L}_n$ with without regression & 17 & -719,309 & 1,438,652 & 1,438,807 \\
$\mathcal{L}_n$ with without penalty & NA & NA & NA & NA \\\hline
$\mathcal{L}_n$ + weak penalty only & 1,524 & -717,969 & 1,438,987 & 1,452,826 \\\hline
$\mathcal{L}_n$ + LASSO penalty w/ pAIC before refit & 809 & -718,312 & 1,438,242 & 1,445,589 \\
$\mathcal{L}_n$ + LASSO penalty w/ pBIC before refit & 42 & -719,139 & 1,438,362 & \textbf{1,438,743} \\
$\mathcal{L}_n$ + LASSO penalty w/ CV before refit & 112 & -719,029 & 1,438,282 & 1,439,299 \\
$\mathcal{L}_n$ + LASSO penalty w/ CV after refit & 112 & -718,779 & \textbf{1,437,781} & \textbf{1,438,798} \\\hline
$\mathcal{L}_n$ + SCAD penalty w/ pAIC before refit & 613 & -718,324 & 1,437,873 & 1,443,439 \\
$\mathcal{L}_n$ + SCAD penalty w/ pBIC before refit & 17 & -719,309 & 1,438,652 & 1,438,807 \\
$\mathcal{L}_n$ + SCAD penalty w/ CV before refit & 197 & -718,925 & 1,438,244 & 1,440,033 \\
$\mathcal{L}_n$ + SCAD penalty w/ CV after refit & 197 & -718,925 & 1,438,244 & 1,440,033 \\ \hline
\end{tabular}
\caption{Summary of regression model selection and performance across various settings.}
\label{tab:reg:summary_performance}
\end{table}

On the other hand, using CV with a one standard deviation rule provides fitted models with more reasonable complexity ($\mathcal{N}=112$ under LASSO or $\mathcal{N}=197$ under SCAD). Both LASSO and SCAD penalties suggest that there
are not any systematic effects in the tails that are explained by the available variables. On the other hand, both penalty functions reveal similar sets of variables important to explain the body and subgroup probability parts. The higher model complexity under SCAD is mainly due to more granular mergers among different levels of some variables (such as driver's age). Under LASSO, the resulting AIC under the CV approach is close to that under the corresponding pAIC approach, while the BIC is just slightly inferior to the pBIC approach. 

Table \ref{tab:reg:summary_performance} also shows the performance of the LASSO and SCAD CV-selected models after 
the model refit procedure. Recall from Section \ref{sec:est:tuning} that the refitting procedure involves re-estimation of parameters for the shrinked model with regularization terms excluded to reduce biasedness in the estimated parameters. For the LASSO penalty, the improvements of the log-likelihood, AIC and BIC are all expected after refitting. For the SCAD penalty, since the concavity of SCAD penalty function already mitigates the biasedness of estimated parameters (\cite{fan2001variable}), there is no apparent improvement of the fitting performance after performing the refitting procedure. After refitting, the LASSO penalty approach results to superior fitting performance compared to the SCAD approach, as evidenced by lower AICs and BICs. As a result for conciseness concern, we focus solely on the CV approach with LASSO penalty as model selection criterion in the following analysis. 

The final refitted model suggests that the subgroup probabilities $\pi_j(\bm{x};\bm{\alpha})$, $1\le j \le g+1$,
are influenced by the variables as follows.
\begin{itemize}
\item Driver's age: The model merges this variable into 6 categories -- $\{18-30,31-34,35-41,42-51,52-72,73+\}$.
\item Car cubism: 3 categories -- $\{0-12,13-14,15-18\}$.
\item Policy type: Expensive type causes higher probability of a claim falling into the tail component.
\item Horsepower: 2 categories -- $\{1-3,4-13\}$. Larger tail probability for higher horsepower.
\item Payment way: Cash payment results in higher tail probability.
\item Region: 4 regions (Regions 4, 8, 9, 12) differs the subgroup probabilities from other regions.
\item Car brand, insurance duration, vehicle age and sum insured: No significant impacts.
\end{itemize}

\begin{figure}[!h]
\begin{center}
\begin{subfigure}[h]{0.49\linewidth}
\includegraphics[width=\linewidth]{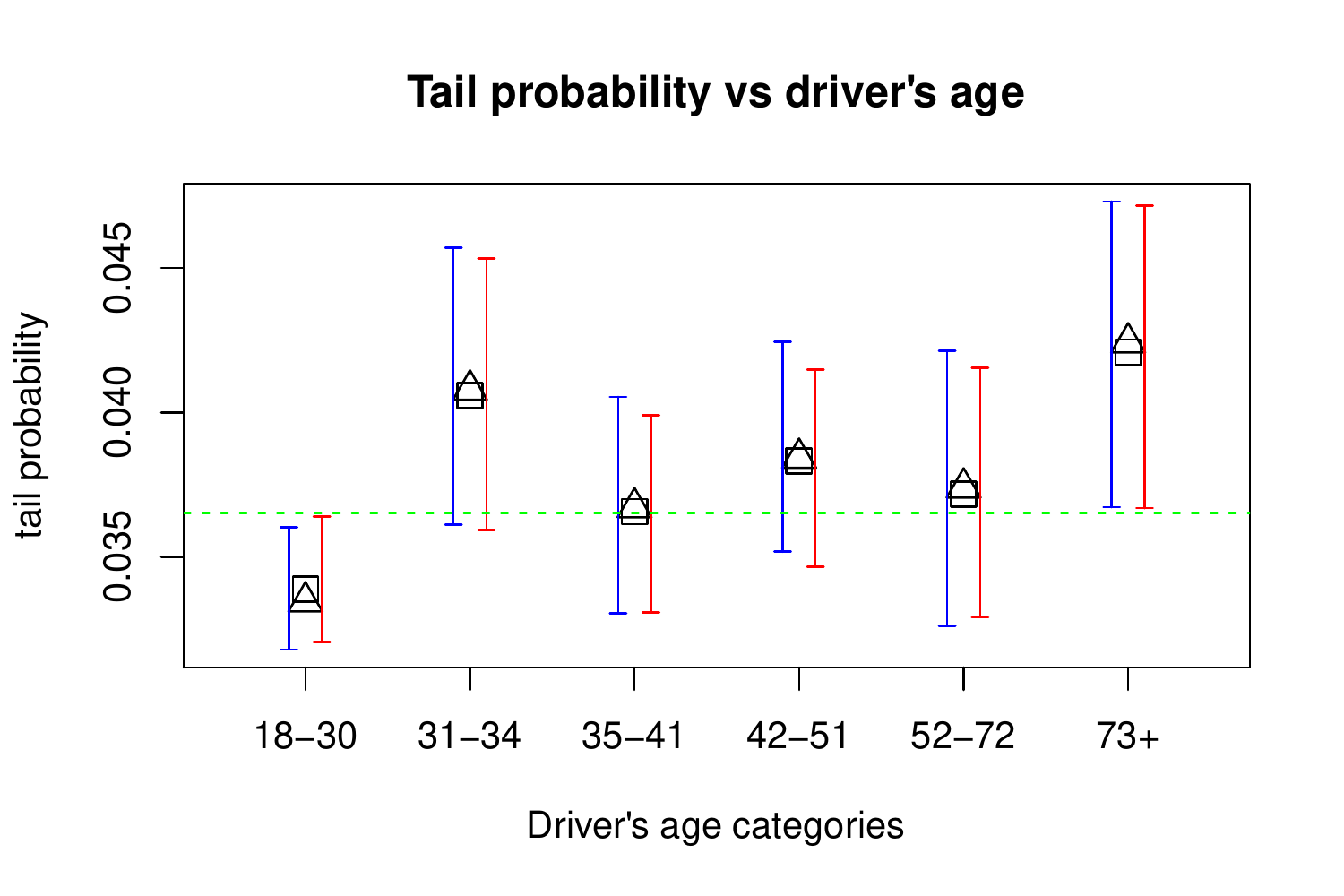}
\end{subfigure}
\begin{subfigure}[h]{0.49\linewidth}
\includegraphics[width=\linewidth]{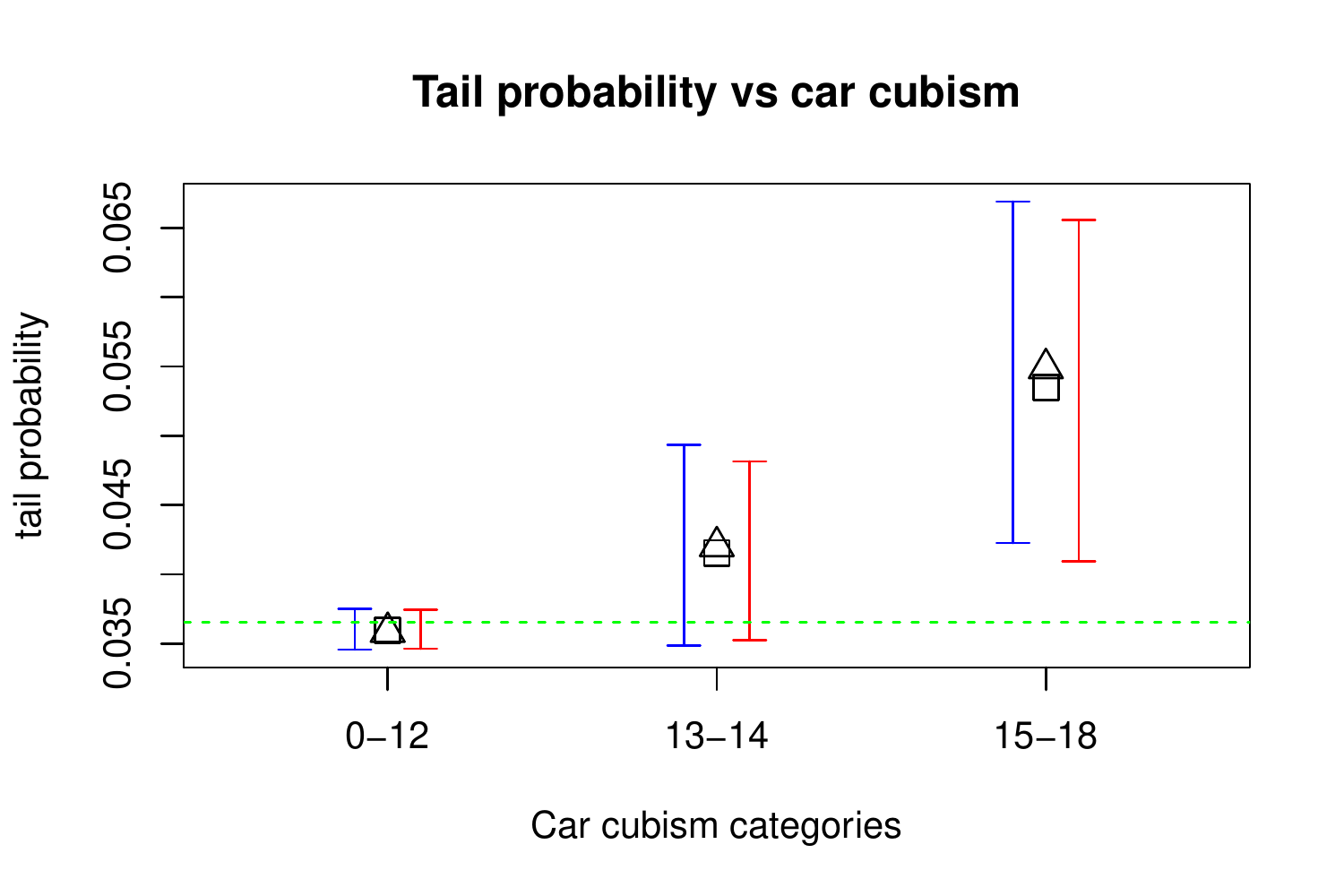}
\end{subfigure}
\begin{subfigure}[h]{0.49\linewidth}
\includegraphics[width=\linewidth]{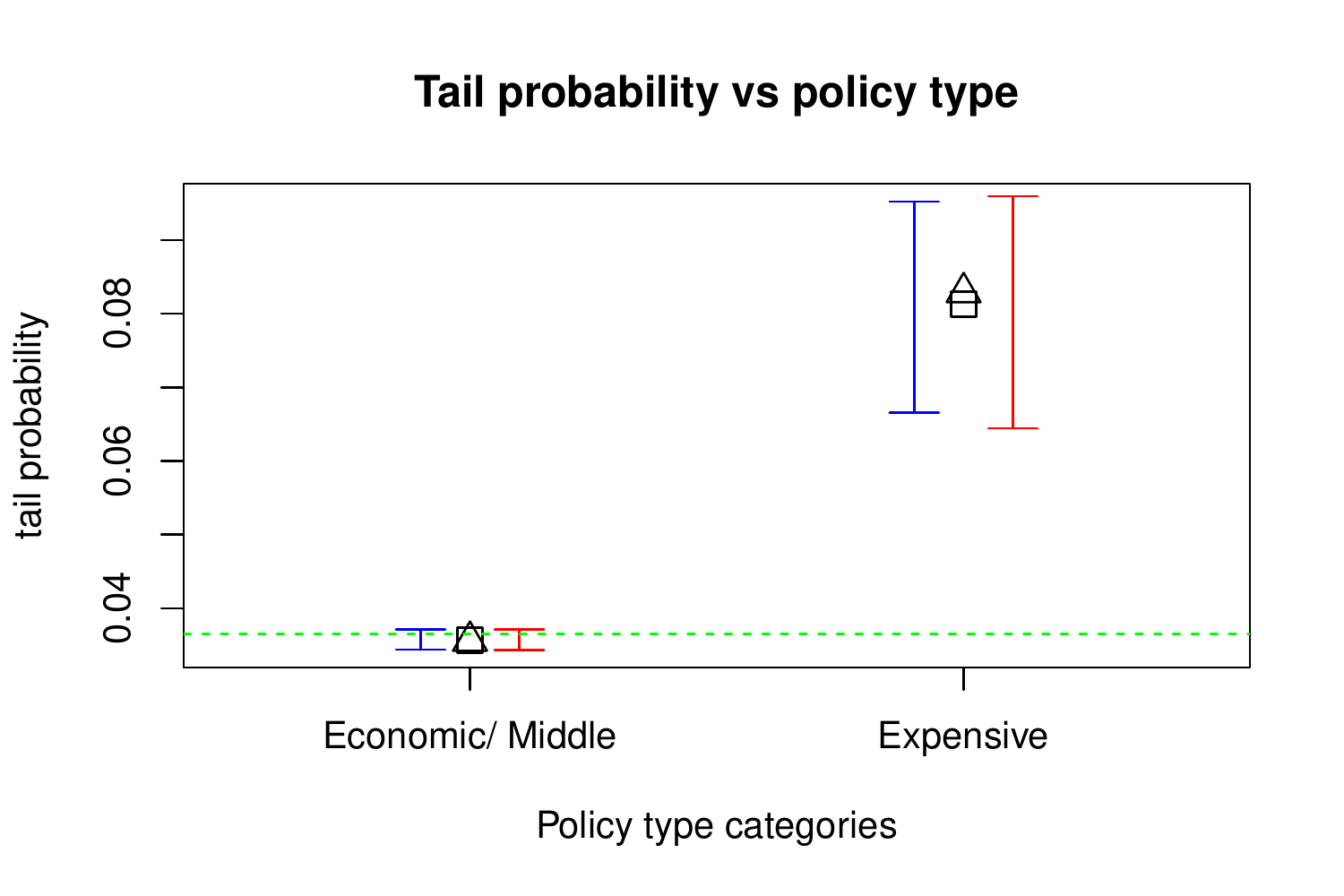}
\end{subfigure}
\begin{subfigure}[h]{0.49\linewidth}
\includegraphics[width=\linewidth]{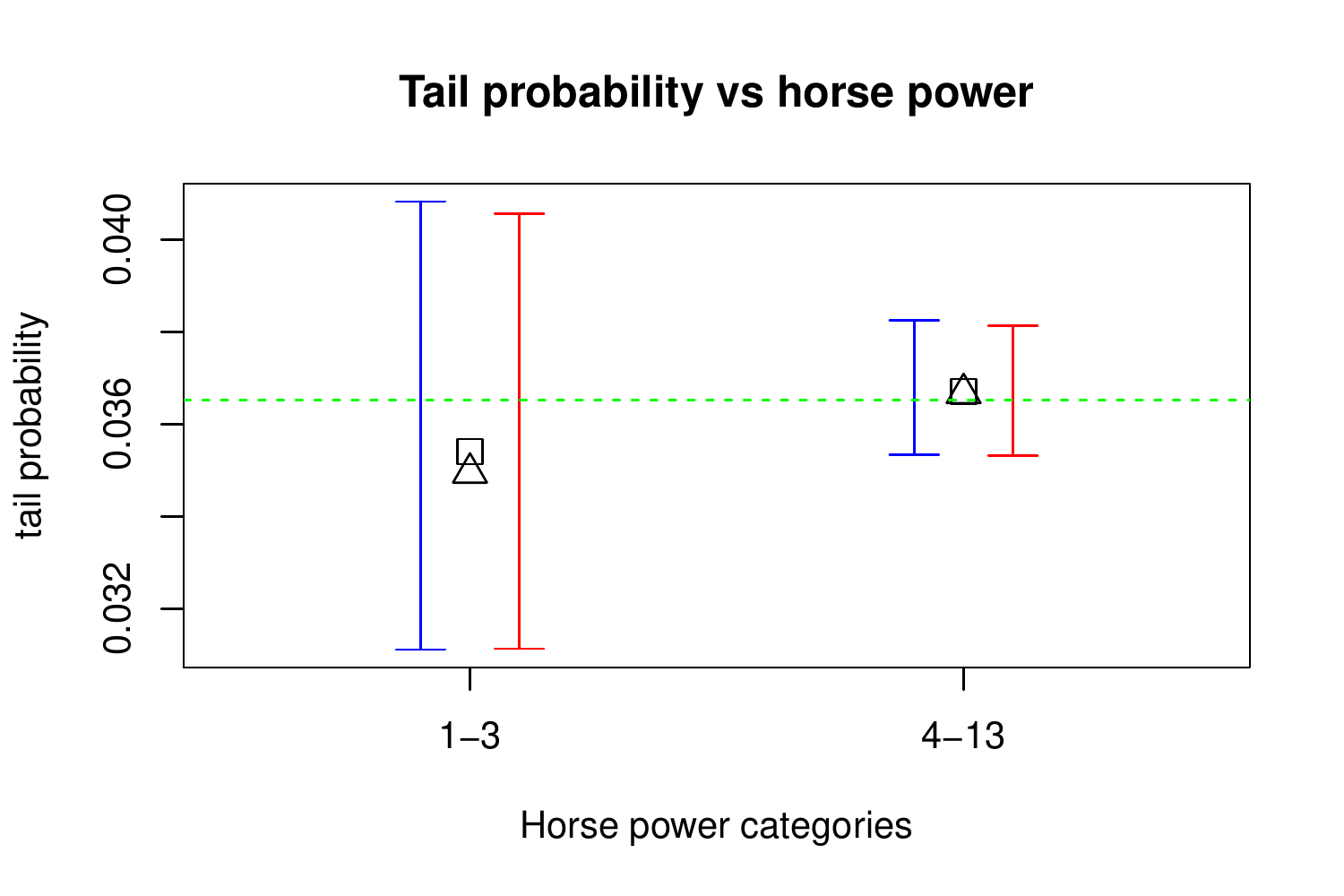}
\end{subfigure}
\begin{subfigure}[h]{0.49\linewidth}
\includegraphics[width=\linewidth]{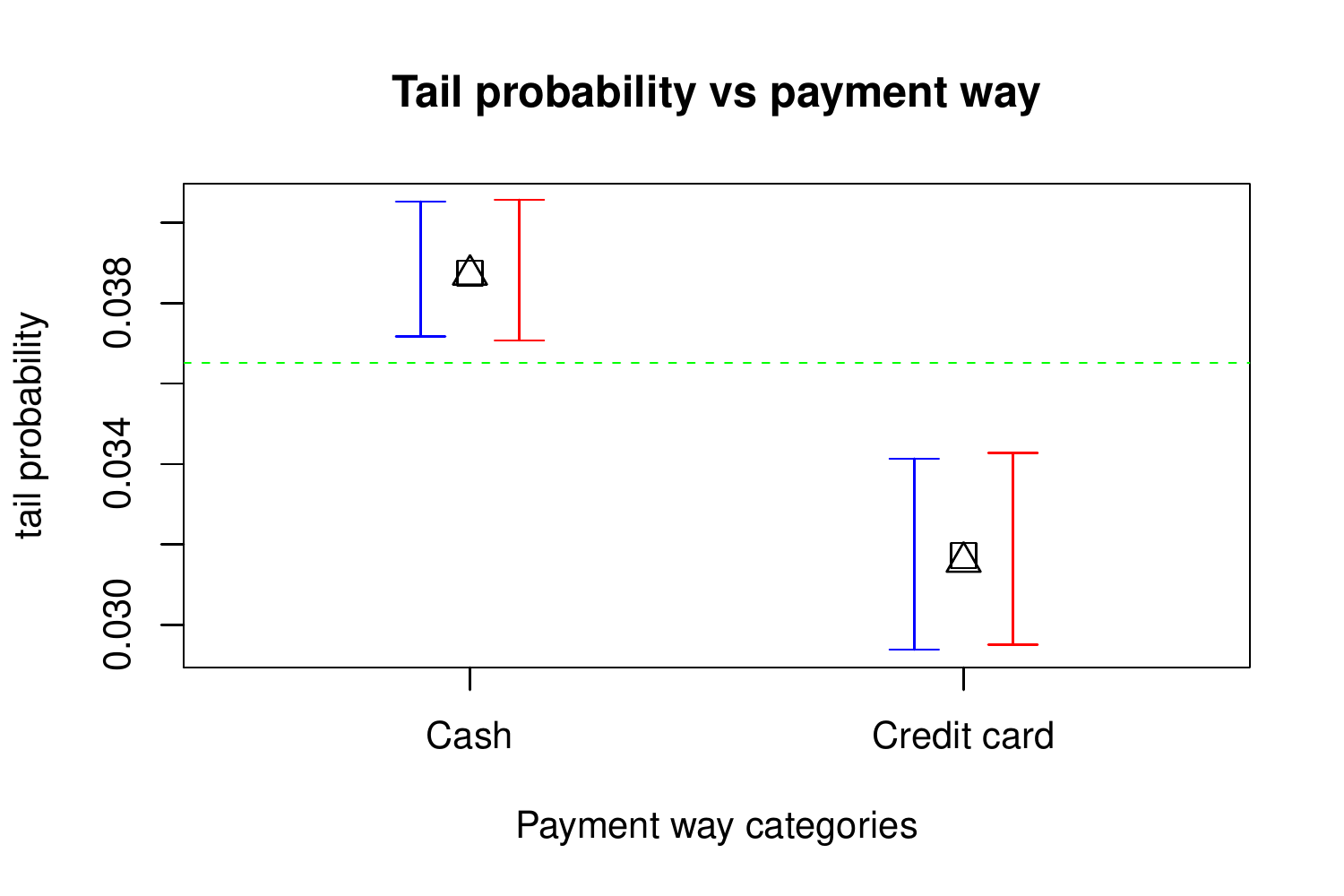}
\end{subfigure}
\begin{subfigure}[h]{0.49\linewidth}
\includegraphics[width=\linewidth]{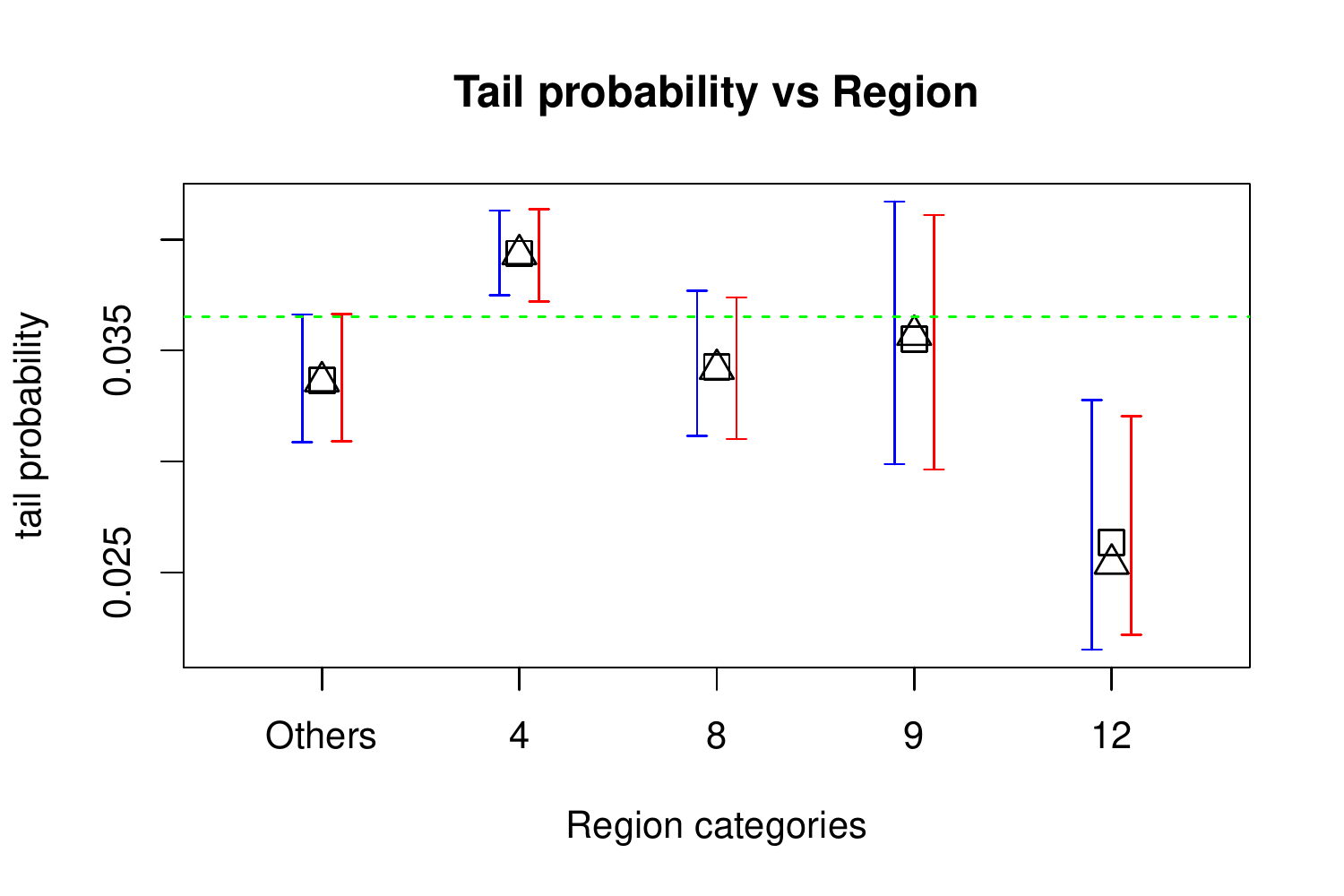}
\end{subfigure}
\end{center}
\caption{Tail probabilities vs.~several variables.}
\label{fig:reg:probt}
\end{figure}

These results are also presented by plots in Figure \ref{fig:reg:probt}, which display the probability being classified to tail component versus various variables. The points indicated as triangle ($\triangle$) and square ($\square$) correspond to the fitted and empirical probabilities, respectively. Details on visualizing covariate influences through non-parametric approaches are discussed by \cite{FUNG2019MoEApplication}. As we can see from the figure, conditioned on any categories/ levels of any explanatory variables, the fitted and empirical probabilities match very well, reflecting the ability of the proposed regression model to capture well the covariates influence. The green dotted line is the overall empirical tail probability across all observations. The blue and red intervals are respectively the 95\% Wald-type and Efron bootstrap CIs presented in Section \ref{sec:asym}. The CIs generated by the two approaches reconcile well.

The model chooses a smaller set of variables which are important in explaining the body distributions $f(y_i;\exp\{\bm{\beta}_j^T\bm{x}_i,\phi_j)\})$, reflecting more heterogeneity among subgroup probabilities than within-subgroup average claim sizes:
\begin{itemize}
\item Driver's age: 3 categories -- $\{18-28,29-69,70+\}$.
\item Car cubism: 2 categories -- $\{0-15,16-18\}$.
\item Payment way: Cash payment results in a generally higher within-subgroup mean claim severity.
\item Region: 2 categories -- $\{\text{Region }4,\text{Others}\}$.
\item Other variables are excluded.
\end{itemize}

\begin{figure}[!h]
\begin{center}
\begin{subfigure}[h]{0.49\linewidth}
\includegraphics[width=\linewidth]{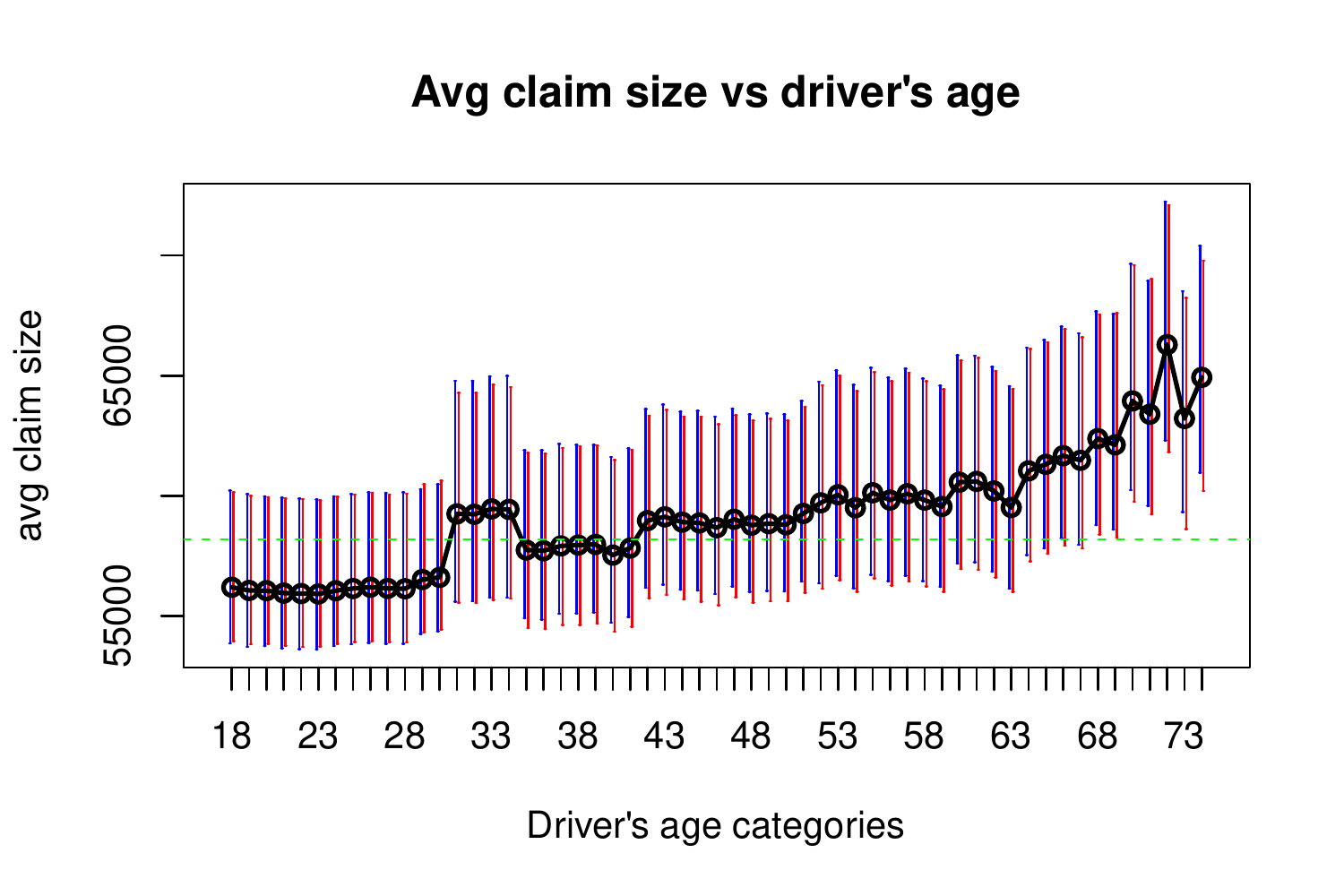}
\end{subfigure}
\begin{subfigure}[h]{0.49\linewidth}
\includegraphics[width=\linewidth]{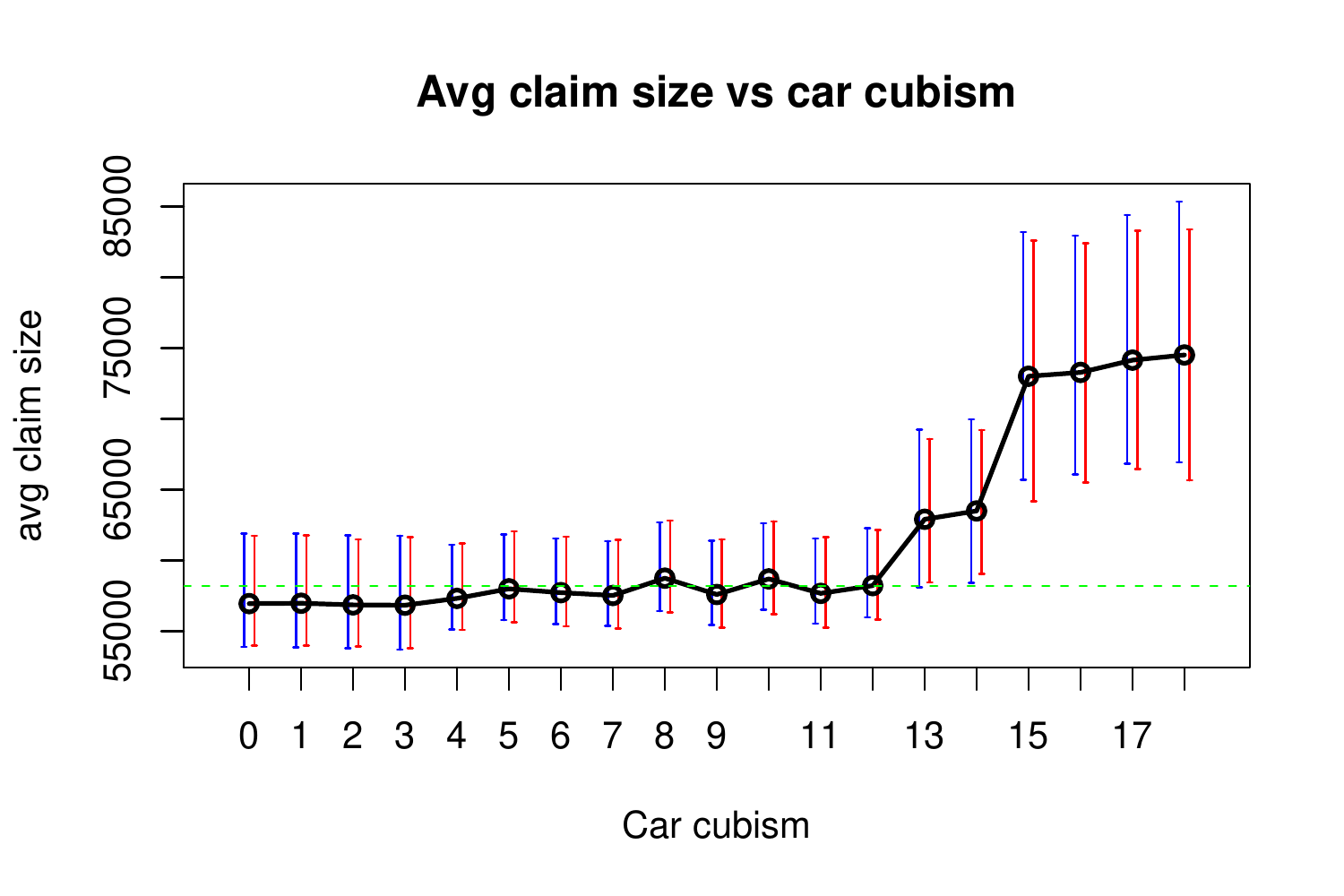}
\end{subfigure}
\begin{subfigure}[h]{0.49\linewidth}
\includegraphics[width=\linewidth]{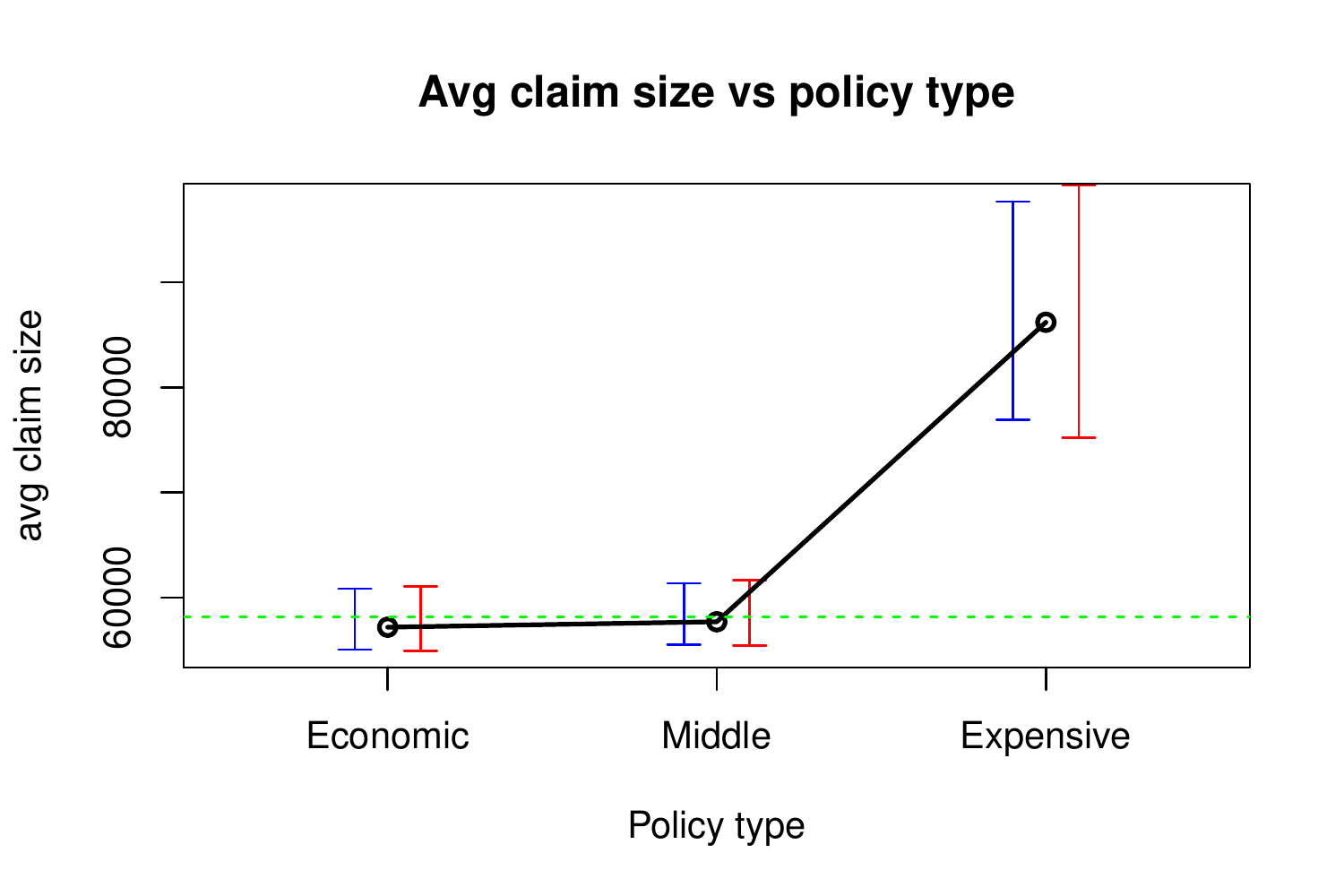}
\end{subfigure}
\begin{subfigure}[h]{0.49\linewidth}
\includegraphics[width=\linewidth]{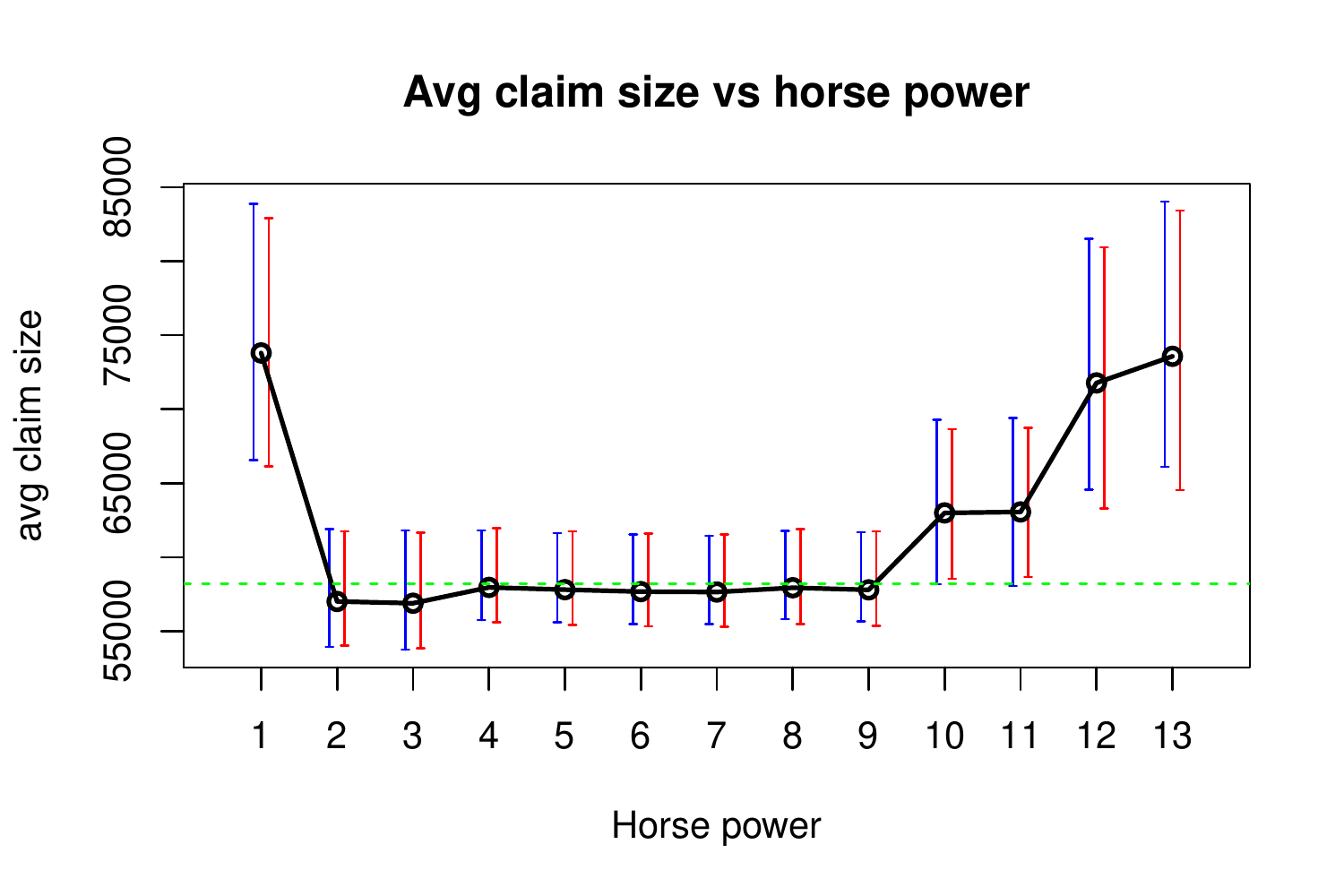}
\end{subfigure}
\begin{subfigure}[h]{0.49\linewidth}
\includegraphics[width=\linewidth]{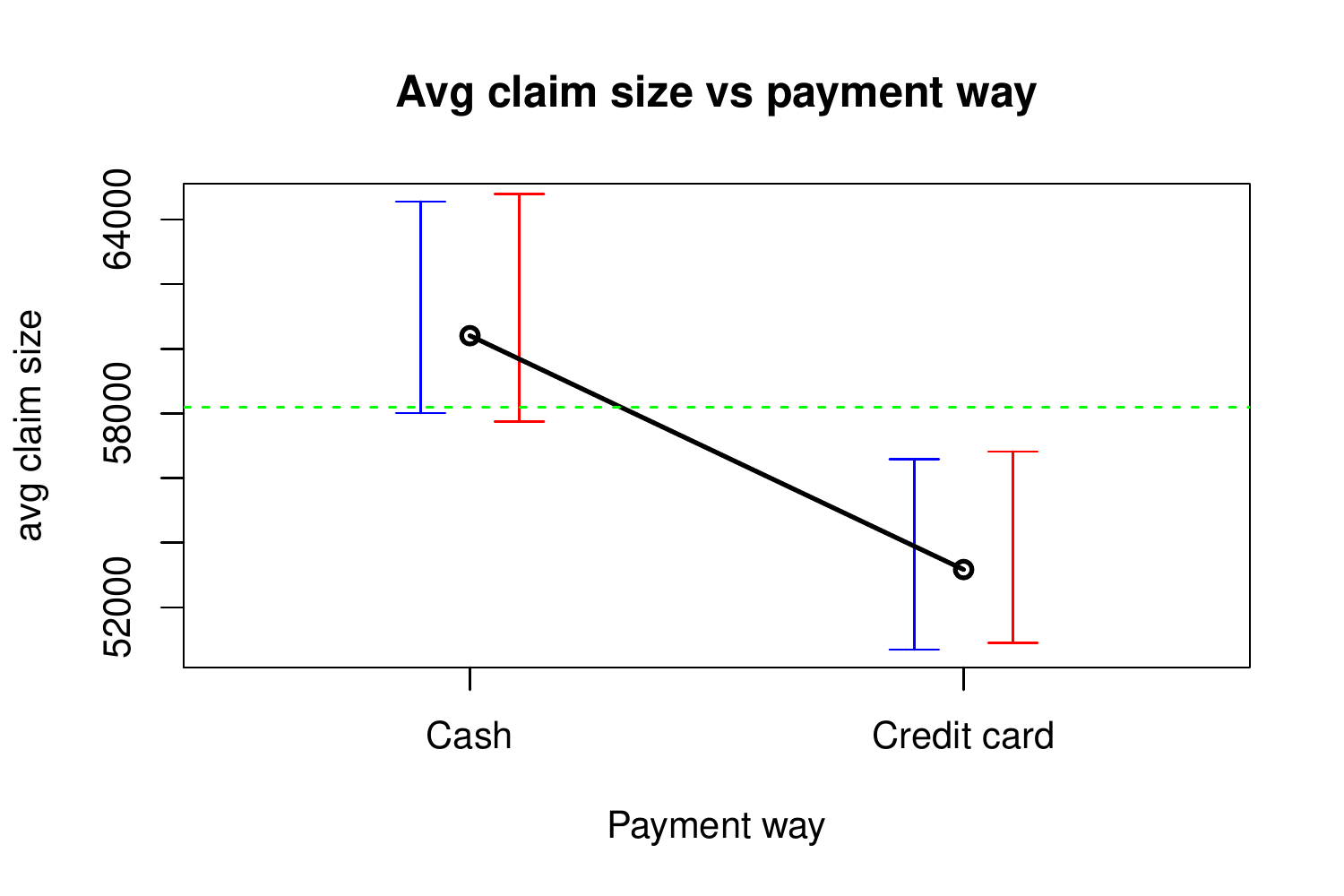}
\end{subfigure}
\begin{subfigure}[h]{0.49\linewidth}
\includegraphics[width=\linewidth]{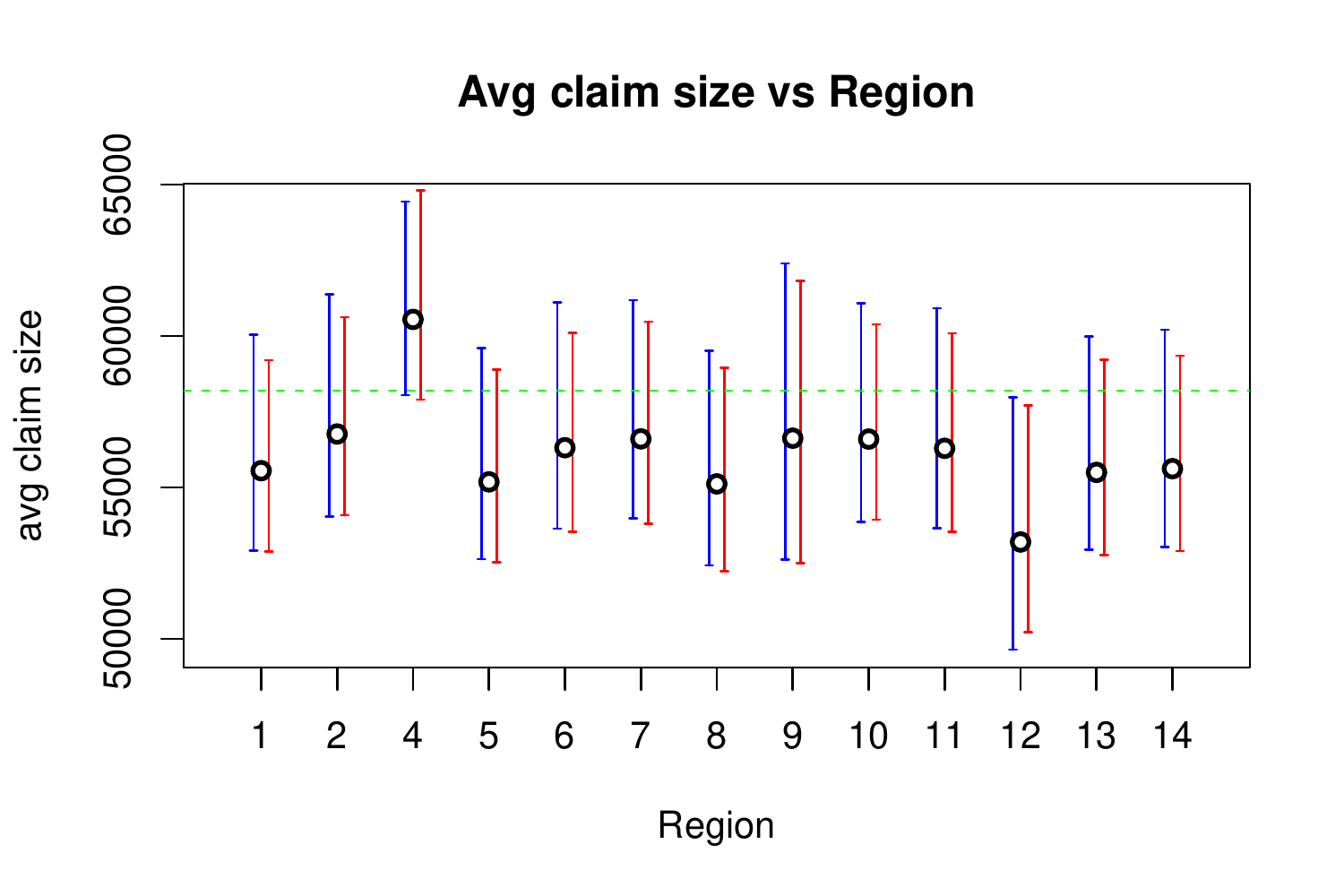}
\end{subfigure}
\end{center}
\caption{Average claim severities vs.~several variables.}
\label{fig:reg:mean}
\end{figure}

Finally, the fitted model suggests that none of the explanatory variables are significantly influential to the tail distribution $h(y_i;\theta,\exp\{\bm{\nu}^T\bm{x}_i\})$. Overall, the effects on various variables to the average claim severity are demonstrated in Figure \ref{fig:reg:mean}.

\subsection{Summary findings}
In this real data analysis, we get a deeper understanding on the influence of policyholder attributes to the claim severity distribution with highly complex structure including multimodality and tail-heaviness. Using the proposed mixture composite modeling framework embedded with a variable selection approach, we find that the explanatory variables most prominently impact the subgroup probabilities of the severity distribution, explaining the unobserved heterogeneity of policyholder risk profiles and/or claim types. Fewer variables explain well the body part of the distribution, reflecting relatively homogeneous claim severity distributions conditioned on the subgroups where each claim is belonging to. This finding is in contrast to many traditional regression models widely adopted in actuarial practice, including GLM and GAM, where regression links are set to capture the systematic effects in distributions instead of the subgroup heterogeneity. 

Further, we do not find any variables significantly influencing the tail-heaviness of the claim severity distribution, which may be the result of scarcity of large claims (only around 2,400 claims exceed the splicing threshold $\tau$) to allow for statistically significant covariates influence to the tail part. This empirically verifies the legitimacy of actuarial practice where covariates influence is often excluded in modeling large claims. In actuarial literature, we refer to \cite{laudage2019severity} who also refrains from incorporating regression in the tail part of their severity distribution.

\section{Discussions}

In this article, we considered a mixture composite regression model for addressing several challenges when modeling claim severities such as multimodality and tail-heaviness of claims, extending the framework of \cite{reynkens2017modelling} who considered the case without covariates. For variables selection, we proposed a group-fused regularization approach. Our covariates may influence the mixture probabilities, the body and the tail of the claim size distribution, in such way that
model interpretability is preserved. This approach enables regularization under multi-type variable settings. For this setup, we developed an asymptotic estimation theory which justified the efficiency of the proposed method. In particular, we showed that the method we presented is: (i) consistent in terms of covariate selection since, when the sample size goes to infinity, it will merge and shrink correctly regression coefficients across all modeling parts, and (ii) the parameters of the reduced model  are asymptotically normal. The implementation was illustrated by a real data application which involved fitting claim size data from a Greek automobile insurance company. Maximum likelihood estimation of the model parameters was achieved through a novel Generalized Expectation-Maximization algorithm that was demonstrated to perform well.

Furthermore, it is worth noting that instead of following a data driven approach for selecting the number of mixture components in the body area based on specification criteria, as is done herein, an interesting direction of further research would be to extend the framework to a non-parametric maximum likelihood estimation approach which can be utilized for automated selection of the number of mixture components.

Finally, it is worth noting that while the proposed composite model mitigates instabilities of tail index estimations inherited by finite mixture models, selection of the splicing threshold is often subjective. Therefore, it would be worth to explore alternative approaches for robust estimation of the tail index. One possible way is to modify the maximum likelihood approach for parameter estimation such that an observation with a larger claim severity has a higher relative importance in determining the model parameters. Another possible way is to explore models which bridge the gap between finite mixture models and composite models, and share the advantages of both model classes.

\bibliographystyle{abbrvnat}
\bibliography{reference}

\begin{appendices}
\section{Supplementary information in Section \ref{sec:asym}} \label{apx:asym}

The proof techniques are in general followed by the arguments of \cite{fan2001variable}, \cite{khalili2007variable} and \cite{khalili2010new}, while we here extend to the setting of the proposed group fused regularization method for multi-type feature selection.

\subsection{Assumptions on the penalty functions} \label{apx:asym_ass}

Denote $\mathcal{Z}_1=\{k:\|\bm{c}_{1k}^T\bm{\alpha}_0\|_2=0\}$, $\mathcal{Z}_2=\{k:\|\bm{c}_{2k}^T\bm{\beta}_0\|_2= 0\}$ and $\mathcal{Z}_3=\{k:|\bm{c}_{3k}^T\bm{\nu}_0|= 0\}$. We define the following quantities which are helpful for presenting the asymptotic results:
\begin{equation*}
b_{1n}=\underset{1\leq k\leq K_1}{\text{max}}\left\{p'_{1n}\left(\|\bm{c}_{1k}^T\bm{\alpha}_0\|_2;\lambda_{1kn}\right)/\sqrt{n}:k\notin\mathcal{Z}_1\right\},
\quad b_{1n}^*=\underset{1\leq k\leq K_1}{\text{max}}\left\{p''_{1n}\left(\|\bm{c}_{1k}^T\bm{\alpha}_0\|_2;\lambda_{1kn}\right)/n:k\notin\mathcal{Z}_1\right\},
\end{equation*}
\begin{equation*}
b_{2n}=\underset{1\leq k\leq K_2}{\text{max}}\left\{p'_{2n}\left(\|\bm{c}_{2k}^T\bm{\beta}_0\|_2;\lambda_{2kn}\right)/\sqrt{n}:k\notin\mathcal{Z}_2\right\},
\quad b_{2n}^*=\underset{1\leq k\leq K_2}{\text{max}}\left\{p''_{2n}\left(\|\bm{c}_{2k}^T\bm{\beta}_0\|_2;\lambda_{2kn}\right)/n:k\notin\mathcal{Z}_2\right\},
\end{equation*}
\begin{equation*}
b_{3n}=\underset{1\leq k\leq K_3}{\text{max}}\left\{p'_{3n}\left(|\bm{c}_{1k}^T\bm{\nu}_0|;\lambda_{3kn}\right)/\sqrt{n}:k\notin\mathcal{Z}_3\right\},
\quad b_{3n}^*=\underset{1\leq k\leq K_3}{\text{max}}\left\{p''_{3n}\left(|\bm{c}_{3k}^T\bm{\nu}_0|;\lambda_{3kn}\right)/n:k\notin\mathcal{Z}_3\right\},
\end{equation*}
where $p'_{ln}(\psi;\eta_n)$ and $p''_{ln}(\psi;\eta_n)$ are the 1st and 2nd derivatives of the penalty functions $p_{ln}(\psi;\eta_n)$ w.r.t.~$\psi$, for $l=1,2,3$. We require the following conditions on the penalty functions $p_{ln}(\psi;\eta_n)$, $l=1,2,3$:

\begin{enumerate}[font={\bfseries},label={H\arabic*.}]
\item For all $n$, $\lambda_{lkn}$ and $k=1,\ldots,K_l$, we have $p_{ln}(0;\lambda_{lkn})=0$;
 and $p_{ln}(\psi;\lambda_{lkn})$ is non-decreasing and twice differentiable in $\psi\in (0,\infty)$ except in a finite set.
\item $b_{ln}^*=o_P(1)$ as $n\rightarrow\infty$.
\item For $T_n=\{\psi:0<\psi\leq n^{-1/2}\log n\}$, we have $\underset{n\rightarrow\infty}{\lim}\underset{\psi\in T_n}{\inf}p'_{ln}(\psi;\lambda_{lkn})/\sqrt{n}=\infty$ for every $k\in\mathcal{Z}_l$.
\item $b_{ln}=O_P(1)$ as $n\rightarrow\infty$. 
\end{enumerate}

It is easy to check that LASSO and SCAD penalties with $\lambda_{lkn}=O_P(n^{-1/2})$ for any $k\notin\mathcal{Z}_l$ ($l=1,2,3$) both satisfy the aforementioned assumptions.

\subsection{Regularity conditions} \label{apx:asym_reg}
Let $h(\bm{v};\bm{\Phi})$ be the density function of $\bm{V}=(Y,\bm{x})$ with parameter space of $\bm{\Phi}\in\bm{\Omega}$. For a more concise presentation on the regularity conditions, we here write $\bm{\Phi}=(\psi_1,\ldots,\psi_Q)$ where $Q$ is the total number of parameters in the model. The regularity conditions are equivalent to \cite{khalili2010new} given by:

\begin{enumerate}[font={\bfseries},label={R\arabic*.}]
\item $h(\bm{v};\bm{\Phi})$ has common support in $\bm{v}$ for all $\bm{\Phi}\in\bm{\Omega}$, $h(\bm{v};\bm{\Phi})$ is identifiable in $\bm{\Phi}$ up to a permutation of mixture components.
\item $h(\bm{v};\bm{\Phi})$ admits third partial derivatives with respect to $\bm{\Phi}$ for each $\bm{\Phi}\in\bm{\Omega}$ and for almost all $\bm{v}$.
\item For all $j_1,j_2=1,\ldots,Q$, the first two derivatives of $h(\bm{v};\bm{\Phi})$ satisfy
\begin{equation}
E\left[\frac{\partial}{\partial\psi_{j_1}}\log h(\bm{v};\bm{\Phi})\right]=0;
\end{equation}
\begin{equation}
E\left[\frac{\partial}{\partial\psi_{j_1}}\log h(\bm{v};\bm{\Phi})\frac{\partial}{\partial\psi_{j_2}}\log h(\bm{v};\bm{\Phi})\right]=E\left[-\frac{\partial^2}{\partial\psi_{j_1}\partial\psi_{j_2}}\log h(\bm{v};\bm{\Phi})\right].
\end{equation}
\item The Fisher information matrix is finite and positive definite at $\bm{\Phi}=\bm{\Phi}_0$:
\begin{equation}
\mathcal{I}(\bm{\Phi})=E\left[\left(\frac{\partial}{\partial\bm{\Phi}}\log h(\bm{v};\bm{\Phi})\right)\left(\frac{\partial}{\partial\bm{\Phi}}\log h(\bm{v};\bm{\Phi})\right)^T\right].
\end{equation}
\item There exists a function $\mathcal{M}(\bm{v})$ such that
\begin{equation}
\hspace{-1cm}
\left|\frac{\partial}{\partial\psi_{j_1}}\log h(\bm{v};\bm{\Phi})\right|\leq \mathcal{M}(\bm{v}),\quad
\left|\frac{\partial^2}{\partial\psi_{j_1}\partial\psi_{j_2}}\log h(\bm{v};\bm{\Phi})\right|\leq \mathcal{M}(\bm{v}),\quad
\left|\frac{\partial^3}{\partial\psi_{j_1}\partial\psi_{j_2}\partial\psi_{j_3}}\log h(\bm{v};\bm{\Phi})\right|\leq \mathcal{M}(\bm{v}).
\end{equation}
\end{enumerate}

\subsection{Proof of Theorem \ref{thm:consistency}}
Let $r_n=n^{-1/2}(1+b_{1n}+b_{2n}+b_{3n})$. It suffices to show that for any $\epsilon>0$, there exists a large constant $M_{\epsilon}$ such that 
\begin{equation} \label{eq:apx:problim}
\underset{n\rightarrow\infty}{\lim}P\left\{\underset{\|\bm{u}\|_2=M_{\epsilon}}{\sup}\mathcal{F}_n(\bm{\Phi}_0+r_n\bm{u})<\mathcal{F}_n(\bm{\Phi}_0)\right\}\geq 1-\epsilon,
\end{equation}
where $\bm{u}=(\bm{u}_{\bm{\alpha}},\bm{u}_{\bm{\beta}},\bm{u}_{\bm{\phi}},\bm{u}_{\theta},\bm{u}_{\bm{\nu}})$ represents a vector of (the change of) all parameters. This corresponds to the existence of local maximizer $\hat{\bm{\Phi}}_n$ with $\|\hat{\bm{\Phi}}_n-\bm{\Phi}_0\|_2=O_P(n^{-1/2}(1+b_{1n}+b_{2n}+b_{3n}))$.

Denote $\mathcal{D}_n(\bm{u}):=\mathcal{F}_n(\bm{\Phi}_0+r_n\bm{u})-\mathcal{F}_n(\bm{\Phi}_0)$ which satisfies the following inequality
\begin{align}
\mathcal{D}_n(\bm{u})
&\leq\left[\mathcal{L}_n(\bm{\Phi}_0+r_n\bm{u})-\mathcal{L}_n(\bm{\Phi}_0)\right] \nonumber\\
&\quad-\sum_{k=m_1+1}^{K_1}\left[p_{1n}\left(\big\|\bm{c}_{1k}^T(\bm{\alpha}_0+r_n\bm{u}_{\bm{\alpha}})\big\|_2;\lambda_{1kn}\right)-p_{1n}\left(\big\|\bm{c}_{1k}^T\bm{\alpha}_0\big\|_2;\lambda_{1kn}\right)\right] \nonumber\\
&\quad-\sum_{k=m_2+1}^{K_2}\left[p_{2n}\left(\big\|\bm{c}_{2k}^T(\bm{\beta}_0+r_n\bm{u}_{\bm{\beta}})\big\|_2;\lambda_{2kn}\right)-p_{2n}\left(\big\|\bm{c}_{2k}^T\bm{\beta}_0\big\|_2;\lambda_{2kn}\right)\right] \nonumber\\
&\quad-\sum_{k=m_3+1}^{K_3}\left[p_{3n}\left(\big|\bm{c}_{3k}^T(\bm{\nu}_0+r_n\bm{u}_{\bm{\nu}})\big|;\lambda_{3kn}\right)-p_{3n}\left(\big|\bm{c}_{3k}^T\bm{\nu}_0\big|;\lambda_{3kn}\right)\right] \nonumber\\
&:=\mathcal{D}_{1n}(\bm{u})-\mathcal{D}_{2n}(\bm{u})-\mathcal{D}_{3n}(\bm{u})-\mathcal{D}_{4n}(\bm{u}),
\end{align}
where $\mathcal{D}_{1n}(\bm{u}), \mathcal{D}_{2n}(\bm{u}), \mathcal{D}_{3n}(\bm{u}), \mathcal{D}_{4n}(\bm{u})$ are the four corresponding terms expressed in $\mathcal{D}_n(\bm{u})$ above, and recall that $m_l$ is defined in Section \ref{sec:asym} as the number of linearly independent vectors in a reduced designed matrix. Taylor's expansion and triangular inequality yield
\begin{equation}
\mathcal{D}_{1n}(\bm{u})=n^{-1/2}(1+b_{1n}+b_{2n}+b_{3n})\mathcal{L}'_n(\bm{\Phi}_0)^T\bm{u}
-\frac{1}{2}(1+b_{1n}+b_{2n}+b_{3n})^2\bm{u}^T\mathcal{I}(\bm{\Phi}_0)\bm{u}\left(1+o_P(1)\right),
\end{equation}
and
\begin{align}
\left|\mathcal{D}_{2n}(\bm{u})\right|
&= \Bigg|\sum_{k=m_1+1}^{K_1}\bigg[p'_{1n}\left(\big\|\bm{c}_{1k}^T\bm{\alpha}_0\big\|_2;\lambda_{1kn}\right)\left(\|\bm{c}_{1k}^T\bm{\alpha}_0+r_n\bm{c}_{1k}^T\bm{u}_{\bm{\alpha}}\big\|_2-\|\bm{c}_{1k}^T\bm{\alpha}_0\big\|_2\right) \nonumber\\
&\hspace{5em}+\frac{1}{2}p''_{1n}\left(\big\|\bm{c}_{1k}^T\bm{\alpha}_0\big\|_2;\lambda_{1kn}\right)\left(\|\bm{c}_{1k}^T\bm{\alpha}_0+r_n\bm{c}_{1k}^T\bm{u}_{\bm{\alpha}}\big\|_2-\|\bm{c}_{1k}^T\bm{\alpha}_0\big\|_2\right)^2\left(1+o(1)\right)\bigg]\Bigg|\\
&\leq (K_1-m_1)b_{1n}(1+b_{1n}+b_{2n}+b_{3n})\underset{m_1+1\leq k\leq K_1}{\max}\|\bm{c}_{1k}^T\bm{u}_{\bm{\alpha}}\big\|_2 \nonumber\\
&\qquad +(K_1-m_1)\frac{1}{2}b_{1n}^*(1+b_{1n}+b_{2n}+b_{3n})^2\underset{m_1+1\leq k\leq K_1}{\max}\|\bm{c}_{1k}^T\bm{u}_{\bm{\alpha}}\big\|_2^2\\
&\leq (K_1-m_1)\left[b_{1n}(1+b_{1n}+b_{2n}+b_{3n})gC_{\max}\|\bm{u}_{\bm{\alpha}}\|_2+\frac{1}{2}b_{1n}^*(1+b_{1n}+b_{2n}+b_{3n})^2g^2C_{\max}^2\|\bm{u}_{\bm{\alpha}}\|_2^2\right],
\end{align}
where $C_{\max}$ is a fixed constant determined by the design matrix $\bm{C}_1$. By similar arguments,
\begin{equation}
\left|\mathcal{D}_{3n}(\bm{u})\right|
\leq (K_2-m_2)\left[b_{2n}(1+b_{1n}+b_{2n}+b_{3n})gC_{\max}\|\bm{u}_{\bm{\beta}}\|_2+\frac{1}{2}b_{2n}^*(1+b_{1n}+b_{2n}+b_{3n})^2g^2C_{\max}^2\|\bm{u}_{\bm{\beta}}\|_2^2\right],
\end{equation}
and
\begin{equation}
\left|\mathcal{D}_{4n}(\bm{u})\right|
\leq (K_3-m_3)\left[b_{3n}(1+b_{1n}+b_{2n}+b_{3n})gC_{\max}\|\bm{u}_{\bm{\nu}}\|_2+\frac{1}{2}b_{3n}^*(1+b_{1n}+b_{2n}+b_{3n})^2g^2C_{\max}^2\|\bm{u}_{\bm{\nu}}\|_2^2\right].
\end{equation}
Performing an order analysis while keeping in mind that penalty function conditions \textbf{H1}-\textbf{H2} and regularity conditions \textbf{R1}-\textbf{R5} hold, we know that $\mathcal{L}'_n(\bm{\Phi}_0)=\bm{O}_P(n^{1/2})$ and that
\begin{equation}
-\frac{1}{2}(1+b_{1n}+b_{2n}+b_{3n})^2\bm{u}^T\mathcal{I}(\bm{\Phi}_0)\bm{u}\left(1+o_P(1)\right)<0
\end{equation}
is the sole leading term in $\mathcal{D}_n(\bm{u})$ after choosing large enough $M_{\epsilon}$. This shows that Equation (\ref{eq:apx:problim}) holds and hence the result follows.

\subsection{Proof of Theorem \ref{thm:oracle}} \label{apx:asym:proof}

We first start with the following lemma:

\begin{lemma}\label{lm:apx:oracle}
Under the conditions of Theorem \ref{thm:oracle}, for any $\bm{\Phi}^*$ satisfying $\|\bm{\Phi}^*-\bm{\Phi}^*_0\|_2=O(n^{-1/2})$, we have $P\{\mathcal{F}_n^{*}(\{\bm{\Phi}^{*}_{\text{red}},\bm{\Phi}^{*}_{\text{ind}}\})<\mathcal{F}_n^{*}(\{\bm{0},\bm{\Phi}^{*}_{\text{ind}}\})\}\rightarrow 1$ as $n\rightarrow\infty$.
\end{lemma}

\begin{proof}
We first notice that
\begin{align}\nonumber
&\mathcal{F}_n^{*}(\{\bm{\Phi}^{*}_{\text{red}},\bm{\Phi}^{*}_{\text{ind}}\})-\mathcal{F}_n^{*}(\{\bm{0},\bm{\Phi}^{*}_{\text{ind}}\})
\\&\quad\label{eq:apx:loglik_diff}
=\left[\mathcal{L}_n^{*}(\{\bm{\Phi}^{*}_{\text{red}},\bm{\Phi}^{*}_{\text{ind}}\})-\mathcal{L}_n^{*}(\{\bm{0},\bm{\Phi}^{*}_{\text{ind}}\})\right]
-\left[\mathcal{P}_n^{*}(\{\bm{\Phi}^{*}_{\text{red}},\bm{\Phi}^{*}_{\text{ind}}\})-\mathcal{P}_n^{*}(\{\bm{0},\bm{\Phi}^{*}_{\text{ind}}\})\right].
\end{align}
Following the proof techniques by Theorem 2 of \cite{khalili2007variable} and Lemma 2 of \cite{khalili2010new}, by 
the mean value theorem and using condition \textbf{R5}, we have
\begin{equation}
\mathcal{L}_n^{*}(\{\bm{\Phi}^{*}_{\text{red}},\bm{\Phi}^{*}_{\text{ind}}\})-\mathcal{L}_n^{*}(\{\bm{0},\bm{\Phi}^{*}_{\text{ind}}\})=\left[\frac{\partial\mathcal{L}_n^{*}(\{\bm{\xi}_n,\bm{\Phi}^{*}_{\text{ind}}\})}{\partial\bm{\Phi}^{*}_{\text{red}}}\right]^T\bm{\Phi}^{*}_{\text{red}},
\end{equation}
for some $\bm{\xi}_n$ satisfying $\|\bm{\xi}_n\|_2\leq\|\bm{\Phi}^{*}_{\text{red}}\|_2=O(n^{-1/2})$ and
\begin{equation}
\left\|\frac{\partial\mathcal{L}_n^{*}(\{\bm{\xi}_n,\bm{\Phi}^{*}_{\text{ind}}\})}{\partial\bm{\Phi}^{*}_{\text{red}}}-\frac{\partial\mathcal{L}_n^{*}(\{\bm{0},\bm{\Phi}^{*}_{\text{ind}}\})}{\partial\bm{\Phi}^{*}_{\text{red}}}\right\|_2=O_P(n^{1/2}).
\end{equation}
Combining the above two equations, we have
\begin{equation}
\mathcal{L}_n^{*}(\{\bm{\Phi}^{*}_{\text{red}},\bm{\Phi}^{*}_{\text{ind}}\})-\mathcal{L}_n^{*}(\{\bm{0},\bm{\Phi}^{*}_{\text{ind}}\})=O_P(n^{1/2})\times\|\bm{\Phi}^{*}_{\text{red}}\|_2=O_P(1).
\end{equation}
On the other hand, for the penalty terms we have
\begin{align} \nonumber
&\mathcal{P}_n^{*}(\{\bm{\Phi}^{*}_{\text{red}},\bm{\Phi}^{*}_{\text{ind}}\})-\mathcal{P}_n^{*}(\{\bm{0},\bm{\Phi}^{*}_{\text{ind}}\})
\\&\quad=\sum_{k=1}^{K_1}\left[p_{1n}(\|\tilde{\bm{c}}_{1k}^T({\bm{\alpha}^*_{\text{red}}}^T,{\bm{\alpha}^*_{\text{ind}}}^T)^T\|_2;\lambda_{1kn})-p_{1n}(\|\tilde{\bm{c}}_{1k}^T(\bm{0}^T,{\bm{\alpha}^*_{\text{ind}}}^T)^T\|_2;\lambda_{1kn})\right]\nonumber\\
&\qquad +\sum_{k=1}^{K_2}\left[p_{2n}(\|\tilde{\bm{c}}_{2k}^T({\bm{\beta}^*_{\text{red}}}^T,{\bm{\beta}^*_{\text{ind}}}^T)^T\|_2;\lambda_{2kn})-p_{2n}(\|\tilde{\bm{c}}_{2k}^T(\bm{0}^T,{\bm{\beta}^*_{\text{ind}}}^T)^T\|_2;\lambda_{2kn})\right]\nonumber\\\label{eq:apx:penalty_diff_full}
&\qquad +\sum_{k=1}^{K_3}\left[p_{3n}(|\tilde{\bm{c}}_{3k}^T({\bm{\nu}^*_{\text{red}}}^T,{\bm{\nu}^*_{\text{ind}}}^T)^T|;\lambda_{3kn})-p_{3n}(|\tilde{\bm{c}}_{3k}^T(\bm{0}^T,{\bm{\nu}^*_{\text{ind}}}^T)^T|;\lambda_{3kn})\right].
\end{align}
We now perform an order analysis on the first term of the right hand side of the above equation as follows:
\begin{align} \label{eq:apx:penalty_diff}
&p_{1n}(\|\tilde{\bm{c}}_{1k}^T({\bm{\alpha}^*_{\text{red}}}^T,{\bm{\alpha}^*_{\text{ind}}}^T)^T\|_2;\lambda_{1kn})-p_{1n}(\|\tilde{\bm{c}}_{1k}^T(\bm{0}^T,{\bm{\alpha}^*_{\text{ind}}}^T)^T\|_2;\lambda_{1kn})\nonumber\\
&\quad=p'_{1n}(\|\tilde{\bm{c}}_{\text{ind},1k}^T{\bm{\alpha}^*_{\text{ind}}}+\bm{\xi}_n\|_2;\lambda_{1kn})\frac{\tilde{\bm{c}}_{\text{ind},1k}^T{\bm{\alpha}^*_{\text{ind}}}+\bm{\xi}_n}{\|\tilde{\bm{c}}_{\text{ind},1k}^T{\bm{\alpha}^*_{\text{ind}}}+\bm{\xi}_n\|_2}(\tilde{\bm{c}}_{\text{red},1k}^T{\bm{\alpha}^*_{\text{red}}})^T\nonumber\\
&\quad=\frac{p'_{1n}(\|\tilde{\bm{c}}_{\text{ind},1k}^T{\bm{\alpha}^*_{\text{ind}}}+\bm{\xi}_n\|_2;\lambda_{1kn})}{\sqrt{n}}\times n^{1/2}\times O_P(1)\times O_P(n^{-1/2})\nonumber\\
&\quad=:\frac{p'_{1n}(\psi_{n,k};\lambda_{1kn})}{\sqrt{n}}\times O_P(1),
\end{align}
where we have decomposed $\tilde{\bm{c}}_{1k}=(\tilde{\bm{c}}_{\text{red},1k},\tilde{\bm{c}}_{\text{ind},1k})$ such that $\tilde{\bm{c}}_{1k}^T({\bm{\alpha}^*_{\text{red}}}^T,{\bm{\alpha}^*_{\text{ind}}}^T)^T=\tilde{\bm{c}}_{\text{red},1k}^T{\bm{\alpha}^*_{\text{red}}}+\tilde{\bm{c}}_{\text{ind},1k}^T{\bm{\alpha}^*_{\text{ind}}}$, for some $\|\bm{\xi}_n\|_2\leq\|\tilde{\bm{c}}_{\text{red},1k}^T{\bm{\alpha}^*_{\text{red}}}\|_2=O_P(n^{-1/2})$. Note from Equation (\ref{eq:apx:penalty_diff}) that $\psi_{n,k}=O_P(n^{-1/2})$ for $k=1,\ldots,m_1$ and $\psi_{n,k}=O_P(1)$ for $k=m_1+1,\ldots,K_1$. As a result, under condition \textbf{H3}, Equation (\ref{eq:apx:penalty_diff}) has an order greater than $O_P(1)$ for $k=1,\ldots,m_1$, while, under condition \textbf{H4}, Equation (\ref{eq:apx:penalty_diff}) has an order equal to $O_P(1)$ for $k=m_1+1,\ldots,K_1$. Applying similar arguments to above for the second and third terms of the right hand side of Equation (\ref{eq:apx:penalty_diff_full}), by comparing the orders it is clear that
\begin{align}
-&\sum_{k=1}^{m_1}\left[p_{1n}(\|\tilde{\bm{c}}_{1k}^T({\bm{\alpha}^*_{\text{red}}}^T,{\bm{\alpha}^*_{\text{ind}}}^T)^T\|_2;\lambda_{1kn})-p_{1n}(\|\tilde{\bm{c}}_{1k}^T(\bm{0}^T,{\bm{\alpha}^*_{\text{ind}}}^T)^T\|_2;\lambda_{1kn})\right]\nonumber\\
&\quad -\sum_{k=1}^{m_2}\left[p_{2n}(\|\tilde{\bm{c}}_{2k}^T({\bm{\beta}^*_{\text{red}}}^T,{\bm{\beta}^*_{\text{ind}}}^T)^T\|_2;\lambda_{2kn})-p_{2n}(\|\tilde{\bm{c}}_{2k}^T(\bm{0}^T,{\bm{\beta}^*_{\text{ind}}}^T)^T\|_2;\lambda_{2kn})\right]\nonumber\\
&\quad -\sum_{k=1}^{m_3}\left[p_{3n}(|\tilde{\bm{c}}_{3k}^T({\bm{\nu}^*_{\text{red}}}^T,{\bm{\nu}^*_{\text{ind}}}^T)^T|;\lambda_{3kn})-p_{3n}(|\tilde{\bm{c}}_{3k}^T(\bm{0}^T,{\bm{\nu}^*_{\text{ind}}}^T)^T|;\lambda_{3kn})\right]<0
\end{align}
is the dominant term of $\mathcal{F}_n^{*}(\{\bm{\Phi}^{*}_{\text{red}},\bm{\Phi}^{*}_{\text{ind}}\})-\mathcal{F}_n^{*}(\{\bm{0},\bm{\Phi}^{*}_{\text{ind}}\})$ in Equation (\ref{eq:apx:loglik_diff}). Note that this term must be negative because, for example, $p_{1n}(\|\tilde{\bm{c}}_{1k}^T(\bm{0}^T,{\bm{\alpha}^*_{\text{ind}}}^T)^T\|_2;\lambda_{1kn})=0$ for $k=1,\ldots,m_1$ by construction. Hence, the result follows.
\end{proof}

\medskip

For part (a) of Theorem \ref{thm:oracle}, it suffices to show that $\mathcal{F}_n^{*}(\{\bm{\Phi}^{*}_{\text{red}},\bm{\Phi}^{*}_{\text{ind}}\})-\mathcal{F}_n^{*}(\{\bm{0},\hat{\bm{\Phi}}^{*}_{\text{ind},n}\})<0$ in probability for any $\bm{\Phi}^*:=(\bm{\Phi}^{*}_{\text{red}},\bm{\Phi}^{*}_{\text{ind}})$ such that $\|\bm{\Phi}^*-\bm{\Phi}^*_0\|_2=O_P(n^{-1/2})$. Following the arguments of Theorem 2 of \cite{khalili2007variable} and Theorem 3 of \cite{khalili2010new}, note that we have $\mathcal{F}_n^{*}(\{\bm{\Phi}^{*}_{\text{red}},\bm{\Phi}^{*}_{\text{ind}}\})-\mathcal{F}_n^{*}(\{\bm{0},\hat{\bm{\Phi}}^{*}_{\text{ind},n}\})\leq \mathcal{F}_n^{*}(\{\bm{\Phi}^{*}_{\text{red}},\bm{\Phi}^{*}_{\text{ind}}\})-\mathcal{F}_n^{*}(\{\bm{0},\bm{\Phi}^{*}_{\text{ind}}\})<0$ in probability using Lemma \ref{lm:apx:oracle}.

\medskip

For part (b) of Theorem \ref{thm:oracle}, the arguments are completely identical to Theorem 2 of \cite{khalili2007variable} and Theorem 3 of \cite{khalili2010new} after establishing the results of Lemma \ref{lm:apx:oracle}, so the proof is omitted.

\begin{remark} \label{apx:rmk:asym}
Note that the second derivative ${\mathcal{P}^*}''(\bm{\Phi}^*_{\text{ind},0})$ in part (b) of Theorem \ref{thm:oracle} is asymptotically negligible due to condition \textbf{H2}. On the other hand, we need $b_{ln}=o_P(1)$, which is stronger than condition \textbf{H4}, in order for the bias term ${\mathcal{P}^*}'(\bm{\Phi}^*_{\text{ind},0})$ to be negligible. For SCAD penalty, this is not a problem by choosing $\lambda_{lkn}=O_P(n^{-1/2})$. For LASSO penalty, we may use an adaptive approach by choosing $\lambda_{1kn}=o_P(n^{-1/2})/\|\bm{c}_{1k}^T\hat{\bm{\alpha}}_0\|_2$ (analogously for $\lambda_{2kn}$ and $\lambda_{3kn}$), where $\hat{\bm{\alpha}}_0$ is the MLE without penalty. Note that $\|\bm{c}_{1k}^T\hat{\bm{\alpha}}_0\|_2=O_P(n^{-1/2})$ for $k\in\mathcal{Z}_1$ and $\|\bm{c}_{1k}^T\hat{\bm{\alpha}}_0\|_2=O_P(1)$ for $k\notin\mathcal{Z}_1$. This allows that the conditions of $b_{ln}=o_P(1)$ and \textbf{H1}-\textbf{H4} still hold simultaneously to preserve both consistency and asymptotic normality.
\end{remark}

\end{appendices}

\end{document}


\title{Supplementary materials for ``Mixture composite regression model with multi-type feature selection"}

\author{Tsz Chai Fung 
\and George Tzougas\\
\and Mario V. W\"{u}thrich}

\maketitle
\section{Supplementary material for Section 6}
\subsection{Data description}
\begin{figure}[!h]
\begin{center}
\begin{subfigure}[h]{0.49\linewidth}
\includegraphics[width=\linewidth]{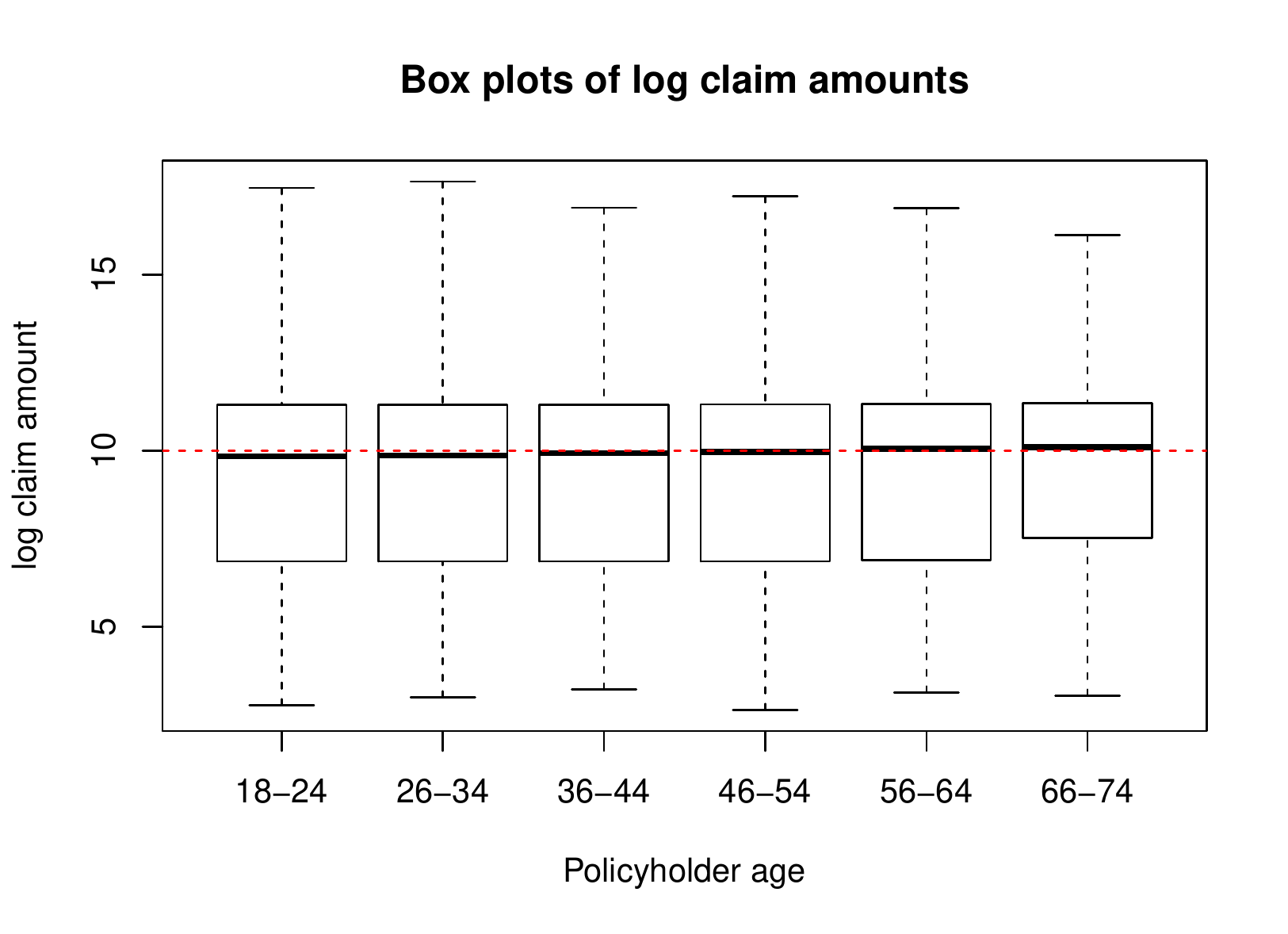}
\end{subfigure}
\begin{subfigure}[h]{0.49\linewidth}
\includegraphics[width=\linewidth]{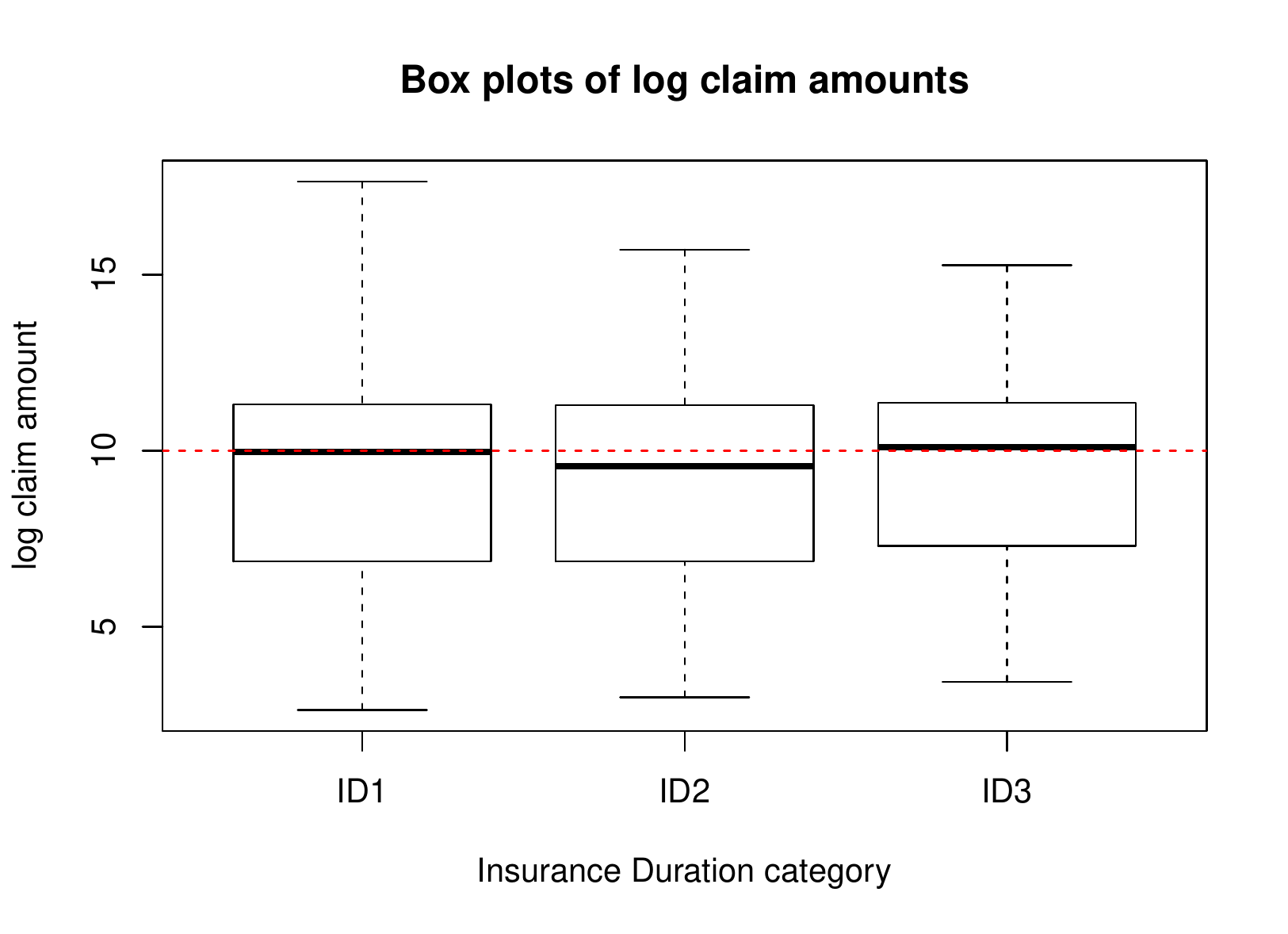}
\end{subfigure}
\begin{subfigure}[h]{0.49\linewidth}
\includegraphics[width=\linewidth]{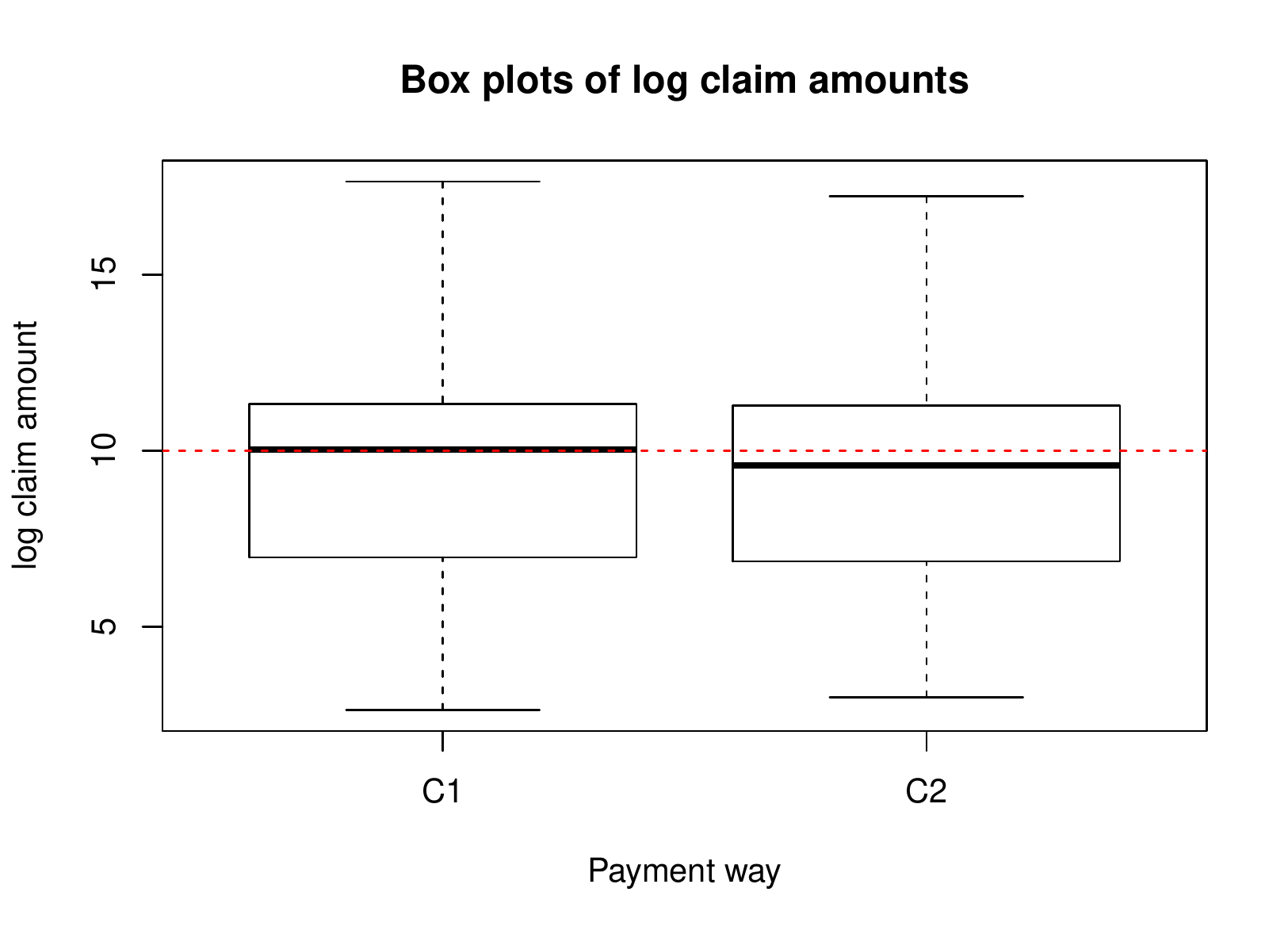}
\end{subfigure}
\begin{subfigure}[h]{0.49\linewidth}
\includegraphics[width=\linewidth]{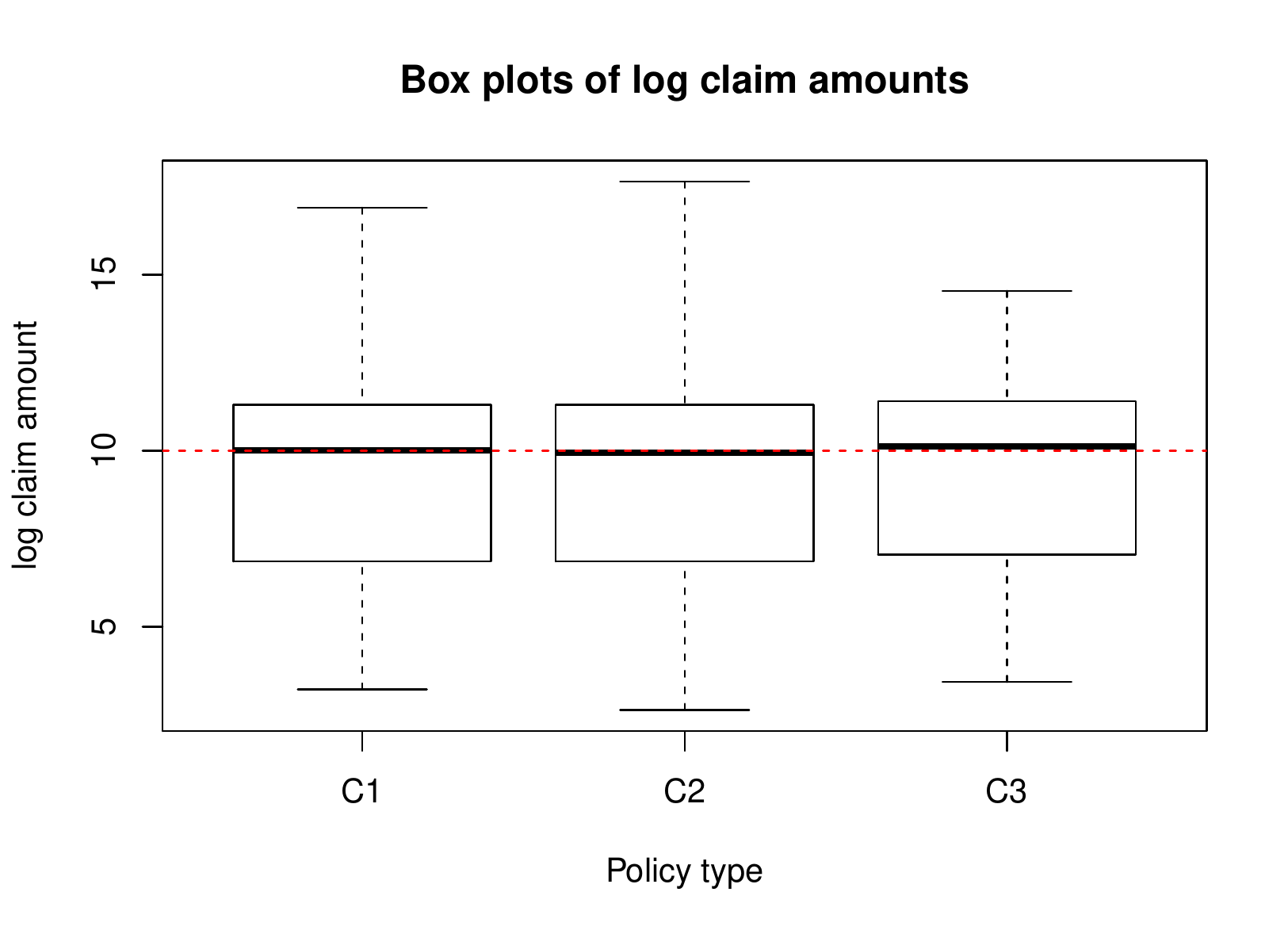}
\end{subfigure}
\end{center}
\caption{Box plots of the log claim amounts across various explanatory variables.}
\label{fig:boxplot1}
\end{figure}

The preliminary analysis of the Greek dataset revealed several challenges.  In this supplementary material we further examine the influence of the covariates on the claim sizes. Firstly, we see that the covariates influences to the log claim amounts are hardly observed solely based on the box plots in Figure \ref{fig:boxplot1} and none of the variables have big and predominant effects on the log claim amounts. This may reflect that, provided that an observation belongs to a particular node or cluster, the influences of variables to the claim costs are weak. Secondly, regarding the influence of explanatory variables to the body and tail of the distribution, as an illustration, we compute the average claim and average log-claim for each administrative region of Greece. The results presented in Table \ref{tab:mean_region} reveal some incoherences across regions. For example, a relatively large average claim is observed in Region 10, but the average log claim for that region is the smallest. This shows that geographical locations may have opposite impacts on the body and tail of the claim distribution, yet based solely on this table we cannot draw any conclusions on the significance of impacts.

\begin{table}[!h]
\centering
\begin{tabular}{lrr}
\multicolumn{1}{c}{Region} & \multicolumn{1}{c}{Avg claim} & \multicolumn{1}{c}{Avg log claim} \\ \hline
1 & 49,720 & 9.1134 \\
2 & 71,973 & 9.1386 \\
4 & 59,810 & 9.1273 \\
5 & 52,093 & 9.1736 \\
6 & 51,082 & 9.2407 \\
7 & 84,274 & 9.3044 \\
8 & 55,597 & 9.1629 \\
9 & 55,022 & 9.2472 \\
10 & 78,824 & 9.0902 \\
11 & 59,155 & 9.2569 \\
12 & 50,106 & 9.2252 \\
13 & 62,989 & 9.3309 \\
14 & 74,803 & 9.2348
\end{tabular}
\caption{Average claim and average log-claim for each administrative region of Greece.}
\label{tab:mean_region}
\end{table}

\subsection{Preliminary model comparisons}
In the research paper we want to compare the distributional fitting results among various models. The descriptions of several distributions omitted in the paper are presented as follows:
\begin{itemize}
    \item Weibull (WEI) distribution: The density of $y$ given the scale parameter $\lambda$ and shape parameter $k$ is given by
    \begin{equation} \label{eq:WEI}
    f(y)=\frac{k}{\lambda}\left(\frac{y}{\lambda}\right)^{k-1}\exp^{(y/\lambda)^k},\quad y>0.
    \end{equation}
    \item Generalized Gamma (GG) distribution: The density of $y$ given the parameters $\mu$, $\phi$ and $p$ is given by
    \begin{equation} \label{eq:GG}
    f(y)=\frac{p(\phi\mu)^{-1/\phi}}{\Gamma(1/(p\phi))}y^{1/\phi-1}e^{-(y/(\phi\mu))^p},\quad y>0.
    \end{equation}
    \item Generalized Pareto (GP) distribution: The density of $y$ given the parameters $\mu$, $\sigma$ and $\xi$ is given by
    \begin{equation} \label{eq:GP}
    f(y)=\frac{1}{\sigma}\left(1+\frac{\xi(y-\mu)}{\sigma}\right)^{-1/\xi-1},\quad y>0.
    \end{equation}
    \item Non-parametric maximum likelihood estimations (NPMLE) for Exponential mixtures: Assume that $y$ is distributed according to a mixed Exponential distribution with pdf given by
    \begin{equation} 
f\left( y\right) =\int\limits_{\mathbb{R}^{+}}\frac{e^{-\frac{x}{\lambda }}%
}{\lambda }F_{\Lambda _{0}}\left( d\lambda \right) ,
\end{equation}%
where $k\in \mathbb{N}$ and $\lambda $ is the observed value of a random
variable $\Lambda $ whose support is $\mathbb{R}^{+}$ and where $F_{\Lambda
_{0}}$ is the mixing distribution, called the structure function, which is
unknown but we assume that its support is in $\mathbb{R}^{+}$. Two kinds of models can be distinguished for the choice of the structure function, the parametric and nonparametric cases. The former consists of families where $F_{\Lambda _{0}}$ is approximated by some well known parametric distribution and the latter consists of choosing to estimate $F_{\Lambda _{0}}$ nonparametrically. 
Hereafter, let $\hat{F}_{\Lambda }$ be the NPMLE of $F_{\Lambda _{0}}$. 
In the setup we adopt, $\hat{F}_{\Lambda }$\ will be
attained for a discrete distribution function $F_{\Lambda _{0}}$
with a maximum number $\hat{q}$\ of support points that maximize
the log-likelihood. Then, the NPMLE of $f(y)$ is the mixture given by

\begin{equation} \label{eq:NPMLE}
f(y)=\sum\limits_{z=1}^{\hat{q}}\hat{p}_{z}\frac{e^{-\frac{y}{\lambda }}}{%
\lambda },
\end{equation}%
for $k\in \mathbb{N}$, where $p_{q}\geq 0$ and where $\sum\limits_{z=1}^{%
\hat{q}}\hat{p}_{z}=1.$ Equation (\ref{eq:NPMLE}) gives the pdf of
a finite Exponential mixture model. Note that a variant of the EM algorithm could be used towards the derivation of the NPMLE, see, for instance, \cite{seidel2006efficient}. For our portfolio, the NPMLE was found to have at most $\hat{q}=5$ support points leading to a five component exponential mixture model.
\end{itemize}

\section{Supplementary material for Section 5}

\subsection{Details of the GEM algorithm}
In this section, we will provide the missing details of the GEM algorithm presented in the paper.

\subsubsection{E-step}
Recall that in the $l^{\text{th}}$ iteration, the expectation of the complete data $\epsilon$-perturbed penalized log-likelihood is computed as
\begin{align}
Q_{\epsilon}(\bm{\Phi};\bm{y},\bm{x},\bm{\Phi}^{(l-1)})
&=\sum_{i=1}^n\sum_{j=1}^{g+1}z_{ij}^{(l)}\Big(\log\pi_j(\bm{x}_i;\bm{\alpha})+\log f(y_i;\exp\{\bm{\beta}_j^T\bm{x}_i\},\phi_j)1\{y_i\leq\tau\}\nonumber\\
&\hskip5em+\log\frac{h(y_i;\theta,\exp\{\bm{\nu}^T\bm{x}_i\})}{1-H(\tau;\theta,\exp\{\bm{\nu}^T\bm{x}_i\})}1\{y_i>\tau\}\Big)\nonumber\\
&\quad+\sum_{i=1}^n\sum_{j=1}^{g}k_{ij}^{(l)}z_{ij}^{(l)}\log \tilde{f}(\widehat{y_{ijk}'}^{(l)},\widehat{\log y_{ijk}'}^{(l)};\exp\{\bm{\beta}_j^T\bm{x}_i\},\phi_j)-\mathcal{P}_{\epsilon}(\bm{\Phi}).
\end{align}

Here, the updated quantities are given by the following formulas:

\begin{align} \label{eq:e:z}
z_{ij}^{(l)}
&=P(Z_{ij}=1|\bm{y},\bm{x},\bm{\Phi}^{(l-1)})\nonumber\\
&=\frac{\pi_j(\bm{x}_i;\bm{\alpha}^{(l-1)})}{h_Y(y_i;\bm{x}_i,\bm{\Phi}^{(l-1)})}\left[\frac{f(y_i;\exp\{\bm{\beta}_j^{T(l-1)}\bm{x}_i\},\phi_j^{(l-1)})}{F(\tau;\exp\{\bm{\beta}_j^{T(l-1)}\bm{x}_i\},\phi_j^{(l-1)})}1\{y_i\leq\tau\}+\frac{h(y_i;\theta^{(l-1)},\exp\{\bm{\nu}^{T(l-1)}\bm{x}_i\})}{1-H(\tau;\theta^{(l-1)},\exp\{\bm{\nu}^{T(l-1)}\bm{x}_i\})}1\{y_i>\tau\}\right],
\end{align}
\begin{equation}
k_{ij}^{(l)}=E[K_{ij}|\bm{y},\bm{x},\bm{\Phi}^{(l-1)}]
=\frac{1-F(\tau;\exp\{\bm{\beta}_j^{T(l-1)}\bm{x}_i\},\phi_j^{(l-1)})}{F(\tau;\exp\{\bm{\beta}_j^{T(l-1)}\bm{x}_i\},\phi_j^{(l-1)})},
\end{equation}
\begin{align}
\log \tilde{f}(\widehat{y_{ij}'}^{(l)},\widehat{\log y_{ij}'}^{(l)};\exp\{\bm{\beta}_j^T\bm{x}_i\},\phi_j)
&=\frac{1}{\phi_j}\left(\widehat{\log y_{ij}'}^{(l)}-\log\phi_j-\bm{\beta}_j^T\bm{x}_i-\widehat{y_{ij}'}^{(l)}\exp\{-\bm{\beta}_j^T\bm{x}_i\}\right)-\log\Gamma(\frac{1}{\phi_j}),
\end{align}
\begin{align}
\widehat{y_{ij}'}^{(l)}=E[Y_{ij}'|\bm{y},\bm{x},\bm{\Phi}^{(l-1)},Z_{ij}=1]=\frac{1-F(\tau;\exp\{\bm{\beta}_j^{T(l-1)}\bm{x}_i\},\phi_j^{(l-1)}/(1+\phi_j^{(l-1)}))}{1-F(\tau;\exp\{\bm{\beta}_j^{T(l-1)}\bm{x}_i\},\phi_j^{(l-1)})}\exp\{\bm{\beta}_j^{T(l-1)}\bm{x}_i\},
\end{align}
\begin{align} \label{eq:e:logy}
\widehat{\log y_{ij}'}^{(l)}=E[\log Y_{ij}'|\bm{y},\bm{x},\bm{\Phi}^{(l-1)},Z_{ij}=1].
\end{align}

\subsubsection{M-step}

The equations with regards to the derivatives under the IRLS approach in Equations (5.19), (5.20) and (5.22) of the paper are showcased as follows:

\begin{equation} \label{eq:irls_alpha_d}
\frac{\partial S^{(l)}(\bm{\alpha})}{\partial\bm{\alpha}_j}=\sum_{i=1}^{n}\Big[z_{ij}^{(l)}-\frac{\exp\{\bm{\alpha}_j^T\bm{x}_i\}}{\sum_{j'=1}^{g+1}\exp\{\bm{\alpha}_{j'}^T\bm{x}_i\}}\Big]\bm{x}_i
-\sum_{k=1}^{K_1}p_{1k}'\left(\big\|\bm{c}_{1k}^T\bm{\alpha}^{(l-1)}\big\|_{2,\epsilon};\lambda_{1kn}\right)\frac{\bm{c}_{1k}\bm{c}_{1k}^T}{\|\bm{c}_{1k}^T\bm{\alpha}^{(l-1)}\|_{2,\epsilon}}\bm{\alpha}_j,
\end{equation}
\begin{align} \label{eq:irls_alpha_dd}
\frac{\partial^2S^{(l)}(\bm{\alpha})}{\partial\bm{\alpha}_j\partial\bm{\alpha}_j^T}
&=\sum_{i=1}^{n}\frac{(\exp\{\bm{\alpha}_{j}^T\bm{x}_i\}-\sum_{j'=1}^{g+1}\exp\{\bm{\alpha}_{j'}^T\bm{x}_i\})\exp\{\bm{\alpha}_{j}^T\bm{x}_i\}}{(\sum_{j'=1}^{g+1}\exp\{\bm{\alpha}_{j'}^T\bm{x}_i\})^2}\bm{x}_i\bm{x}_i^T\nonumber\\
&\qquad-\sum_{k=1}^{K_1}p_{1k}'\left(\big\|\bm{c}_{1k}^T\bm{\alpha}^{(l-1)}\big\|_{2,\epsilon};\lambda_{1kn}\right)\frac{\bm{c}_{1k}\bm{c}_{1k}^T}{\|\bm{c}_{1k}^T\bm{\alpha}^{(l-1)}\|_{2,\epsilon}},
\end{align}

\begin{align} \label{eq:irls_beta_d}
\frac{\partial T^{(l)}(\bm{\beta},\bm{\phi}_j^{(l-1)})}{\partial\bm{\beta}_j}
&=\frac{1}{\phi_j^{(l-1)}}\sum_{i=1}^{n}z_{ij}^{(l)}\left[\left(y_i\exp\{-\bm{\beta}_j^T\bm{x}_i\}-1\right)+k_{ij}^{(l)}\left(\widehat{y_{ij}'}^{(l)}\exp\{-\bm{\beta}_j^T\bm{x}_i\}-1\right)\right]\bm{x}_i\nonumber\\
&\qquad -\sum_{k=1}^{K_2}p_{2k}'\left(\big\|\bm{c}_{2k}^T\bm{\beta}^{(l-1)}\big\|_{2,\epsilon};\lambda_{2kn}\right)\frac{\bm{c}_{2k}\bm{c}_{2k}^T}{\|\bm{c}_{2k}^T\bm{\beta}^{(l-1)}\|_{2,\epsilon}}\bm{\beta}_j,
\end{align}
\begin{align} \label{eq:irls_beta_dd}
\frac{\partial^2T^{(l)}(\bm{\beta},\bm{\phi}^{(l-1)})}{\partial\bm{\beta}_j\partial\bm{\beta}_j^T}
&=-\frac{1}{\phi_j^{(l-1)}}\sum_{i=1}^{n}z_{ij}^{(l)}\left[y_i+k_{ij}^{(l)}\widehat{y_{ij}'}^{(l)}\right]\exp\{-\bm{\beta}_j^T\bm{x}_i\}\bm{x}_i\bm{x}_i^T\nonumber\\
&\qquad -\sum_{k=1}^{K_2}p_{2k}'\left(\big\|\bm{c}_{2k}^T\bm{\beta}^{(l-1)}\big\|_{2,\epsilon};\lambda_{2kn}\right)\frac{\bm{c}_{2k}\bm{c}_{2k}^T}{\|\bm{c}_{2k}^T\bm{\beta}^{(l-1)}\|_{2,\epsilon}},
\end{align}

\begin{align} \label{eq:irls_nu_d}
\frac{\partial V^{(l)}(\theta^{(l-1)},\bm{\nu})}{\partial\bm{\nu}}
&=\sum_{i=1}^{n}\left[1-\log\left(\frac{y_i-\tau+\theta^{(l-1)}}{\theta^{(l-1)}}\right)\right]\exp\{\bm{\nu}^T\bm{x}_i\}1\{y_i>\tau\}\bm{x}_i\nonumber\\
&\qquad -\sum_{k=1}^{K_3}p_{3k}'\left(\big|\bm{c}_{3k}^T\bm{\nu}^{(l-1)}\big|_{\epsilon};\lambda_{3kn}\right)\frac{\bm{c}_{3k}\bm{c}_{3k}^T}{|\bm{c}_{3k}^T\bm{\nu}^{(l-1)}|_{\epsilon}}\bm{\nu},
\end{align}

\begin{align} \label{eq:irls_nu_dd}
\frac{\partial^2V^{(l)}(\theta^{(l-1)},\bm{\nu})}{\partial\bm{\nu}\partial\bm{\nu}^T}
&=-\sum_{i=1}^{n}\log\left(\frac{y_i-\tau+\theta^{(l-1)}}{\theta^{(l-1)}}\right)\exp\{\bm{\nu}^T\bm{x}_i\}1\{y_i>\tau\}\bm{x}_i\bm{x}_i^T\nonumber\\
&\qquad -\sum_{k=1}^{K_3}p_{3k}'\left(\big|\bm{c}_{3k}^T\bm{\nu}^{(l-1)}\big|_{\epsilon};\lambda_{3kn}\right)\frac{\bm{c}_{3k}\bm{c}_{3k}^T}{|\bm{c}_{3k}^T\bm{\nu}^{(l-1)}|_{\epsilon}}.
\end{align}

\newpage
\subsection{Automatic adjustment algorithm} \label{supp:aaa}
The automatic adjustment algorithm is outlined in Algorithm \ref{alg:aaa}, where in practice we find that setting the hyperparameters $\delta=10^{-5}$, $\eta=0.3$ and $\xi=1$ defined in the algorithm works well. 

\begin{algorithm}[H]
    \caption{Automatic adjustment algorithm}
  \begin{algorithmic}[1]
    \REQUIRE Reduce complexity of fitted model to achieve variable selection.
    \INPUT Fitted model parameters $\hat{\bm{\Phi}}$, observed data $(\bm{y},\bm{X})$, hyperparameters $(\bm{\lambda}_1,\bm{\lambda}_2,\bm{\lambda}_3)$, fitted latent variables ($\hat{z}_{ij}$, $\hat{k}_{ij}$, $\widehat{y_{ijk}'}$ and $\widehat{\log y_{ijk}'}$), initial error tolerance $\delta$, increment of error tolerance $\eta$ and log-likelihood error tolerance $\xi$.
    \OUTPUT The adjusted fitted model parameters $\hat{\bm{\Phi}}^{\text{adj}}$.
    \STATE $\text{run}_{\bm{\alpha}}\leftarrow \text{TRUE}$
    \WHILE {$\text{run}_{\bm{\alpha}}=\text{TRUE}$}
    	  \STATE $\hat{\bm{\alpha}}\leftarrow\hat{\bm{\alpha}}'$
      \FOR {$2\leq d_1< d_2 \leq D$}
      	\IF {$\|\hat{\bm{\alpha}}_{d_1}'-\hat{\bm{\alpha}}_{d_2}'\|_{2}<\delta$} $\hat{\bm{\alpha}}_{d_2}'\leftarrow\hat{\bm{\alpha}}_{d_1}'$ \quad \COMMENT{//Merge levels when the fitted parameters are close}
      	\ENDIF
      \ENDFOR
      \FOR {$2\leq d \leq D$}
      	\IF {$\|\hat{\bm{\alpha}}_{d}'\|_{2}<\delta$} $\hat{\bm{\alpha}}_{d}'\leftarrow 0$ \quad \COMMENT{//Force parameters to become zero if they are close to zero}
      	\ENDIF
      \ENDFOR
      \IF {$S(\hat{\bm{\alpha}}')-S(\hat{\bm{\alpha}})<-\xi$} $\text{run}_{\bm{\alpha}}\leftarrow \text{FALSE}$ \quad \COMMENT{//Stop when parameter adjustment decreases log-likelihood by a lot}
      \ENDIF
      \STATE $\delta\leftarrow\delta(1+\eta)$
    \ENDWHILE
    \STATE \textbf{return} $\hat{\bm{\alpha}}$ \quad \COMMENT{//Return the adjusted parameters}
    \STATE Similarly, repeat the above lines to update the adjusted values for $\hat{\bm{\beta}}$ and $\hat{\bm{\nu}}$, with $S$ replaced to $T$ and $V$ respectively, and $\bm{\alpha}$ replaced to $\bm{\beta}$ and $\bm{\nu}$ respectively.
  \end{algorithmic}
\label{alg:aaa}
\end{algorithm}

\bibliographystyle{abbrvnat}
\bibliography{reference}